\documentclass[11pt,document,nofootinbib,superscriptaddress,onecolumn,preprintnumbers,balancelastpage]{article}
\pdfoutput=1
\usepackage{jheppub}
\usepackage{graphicx}
\usepackage{epstopdf}
\usepackage{dcolumn}
\usepackage{bm}
\usepackage{hyperref}
\usepackage{booktabs}
\usepackage[dvipsnames]{xcolor}
\usepackage{amsmath}
\usepackage{cancel}
\usepackage{xpatch}
\usepackage{mathtools}
\definecolor{c1}{rgb}{0.368417, 0.506779, 0.709798}
\definecolor{c2}{rgb}{0.880722, 0.611041, 0.142051}
\definecolor{c3}{rgb}{0.560181, 0.691569, 0.194885}
\definecolor{c4}{rgb}{0.922526, 0.385626, 0.209179}
\definecolor{c5}{rgb}{0.528488, 0.470624, 0.701351}
\definecolor{c6}{rgb}{0.772079, 0.431554, 0.102387}
\definecolor{c7}{rgb}{0.363898, 0.618501, 0.782349}
\definecolor{turq}{rgb}{0.181,0.638,0.594}
\definecolor{pink}{rgb}{1.000,0.54,0.8}
\definecolor{purple}{RGB}{155,100,155}
\definecolor{gray}{RGB}{128,128,128}
\definecolor{lightBlue}{RGB}{148,179,229}
\definecolor{lightRed}{RGB}{213,157,131}
\definecolor{violet}{RGB}{130,121,173}
\definecolor{gold}{RGB}{255,191,0}

\def\beq{\begin{align}}
\def\eeq{\end{align}}
\newcommand{\gsim}{ \mathop{}_{\textstyle \sim}^{\textstyle >} }

\newcommand{\GEV}{ {\rm GeV} }

\def\GeV{{\rm GeV}}
\def\MeV{{\rm MeV}}

\makeatletter
\g@addto@macro\bfseries{\boldmath}
\makeatother

\title{
Dark Radiation Constraints on Heavy QCD Axions
}

\author[1,2]{David I. Dunsky}
\author[1,2]{Lawrence J. Hall}
\author[3]{Keisuke Harigaya}
\affiliation[1]{Department of Physics, University of California, Berkeley, California 94720, USA}
\affiliation[2]{Theoretical Physics Group, Lawrence Berkeley National Laboratory, Berkeley, California 94720, USA}
\affiliation[3]{Theoretical Physics Department, CERN, Geneva, Switzerland}

\abstract{The naturalness problem of PQ symmetry motivates study of the heavy QCD axion, with masses $m_a >$ 1 MeV generated at scales above the QCD scale, and low values of the PQ symmetry breaking scale, $f_a$. We compute the abundance of such axions in a model-independent way, assuming only that they freeze-out after reheating from inflation, and are not subsequently diluted by new physics. If these axions decay between neutrino decoupling and the last scatter era of the Cosmic Microwave Background (CMB), they dilute the neutrinos and their abundance is constrained by CMB measurements of the energy density in dark radiation, $N_{\rm eff}$. We accurately compute this bound using a numerical code to evolve the axion momentum distribution, including many key processes and effects previously ignored. We assume that the only relevant axion decays are to final states involving Standard Model particles. We determine regions of $(m_a, f_a)$ that will give a signal in $N_{\rm eff}$ at CMB Stage 4 experiments. We similarly compute the $N_{\rm eff}$ bound and CMB Stage 4 signal for heavy axions that can decay to light mirror photons. Finally, we compute the bounds on heavy axions with mass below 1 MeV that decay after the era of CMB last scatter, from their contribution to cold or hot dark matter or $N_{\rm eff}$ at this era. 
}

\date{\today}

\begin{document}
\maketitle
\flushbottom

\newpage

\section{Introduction}
The smallest dimensionless parameter of the Standard Model is the strong CP parameter $\bar{\theta} \lesssim 10^{-10}$. This small parameter can be understood as resulting from a discrete spacetime symmetry, $CP$ \cite{Nelson:1983zb, Barr:1984qx} or $P$ \cite{Beg:1978mt,Mohapatra:1978fy,Babu:1989rb}, or from a global Abelian Peccei-Quinn (PQ) symmetry \cite{Peccei:1977hh, Peccei:1977ur}. Imposing a PQ symmetry appears odd, as the symmetry is necessarily broken by the QCD anomaly; but it may appear as an approximate accidental symmetry at low energies~\cite{Georgi:1981pu}.
In such scenarios, the PQ symmetry is expected to be explicitly broken by higher-dimensional operators, typically preventing sufficient dynamical relaxation of $\bar{\theta}$ towards zero even if they are suppressed by powers of the Planck mass, $M_{\rm Pl}$.  

For example, if $\phi$ is the field that spontaneously breaks the PQ symmetry at scale $f_a$, in standard axion theories the interactions $\lambda_n \phi^{n+4}/M_{\rm Pl}^n$ must be suppressed to solve the strong CP problem,
\begin{align}
        \lambda_n \; < \; 10^{ (-46 + 10n)} \left( \frac{10^8 \, \GeV}{f_a} \right)^{4+n}.
        \label{eq:quality}
\end{align}
Even for the lowest values of $f_a$ allowed by observations, of order $10^8 \, \GeV$, operators of dimension 5 through 8 are highly constrained; and the problem gets worse rapidly as $f_a$ is increased. It is non-trivial to find theories where an accidental symmetry is protected to such high order, typically requiring significant additions to the theory. Even if one simply imposes the PQ symmetry as a classical symmetry, it may be broken by quantum gravity \cite{Harlow:2018tng, Banks:2010zn}, reintroducing the quality problem~\cite{Holman:1992us,Barr:1992qq,Kamionkowski:1992mf,Dine:1992vx}. Since the PQ symmetry must have a QCD anomaly, this PQ quality problem cannot be avoided by promoting it to a gauge symmetry.  
On the other hand, $P$ and $CP$ can be embedded in higher-dimensional gauge symmetries \cite{Choi:1992xp,Dine:1992ya} making them attractive avenues for the strong CP problem.

The severity of this PQ quality problem, shown in (\ref{eq:quality}), applies to the standard QCD axion, where its mass arises from non-perturbative QCD physics at the Fermi scale. It motivates theories with a heavy QCD axion, where the axion mass arises from physics at higher energy scales and is much larger. While the conventional QCD axion mass is less than the eV scale, these theories allow the axion mass to be larger than the MeV scale, removing constraints from stellar cooling and/or beam-dump experiments and allowing greatly reduced symmetry breaking scales.
Removing dimension 5 or lower operators by a gauge symmetry,%
\footnote{In models where the spontaneous PQ breaking occurs by hidden quark condensation~\cite{Choi:1985cb}, this only requires the removal of hidden quark mass terms.}
which may underlie the accidental PQ symmetry, the shift of $\bar{\theta}$ from zero by a dimension 6 operator is sufficiently small if
\begin{align}
        m_a \; \gsim \; \MeV \left( \frac{f_a}{3 \times 10^4 \, \GeV} \right)^2.
        \label{eq:d=6heavy}
\end{align}
In heavy axion theories there is a limit to how heavy the axion can be, and solving the quality problem then motivates low values of $f_a$.  

For $f_a$ of order ($10^4$ - $10^7$) GeV, a strong cosmological limit on the heavy QCD axion arises for masses in the (MeV - GeV) range. Such axions may decay after neutrino decoupling, diluting the neutrino abundance, as found for axion-like particles in \cite{Cadamuro:2011fd}. There is a powerful bound on this dark radiation from measurements of the Cosmic Microwave Background radiation (CMB) by the Planck Collaboration \cite{Planck:2018vyg}, $N_{\rm eff} = 2.96^{+0.34}_{-0.33}$ at 95\% c.l., and a significantly more accurate determination, with uncertainties smaller by almost an order of magnitude, is a key objective of CMB Stage 4 experiments \cite{CMB-S4:2016ple}. In this paper we study this bound on $(m_a,f_a)$ in a model-independent way, including many effects previously ignored, several arising from axion-meson interactions.
See Refs.~\cite{Chang:1993gm,Ferreira:2018vjj,Arias-Aragon:2020qtn,Arias-Aragon:2020shv,Ferreira:2020bpb,DEramo:2021psx,DEramo:2021lgb,DEramo:2021usm} for studies on axions that are light and stable and directly contribute to dark radiation, and Ref.~\cite{Hannestad:2003ye,Giare:2020vzo,Caloni:2022uya} for axions as hot dark matter.

There is a long history of theories with a heavy QCD axion, motivated by both the quality problem and the interest in reducing $f_a$ so that the axion is more visible.  One simple possibility is that QCD, or part of the gauge group in which it is embedded, becomes strong in the UV, so that there is an important contribution to the axion potential from short distance instantons \cite{Dimopoulos:1979pp, Tye:1981zy, Holdom:1982ex, Holdom:1985vx, Flynn:1987rs, Agrawal:2017ksf}. It is important that these instantons do not probe new CP violating phases, so that the new contribution to the potential aligns $\bar{\theta}$ to be sufficiently small.
The growth in the QCD coupling in the UV could also arise from extra spatial dimensions~\cite{Gherghetta:2020keg}.

Another simple way to make the QCD axion heavy is to introduce a $Z_2$ symmetry that transforms the Standard Model (SM) into a mirror sector. The $Z_2$ symmetry is spontaneously or softly broken so that the mirror electroweak scale is much larger than the SM weak scale, $v' \gg v$. The mirror quarks are then much heavier than the SM quarks, so that below the mirror quark masses the QCD$'$ coupling runs faster than the QCD coupling and confines at a scale much above the QCD scale, $\Lambda' \gg \Lambda$.  When introducing a PQ field that is $Z_2$ even, the resulting axion couples with the same strength to SM and mirror gluons, and hence its mass is larger than the conventional QCD axion by a factor of roughly $(\Lambda' /\Lambda)^2$.  The first implementations of this idea \cite{Rubakov:1997vp, Berezhiani:2000gh} used a Weinberg-Wilczek axion \cite{Weinberg:1977ma, Wilczek:1977pj}, with the PQ symmetry spontaneously broken by Higgs vevs.  In this case the axion decay constant $f_a$ is large, of order $v'$, and while these theories ameliorate the quality problem of (\ref{eq:quality}, \ref{eq:d=6heavy}),
solving the problem requires contrived arrangements. On the other hand, in the theories considered in~\cite{Fukuda:2015ana, Hook:2019qoh, Kelly:2020dda} using a KSVZ axion~\cite{Kim:1979if,Shifman:1979if}, the heavy QCD axion mass is
\begin{align}
        m_a \; \sim \; 100 \, \MeV \left(\frac{v'}{10^8 \, \GeV} \right)^{8/11} \left(\frac{10^4 \, \GeV}{f_a} \right)
        \label{eq:ma}
\end{align}
so that the quality problem is solved by taking $f_a \ll v'$. As always, one still needs to understand PQ in operators of dimension $\leq 4$ as an accidental consequence of gauge symmetries~\cite{Georgi:1981pu}. If the mirror photon in these theories is light, the CMB constraints from dark radiation are modified, which we also study.

The constraints from dark radiation on the axion mass and its couplings have been studied in the literature.
Refs.~\cite{Cadamuro:2011fd,Millea:2015qra,Depta:2020wmr} consider an axion-like particle that couples only to photons, and do not consider axion-gluon couplings. As we will see, the axion-gluon coupling, which leads to axion-meson couplings, helps to keep the axion in thermal equilibrium, so that the Boltzmann suppression of the axion abundance is more effective, relaxing the constraint on $(m_a,f_a)$. Ref.~\cite{Fukuda:2015ana} studies the mirror QCD case with an axion-mirror photon coupling but, while the decay of the axion into three pions is taken into account, axion-pion scattering and other axion-meson interactions are not included. 

This paper is organized as follows. Sec.~\ref{sec:EFTaboveQCD} shows the Lagrangian of the theory above and below the QCD scale. Sec.~\ref{sec:DR} describes the computation of the dark radiation abundance with a set of Boltzmann equations and shows the resultant $N_{\rm eff}$. Sec.~\ref{sec:mirrorPhoton} discusses the case with a mirror photon. Sec.~\ref{sec:lightaxion} shows a complementary constraint from dark matter overproduction for a sufficiently light axion that decays after the matter-radiation equality, where the $N_{\rm eff}$ constraint is not applicable. We conclude the paper with Sec.~\ref{sec:conclusions}.

\section{The Effective Theory Above and Below the QCD Scale}
\label{sec:EFTaboveQCD}
In this paper, we study a heavy axion with interactions above the QCD scale given by
\begin{align}
\label{eq:LagrangianAboveTQCD}
{\cal L} = \frac{1}{2}\partial^\mu a \partial_\mu a - \frac{1}{2}m_a^2 a^2 + \frac{g_3^2}{32\pi^2}\frac{a}{f_a} G_{\mu \nu} \tilde{G}^{\mu \nu} + \frac{e^2}{32\pi^2} \frac{E}{N} \frac{a}{f_a} F_{\mu \nu} \tilde{F}^{\mu \nu}.
\end{align}
We assume that the axion couplings with up, down, and electron axial currents are negligible. $E/N$ is the ratio of the electromagnetic and QCD anomalies of the PQ symmetry.  For a KSVZ axion with electrically neutral heavy quarks, $E/N =0$. For complete representations of $SU(5)$, $E/N = 8/3$. 

In models with a large axion mass from mirror QCD, the axion may also couple to mirror photons.  We include the effect of mirror photons on $N_{\rm eff}$ in Sec.~\ref{sec:mirrorPhoton}.
Note that mirror-photon effects are negligible if the PQ symmetry does not have an electromagnetic anomaly and the mirror quarks are much heavier than the mirror QCD scale, which causes the mixing between the axion and mirror mesons composed of mirror quarks to be small. The analysis of Sec.~\ref{sec:EFTaboveQCD} and \ref{sec:DR} is also applicable to the case where the mirror photon is massive and decouples by the QCD phase transition.

After an axial rotation to remove the coupling of the axions to gluons, below the QCD scale, the interactions of the axion with mesons and photons are described by 
the chiral Lagrangian, which, to leading order in $p^2$, is

\begin{align}
    \label{eq:chiralLagrangian}
    {\cal L}_{\rm chiral} &= 
    \frac{f_\pi^2}{4}{\rm Tr}\{D^\mu U^{\dagger} D_{\mu}U\} +
    \frac{f_\pi^2}{4}{\rm Tr}\{2B M_q U + \text{h.c.}\} + 
    \frac{1}{2}\frac{\partial^\mu a}{f_a}{\rm Tr}\{Q_A \lambda^a \} J{_\mu^a} \nonumber \\
    &+\frac{e^2}{32\pi^2}( \frac{E}{N} - \frac{8}{3} Q_u - \frac{2}{3} Q_d- \frac{2}{3} Q_s) \frac{a}{f_a}F_{\mu \nu} \tilde{F}^{\mu \nu},
\end{align}
where $f_\pi = 93$ MeV, $B$ is a strong interaction parameter of order the QCD scale, $J_\mu^a$ is the $SU(3)$ axial current given by
\begin{align}
        J{_\mu^a} &= \frac{i}{4}f_\pi^2{\rm Tr}\{\lambda^a (U D^\mu U^{\dagger} -U^{\dagger} D^\mu U)\},
\end{align}
and 
\begin{align}
    Q_A = 
    \begin{pmatrix}
    Q_u     &   & \\
    &      Q_d  & \\
    &       &  Q_s
\end{pmatrix}
\end{align}
is the quark charge matrix for the transformation that eliminates the axion-gluon coupling with ${\rm Tr} \;Q_A = 1$.

The meson nonet and quark mass matrices are given by
\begin{align}
    U &= \exp{i \frac{\sqrt{2}}{f_\pi}
    \begin{pmatrix}
    \frac{\pi_0}{\sqrt{2}} + \frac{\eta_8}{\sqrt{6}} + \frac{\eta_0}{\sqrt{3}}      &    \pi^+      &   K^+ \\
    \pi^{-}     &       -\frac{\pi_0}{\sqrt{2}} + \frac{\eta_8}{\sqrt{6}} + \frac{\eta_0}{\sqrt{3}}     &   K^0 \\
    K^{-}       &       \bar{K}^0       &   - 2\frac{\eta_8}{\sqrt{6}} + \frac{\eta_0}{\sqrt{3}}
    \end{pmatrix}}, \\
M_q  &=  
\begin{pmatrix}
m_u e^{i \frac{a}{f_a}Q_u}     &   & \\
&       m_d e^{i \frac{a}{f_a}Q_d}  & \\
&       &   m_s  e^{i \frac{a}{f_a}Q_s}  
\end{pmatrix}.
\end{align}
When calculating the effect of axion-meson scattering at energies below $\Lambda_{\rm QCD} \approx 150$ MeV, we limit ourselves to the relevant two-dimensional subspace, $SU(2)_L \times SU(2)_R \to SU(2)_V$, since the only active QCD degrees of freedom are pions. For axion masses above $\Lambda_{\rm QCD}$, the chiral perturbation based on the symmetry $SU(2)_L \times SU(2)_R \to SU(2)_V$ breaks down. In this case, the axion may decay to heavy mesons like $\eta$ and $K$, or for sufficiently large $m_a$, directly to gluons. Consequently, we use the results of~\cite{Aloni:2018vki} for the axion decay rate into mesons, gluons, and photons (including the enhancement in the axion-photon coupling from $a-\eta(\eta')$ mixing) for $m_a > m_\pi$. We discuss this further in the following section, but for now focus on the $SU(2)_V$ subspace which is sufficient for inferring axion-meson scattering in the early Universe.

In this two-dimensional subspace, we take the $Q$ matrix proportional to the identity in isospin space with $Q_u = Q_d = 1/2$, so that kinetic mixing between the axion and pion is absent, though mass mixing is present. Expanding out the chiral Lagrangian \eqref{eq:chiralLagrangian} generates the following axion-pion mass matrix and interactions
\begin{align}
    {\cal L}_{a,\pi} &=
    - \frac{1}{2}
    \begin{pmatrix}
        \pi_0 && a
    \end{pmatrix}
    \begin{pmatrix}
        B (m_u+m_d) &&  B (f_\pi/f_a) \; (Q_u m_u - Q_d m_d) \\
        B (f_\pi/f_a) \; (Q_u m_u - Q_d m_d) &&  m_a^2 + B (f_\pi/f_a)^2 \; (Q_u^2m_u+Q_d^2m_d)
    \end{pmatrix}
    \begin{pmatrix}
        \pi_0 \\ 
        a
    \end{pmatrix} \nonumber \\
    &+ \frac{B (Q_u m_u+Q_d m_d)}{24 f_\pi^2} (\pi_0^4 + 4 \pi_- \pi_+ \pi_0^2) 
    + \frac{B (Q_u m_u-Q_d m_d)}{6 f_\pi}\frac{a}{f_a} (\pi_0^3 + 2 \pi_0 \pi_+ \pi_-) \nonumber \\
    &+ \frac{1}{3f_\pi^2}(\pi_+ \pi_0 \, \partial^\mu \pi_- \partial_\mu \pi_0  +
    \pi_- \pi_0 \, \partial^\mu \pi_+ \partial_\mu \pi_0 - \pi_0^2 \, \partial^\mu \pi_- \partial_\mu \pi_+ -
    \pi_- \pi_+ \, \partial^\mu \pi_0 \partial_\mu \pi_0).
\end{align}
From the $\pi_0-a$ mass matrix in the limit $f_a \gg f_\pi$, we can identify $B = m_\pi^2/(m_u + m_d)$. Moreover, since we study axions heavier than the standard QCD axion, 
$m_a f_a \gg m_\pi f_\pi$,  and we can thus drop the second term 
in the bottom right entry of the mass matrix.

The $\pi_0 - a$ mass matrix is diagonalized by the rotation
\begin{align}
    \begin{pmatrix}
        \pi_0 \\ a
    \end{pmatrix}
    =
    \begin{pmatrix}
        \cos \theta && \sin \theta \\
       - \sin \theta && \cos \theta \\
    \end{pmatrix}
    \begin{pmatrix}
        \hat{\pi}_0 \\ \hat{a}
    \end{pmatrix}
\end{align}
with
\begin{align}
    \label{eq:axionPionMixingAngle}
    \tan 2\theta = \frac{f_\pi}{f_a}\frac{1-z}{1+z}\frac{1}{1-r^2} ,
\end{align}
where $r \equiv m_a/m_\pi$ and $z \equiv m_u/m_d \simeq 0.47$ \cite{ParticleDataGroup:2020ssz}. In terms of the mass eigenstates $\hat{\pi}_0, \hat{a}$, the interaction of the axion with three pions is described by
\begin{align}
    \label{eq:effectivePionLagrangian}
    {\cal L}_{\hat{a},\hat{\pi}} &\supset
    \frac{A}{f_a f_\pi}\frac{1}{1-r^2}\left[\partial^\mu \hat{a}(\pi_- \hat{\pi}_0 \, \partial_\mu \pi_+ + \pi_+ \hat{\pi}_0 \, \partial_\mu \pi_- 
    - 2 \pi_+ \pi_- \, \partial_\mu \hat{\pi}_0) + \frac{r^2}{4}m_\pi^2 \hat{a} (2 \hat{\pi}_0 \pi_- \pi_+ + \hat{\pi}_0^3) \right]
\end{align}
where $A = 
\frac{1}{3}(1-z)/(1+z) \simeq 0.12$.
For the remainder of this paper, we drop the hats and refer to the mass eigenstates as $a$ and $\pi_0$.  Note that we do not consider the case where $m_a$ and $m_\pi$ are so highly degenerate that the axion-pion mixing angle becomes of order unity since this does not occur as long as 
\begin{align}
    \left|\frac{m_\pi - m_a}{m_\pi}\right| \gtrsim \frac{f_\pi}{f_a} \simeq 10^{-4} \left(\frac{10^3 {\, \rm GeV}}{f_a}\right) \qquad
    \Big(\parbox{4.65cm}{\centering
    Axion-Pion \\ Non-Degeneracy Condition}\Big) \, ,
\end{align}
which is only violated for axions that are \textit{extremely} degenerate with pions. 

The coupling of the axion with photons is
\begin{align}
    {\cal L}_{a,\gamma} &=
    \frac{g_\gamma}{4}\frac{a}{f_a} F_{\mu \nu}\tilde{F}^{\mu \nu},
    \\
    & g_\gamma =
    \frac{e^2}{8\pi^2}\left( \frac{E}{N} - \frac{5}{3}  - \mathcal{F}_\theta (m_a)\right)
    \label{eq:g_gamma}
\end{align}
where $E/N$ arises from the UV contribution associated with the anomalies of the PQ symmetry,  $5/3$ from the axial rotation that removes the axion couplings to gluons, and $\mathcal{F}_\theta$ from axion-meson mixing. For $m_a \ll m_\eta$, $\mathcal{F}_\theta$ reduces to $2 \sin \theta f_a/f_\pi$, where $\theta$ is the axion-pion mixing angle, \eqref{eq:axionPionMixingAngle}. Further, the value of $g_{\gamma}$ in the massless axion case is recovered in the limit $m_a \ll m_\pi$ in which the term in parenthesis reduces to the standard result,  $\frac{E}{N} -\frac{2}{3}\frac{4+z}{1+z} \simeq \frac{E}{N} -2.03$ \cite{ParticleDataGroup:2020ssz}. For $m_a > m_\pi$, we extract $\mathcal{F}_\theta$ from the calculations of \cite{Aloni:2018vki} which include $a-\eta$ and $a-\eta'$ mixing. Note that previous considerations of heavy-axion cosmological constraints, except for~\cite{Kelly:2020dda}, neglect the effect of axion-meson mixing on $\mathcal{F}_\theta$ which leads to significantly different values of $g_\gamma$ for $m_a > m_\pi$.  Finally, we shall consider two reference values: $E/N = 8/3$, which is the case for the KSVZ model with $SU(5)$ unification, and $E/N=0$.

\section{Computation of the Dark Radiation Density}
\label{sec:DR}
In this section, we present the numerical results  of the Boltzmann equations describing the cosmological evolution of the heavy QCD axion in the early universe. Unlike the standard QCD axion which is light and very long-lived, the heavy QCD axion, with $m_a \gtrsim 1$ MeV and $f_a$ limited by the quality problem, is cosmologically unstable.  When the heavy axion decays during or after neutrino decoupling,  photons are subsequently heated relative to neutrinos, producing a potentially observable \textit{negative} contribution to $\Delta N_{\rm eff}$. Since the heavy axion can be out of thermal equilibrium around neutrino decoupling and contain a population of axions which have large momenta that decay (dangerously) late, it is crucial to track the momentum space distribution function of the axion, $f_a(\mathbf{p})$, throughout neutrino decoupling.

While neutrino decoupling occurs around the MeV era, the axion often decouples at earlier times and hence its abundance must be traced back to temperatures far above the MeV scale. The dominant interactions between the axion and thermal bath change as the universe cools. For temperatures above the QCD scale, axion-gluon scattering dominates and ensures the axions reach a thermal distribution for sufficiently high temperatures \cite{Salvio:2013iaa} as discussed in Sec. \ref{sec:gluonProduction}. For temperatures below the QCD scale, axion-meson scattering and axion-photon scattering can be effective as discussed in Sec. \ref{sec:boltzmannEqns}.

\subsection{Axion Initial Conditions}
\label{sec:gluonProduction}
At temperatures above the QCD phase transition temperature, $T_{\rm QCD} \approx 150$ MeV, axion-gluon interactions may be strong enough to keep the axion in thermal equilibrium. Generally, this process is UV dominated so that for sufficiently high temperatures, the axion reaches thermal equilibrium. As the universe cools, depending on $f_a$, the axion-gluon interactions may decouple. Likewise, below $T_{\rm QCD}$, axion-pion interactions may be strong enough to keep the axion in thermal equilibrium. In this subsection, we compute the temperature at which the axion scattering rate with strongly coupled particles decouples, $T_{\rm FO}$, and in the following subsection we use this freeze-out temperature to set the initial conditions of the axion distribution function of our Boltzmann code, which evolves the axion phase space distribution from temperatures where first-order chiral perturbation theory is valid, $T_{\chi \rm PT} \equiv 100$ MeV \cite{DiLuzio:2021vjd} to temperatures past neutrino and electron decoupling. We further discuss the region of parameter space in the ($m_a,f_a)$ plane where perturbation theory in gluon and pion descriptions breaks down at axion decoupling, and quantify the resulting uncertainty in the initial axion abundance.
\subsubsection{Equilibrium from Scatterings}

When the temperature $T$ is much greater than $m_a$ and $T_{\rm QCD}$, the axion-gluon interaction $a + g \leftrightarrow g + g$ dominates axion production. The thermally averaged rate of axion-gluon scattering is given by \cite{Salvio:2013iaa,DEramo:2021lgb}
\begin{align}
    \label{eq:agScatteringRate}
    \Gamma_{ag \leftrightarrow gg} \simeq \frac{16}{\pi} \left(\frac{g_3^2}{32\pi^2} \right)^2 \frac{T^3}{f_a^2}\mathcal{F}_g(T) \, ,
\end{align}
where $\mathcal{F}_g(T)$ is a temperature-dependent function that captures the axion production enhancement in the plasma from thermal gluon decays. $\mathcal{F}_g(T)$ is numerically computed in \cite{Salvio:2013iaa,DEramo:2021lgb}, and for $g_3 \lesssim 1$, takes the approximate analytic form $\mathcal{F}_g \approx 2 g_3^2 \ln 1.5/g_3$ \cite{Graf:2010tv,Salvio:2013iaa}.
\begin{figure}[tb]
    \centering
    \includegraphics[width=.95\textwidth]{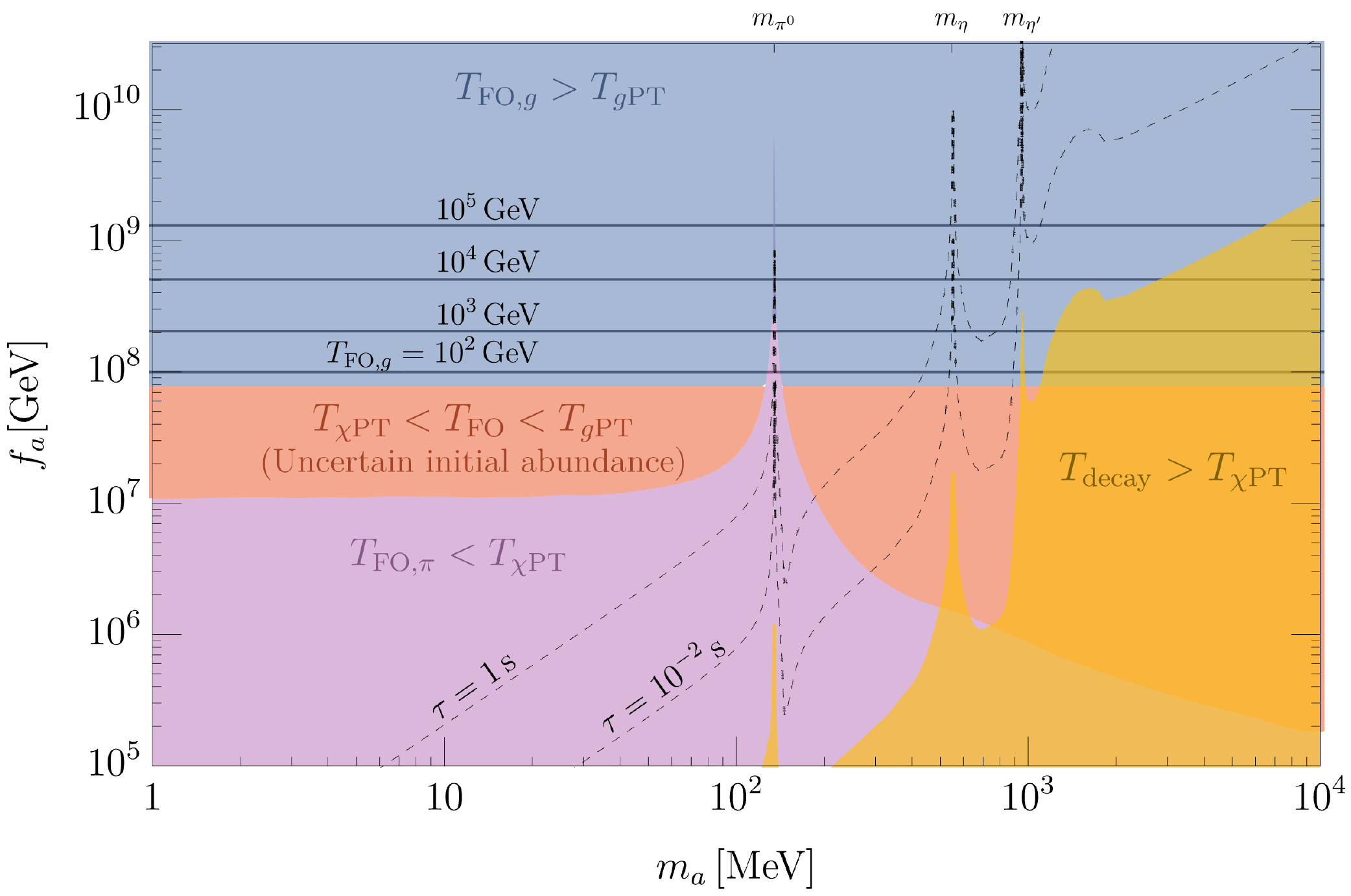} 
    \caption{Overview of the parameter space in the $m_a - f_a$ plane where axion interactions with mesons or gluons thermalize. In the purple region, scatterings with pions keep the axion in equilbrium below $T_{\chi \rm PT}$. In the yellow region, meson or gluon decays and inverse decays keep the axion in equilibrium below $T_{\chi \rm PT}$. In the blue region, scatterings with gluons once kept the axion in equilbrium. In this region, the horizontal contours denote the gluon freeze-out temperature, $T_{\rm FO,g}$. The red region indicates where the axions freeze out at temperatures above the validity of chiral perturbation theory ($T_{\chi \rm PT} \equiv 100$ MeV) but below the validity of gluon peturbation theory ($T_{\rm g PT} \equiv 2$ GeV). In this region, the freeze-out temperature is uncertain. In each region, we choose the appropriate initial condition of the axion distribution at $T_{\chi {\rm PT}}$; see Eq. \ref{eq:Tstar} for details.}
    \label{fig:overviewPlot}
\end{figure}

On the other hand, when $T \lesssim T_{\rm QCD}$, the interaction $a + \pi \leftrightarrow \pi + \pi$ dominates axion production. The thermally averaged rate of axion-pion scattering is given by
\begin{align}
        \label{eq:axionPionScatteringRate}
    \Gamma_{a \pi \leftrightarrow \pi \pi} = \frac{T^5}{f_a^2 f_\pi^2}\frac{A^2}{(1-r^2)^2} \mathcal{F}_\pi(m_a,T)
\end{align}
where $\mathcal{F}_\pi(m_a,T)$ is a temperature and axion mass dependent function that we  compute numerically in Appendix \ref{app:axionPionCollisionTerm}. For reference, Fig.~\ref{fig:fPi} shows $\mathcal{F}_\pi$ as a function of $m_\pi/T$ for a variety of $m_a$.

We define the axion decoupling temperature, $T_{\rm FO}$, when $\Gamma_{ag \leftrightarrow gg} = 3H(T_{\rm FO,g})$ for $T \gg T_{\rm QCD}$, or $\Gamma_{a \pi \leftrightarrow \pi \pi} = 3H(T_{\rm FO,\pi})$ for $T \ll T_{\rm QCD}$, where $H(T_{\rm FO})$ is the Hubble rate at $T_{\rm FO}$. Contours of $T_{{\rm FO},g}$ in the $m_a - f_a$ plane are shown by the horizontal lines in the blue-shaded region of Fig.~\ref{fig:overviewPlot}. Likewise, the purple region indicates where the axion-pion freeze-out temperature, $T_{\rm FO,\pi}$ is less than $T_{\chi \rm PT}$.
The estimation of $T_{\rm FO,g}$ based on Eq.~(\ref{eq:agScatteringRate}) breaks down if $m_a > T_{\rm FO,g}$, but we find that it anyway occurs in the red parameter region which possesses greater uncertainty:  
In the red-shaded region, $T_{\rm FO}$ occurs above the temperature at which chiral perturbation theory breaks down ($T_{\chi \rm PT}$) but below the temperature where the strong coupling constant, $g_3$, becomes non-perturbative ($T_{g \rm PT}$).
We take $T_{\chi \rm PT} \simeq 100$ MeV, the temperature  above which one-loop corrections in chiral perturbation theory become comparable to tree-level results \cite{DiLuzio:2021vjd}. Similarly, $T_{g \rm PT} \simeq 2$ GeV is conservatively associated with the energy scale below which $g_3$ becomes non-perturbative and one-loop corrections become comparable to tree-level results~\cite{Aloni:2018vki}.
In the red-shaded region, we cannot precisely determine $T_{\rm FO}$.
In Sec.~\ref{sec:initial}, we evaluate the uncertainty in $N_{\rm eff}$ arising from this uncertainty in $T_{\rm FO}$.

\subsubsection{Equilibrium from Decays and Inverse Decays}

Even when $T_{\chi \rm PT} < T_{\rm FO} < T_{g \rm PT}$, and hence the freeze-out temperature is uncertain, it is still possible to infer the axion abundance at $T_{\chi \rm PT}$ for sufficiently large $m_a$. Specifically, if the axion \textit{decay} rate is greater than Hubble at $T_{\chi \rm PT}$, then the axion possesses a thermal distribution at $T_{\chi \rm PT}$. For example, when $m_a \gg T$, the decay $a \rightarrow g + g$ can dominate over axion-gluon scattering when $T > T_{g \rm PT}$, or, for example, $a \rightarrow 3\pi$ can dominate at $T = T_{\chi \rm PT}$. The axion decay rate to strongly coupled particles is given by 
\begin{align}
    \label{eq:aMesonDecayRate}
    \Gamma_{a \rightarrow \text{QCD}} = \frac{m_a^3}{f_a^2} \mathcal{F}_{c}(m_a) \, ,
\end{align}
where $\mathcal{F}_c(m_a)$ is an axion mass dependent function that captures the variety of strongly coupled degrees of freedom the axion can decay to. We use $\mathcal{F}_c(m_a)$ as numerically computed in \cite{Aloni:2018vki}, which includes the following meson decay channels: $a \rightarrow 3\pi, \, \pi \pi \gamma, \, \eta \pi \pi, \, KK\pi, \, \eta' \pi \pi, $ $\, \rho \rho, \, \omega \omega, \, K^* \overline{K}^*,$ and $\phi \phi$. For $m_a \geq T_{g \rm PT}$, the axion to gluon decay rate dominates and $\mathcal{F}_c$ smoothly interpolates to the perturbative result \cite{Aloni:2018vki}
\begin{align}
\label{eq:agg}
   \mathcal{F}_c \simeq \frac{2}{\pi} \left(\frac{g_3^2}{32\pi^2}\right)^2 \left(1 + \frac{83 g_3^2}{16 \pi^2}\right) \, .
\end{align}
Last, the axion to photon decay rate is given by 
\begin{align}
    \label{eq:agammagamma}
    \Gamma_{a \rightarrow \gamma\gamma} = \frac{g_\gamma(m_a)^2}{64\pi} \frac{m_a^3}{f_a^2} \, ,
\end{align}
where $g_{\gamma}(m_a)$ is given in \eqref{eq:g_gamma}. Note that $g_{\gamma}(m_a)$ is a function of the axion mass due to the  effects of axion-meson mixing as encoded in the mixing function $\mathcal{F}_\theta(m_a)$.

We  define the axion decay temperature, $T_{\rm decay}$, when $\Gamma_{a \rightarrow \rm QCD} + \Gamma_{a \rightarrow \gamma \gamma} = 3H(T_{\rm decay})$. The yellow region of Fig.~\ref{fig:overviewPlot} shows the region where $T_{\rm decay} > T_{\chi \rm PT}$. In this region, the axion possesses a thermal distribution when we begin our Boltzmann code at $T_{\chi \rm PT}$, even if $T_{\rm FO}$ is uncertain.

Note that if the axion decays far before or after neutrino decoupling, the initial condition of the axion at $T_{\chi \rm PT}$ becomes insensitive to the calculation of $\Delta N_{\rm eff}$. In particular,
below the lower dashed line,
$\Gamma_a = \Gamma_{a \rightarrow \rm QCD} + \Gamma_{a \rightarrow \gamma \gamma}$, is so large that the axion always decays before the universe is $10^{-2}$ seconds old. In this region, the neutrinos are still strongly coupled to the thermal bath when the axion decays so that any effect to $\Delta N_{\rm eff}$ from the axion is erased, regardless of the initial axion abundance at $T_{\chi \rm PT}$. Likewise, 
above the upper dashed line.
$\Gamma_a$ is so small that the axion always decays after the universe is $1$ second old. In this region, neutrinos have long since decoupled from the thermal bath when the axion decays, leading to $\Delta N_{\rm eff} \ll -0.3$, which is already excluded by the observations of CMB~\cite{Planck:2018vyg}.

\subsubsection{Initial Condition and Its Uncertainty}
\label{sec:initial}
\begin{figure}[tb]
    \centering
    \includegraphics[width=.45\textwidth]{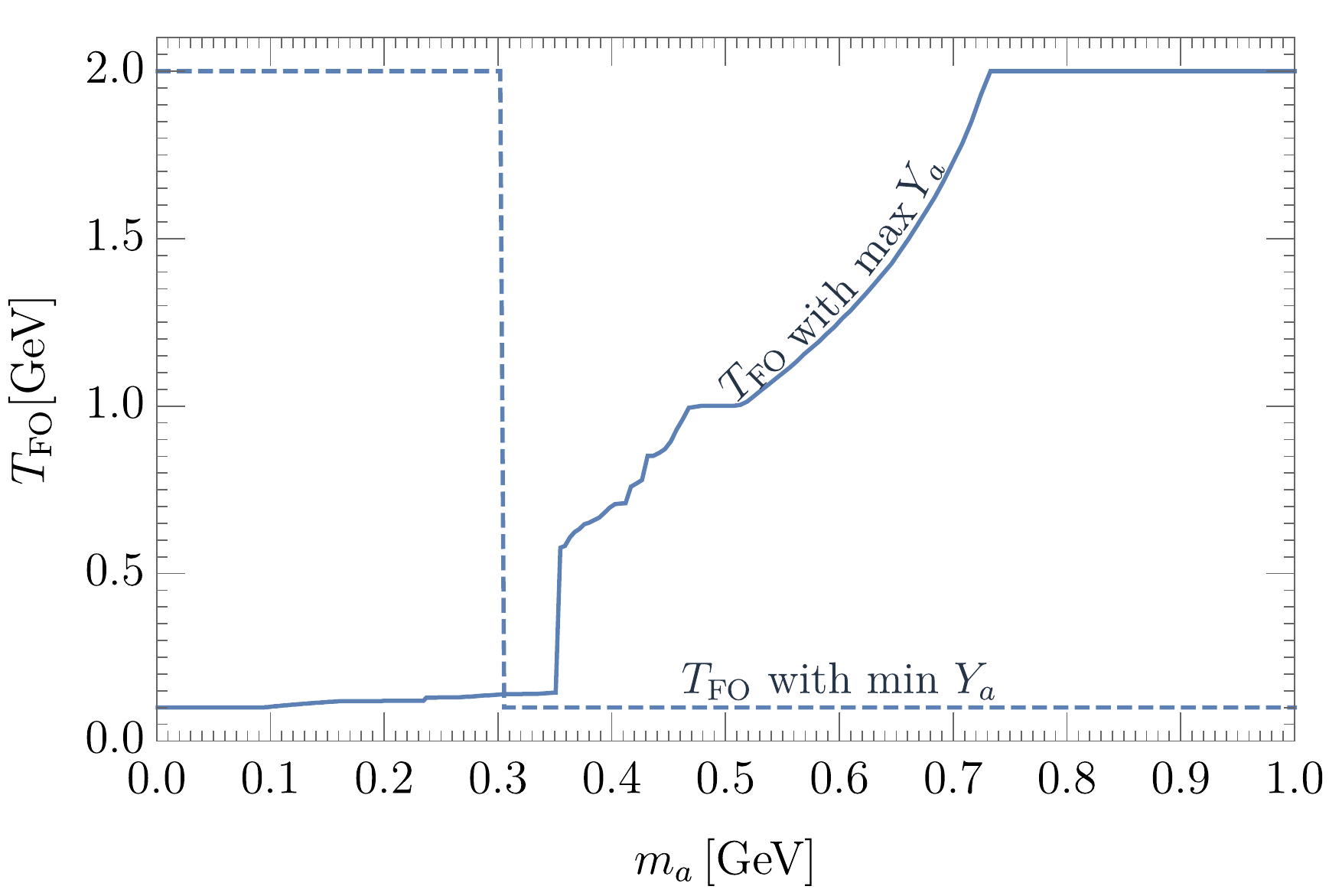} 
    \includegraphics[width=.46\textwidth]{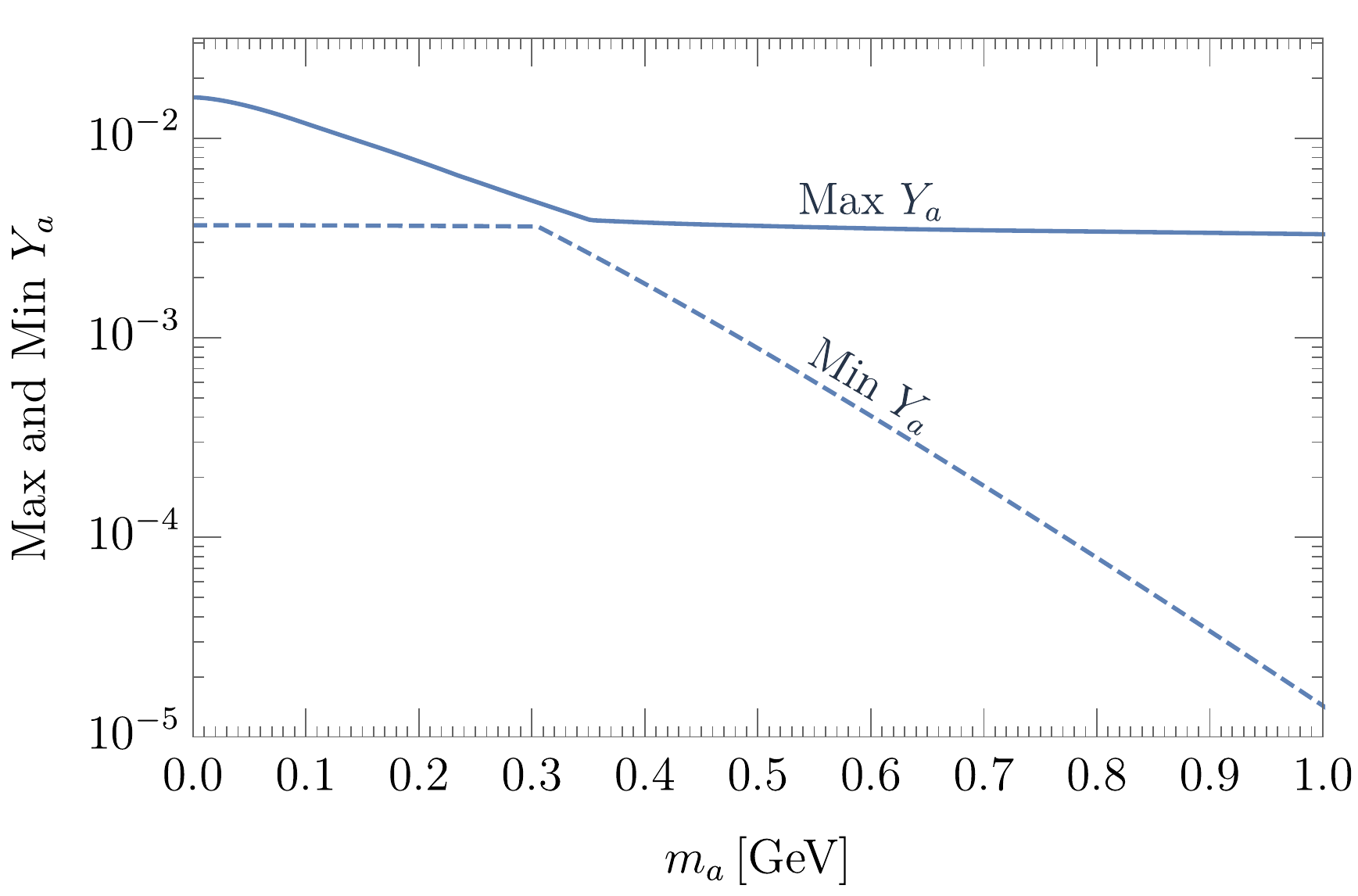} 
    \caption{\small When $T_{\chi \rm PT} < T_{\rm FO} < T_{\rm  g PT}$, the initial axion density at the start of the Boltzmann code is uncertain. The left panel bounds this uncertainty by showing the two extreme $T_{\rm FO} \in (T_{\chi \rm PT}, T_{\rm g PT})$ that give the largest (blue) and smallest (orange) axion densities at the starting temperature of the Boltzmann code.  The right panel shows the corresponding axion yields. The cross-over between the maximum and minimum $T_{\rm FO}$ around $0.3$ GeV arises from the balance between Boltzmann suppression and dilution by $g_*$ across the QCD phase transition.}
    \label{fig:TFOUncertainty}
\end{figure}

For $(m_a,f_a)$ in the purple or yellow regions of Fig. \ref{fig:overviewPlot}, the axion is in thermal equilbrium at $T_{\chi \rm PT}$. Consequently, in these regions, we take the initial distribution function of the axion at $T_{\chi{\rm PT}}$ to be a Bose-Einstein distribution of temperature $T_{\chi{\rm PT}}$. For sufficiently large $f_a$, however, the axion decouples from the bath at $T_{\rm FO}>T_{\chi{\rm PT}}$ (blue region), and we take the initial axion distribution to be a thermal one at $T_{\rm FO}$, red-shifted down to $T_{\chi{\rm PT}}$.
Inside the red-shaded region, however, $T_{\rm FO}$ is uncertain. 
Despite this uncertainty, we can still bound $\Delta N_{\rm eff}$ by running our axion Boltzmann code with both the maximum and minimum possible axion abundance at $T_{\chi \rm PT}$. We scan over possible freeze-out temperatures between $T_{\chi \rm PT}$ and $T_{g \rm PT}$ for all axion masses in the red-shaded region to determine the smallest and largest abundance at $T_{\chi \rm PT}$ as a function of $m_a$. In Fig.~\ref{fig:TFOUncertainty}, the left panel shows the $T_{\rm FO}$ that gives the maximum (blue) and minimum (orange) axion yield, $Y_a = n_a/s$, as a function of $m_a$, and the right panel shows these maximum and minimum yields. Note that the smallest possible axion number density is \textit{not} necessarily that of an axion in thermal equilibrium at $T_{\chi \rm PT}$ and the largest that of a frozen-out abundance at $T_{\rm FO} = T_{g \rm PT}$. This is because large changes in the degrees of freedom of the thermal bath between $T_{\chi \rm PT}$ and $T_{g \rm PT}$ can dilute the previously frozen-out axion. For all future plots, we show the results of $\Delta N_{\rm eff}$ arising from these two possible initial conditions.

In summary, we take the initial axion distribution function at $T_{\chi \rm PT}$ (time $t_{\chi \rm PT}$) to be
\begin{align}
    f_a(\mathbf{p}_a,t = t_{\chi \rm PT}) = \left(\exp{\dfrac{\sqrt{|\mathbf{p_*}|^2 + m_a^2}}{T_*}} - 1\right)^{-1}, \quad |\mathbf{p_*}| = |\mathbf{p}_a|\frac{ a(T_*)}{a(T_{\chi \rm PT})}
\end{align}
where $f_a(\mathbf{p}_a)$ is a Bose-Einstein distribution with momentum $\mathbf{p}_a$ and effective temperature $T_*$ given by 
\begin{align}
    T_* = 
    \begin{cases}
    T_{{\rm FO},g} \quad &\text{if } T_{\rm FO} \geq T_{g \rm PT} 
    \\
    T_{\chi \rm PT} \quad &\text{if } T_{\rm FO} \leq T_{\chi \rm PT} \text{ or } T_{\rm decay} \geq T_{\chi \rm PT} 
    \\
    \text{Max \& Min } T_{\rm FO} \text{ (see Fig.~\ref{fig:TFOUncertainty})} \quad &\text{if } T_{\chi \rm PT} < T_{\rm FO} < T_{g \rm PT}  \text{ and } T_{\rm decay} < T_{\chi \rm PT} 
    \end{cases}
    \label{eq:Tstar}
\end{align}
Note that if $T_{\rm RH} < T_*$ or if there is a source of dilution in the universe after the axion freezes-out, then the initial abundance of axions can be small and the bounds on $N_{\rm eff}$ discussed in this work are weakened.

\subsection{Axion Abundance Below \texorpdfstring{$\Lambda_{\rm QCD}$}{LQCD}: Boltzmann Equations}
\label{sec:boltzmannEqns}
Accurately capturing the effect of heavy axion decoupling and decay on our cosmology requires understanding the phase space evolution of the axion in the primordial thermal bath. The Boltzmann equation describing the evolution of the axion phase space density, $f_a(\mathbf{p}_a)$, is
\begin{align}
    \label{eq:axionPhaseSpace}
    \frac{\partial f_a}{\partial t} - p_a H \frac{\partial f_a}{\partial p_a} &= ( C_\gamma + C_{P} + C_\pi  + C_{\Gamma})(f_{a,\rm eq}-f_a) ,
\end{align}
where $p_a = |\mathbf{p}_a|$ is the magnitude of the axion momentum, $H$ is the Hubble expansion rate, and $C_\gamma$, $C_P$, $C_\pi$, and $C_\Gamma$ are the collision terms for axion-two photon scattering,  axion-Primakoff scattering, axion-pion scattering, and axion-meson decay, respectively. 

Generally, the collision term, $C$, corresponding to the axion interaction $a + A + B + ... \leftrightarrow I + J + ...$, is
\begin{align}
    \label{eq:collisionTerm}
    C(f_{a, \rm eq} - f_a) &=  \frac{1}{2E_a} \int 
    d\Pi_A d\Pi_B ... d\Pi_I ...d\Pi_J...
   S\, |\mathcal{M}|^2 \Lambda \, (2\pi)^4 \delta^4(p_a + p_A + p_B ...-p_I - p_J...)
\end{align}
where $E_a = \sqrt{p_a^2 + m_a^2}$ is the axion energy, $d\Pi = d^3 p/(2\pi)^3 2 E$ is the phase space measure per particle, $|M|^2$ is the matrix element of the interaction, $S = 1/m!$ is the symmetry factor for every $m$ identical particles in the initial or final states, and
\begin{align}
    \label{eq:psFactors}
    \Lambda &= 
    \left[(1 \pm f_A)(1 \pm f_B)...(1 \pm f_a)f_I f_J... - f_A f_B ...f_a(1 \pm f_I)(1 \pm f_J)...\right]
    \\
    &\simeq 
    (f_{a, \rm eq} - f_a)\exp(-E_A/T)\exp(-E_B/T)... \quad \text{(Kinetic equilibrium limit)} \nonumber
\end{align}
is the phase space density factor for all the incoming and outgoing particles interacting with the axion, where the plus sign refers to stimulated emission (boson) and the minus to Fermi blocking (fermion). The second line of Eq.~\eqref{eq:psFactors} shows $\Lambda$ in the limit  where $1 \pm f \simeq 1$ with $f = \exp(-E/T)$ the Boltzmann distribution, which is an excellent approximation for particles in kinetic equilibrium. 

In past literature, only the axion-two photon, $C_\gamma$, and the axion-Primakoff, $C_P$, collision terms have been considered in calculations involving the axion Boltzmann equation. Consequently, these terms have already been computed, and their values are \cite{Cadamuro:2010cz}
\begin{align}
    C_\gamma &\approx  \frac{m_a^2 - 4 m_\gamma^2}{m_a^2} \frac{m_a}{E_a}\left[1 + \frac{2T}{p_a} \log \frac{1 - e^{-(E_a + p_a)/2T}}{1 - e^{-(E_a - p_a)/2T}} \right] \Gamma_{a \rightarrow \gamma \gamma}
    \\
    C_P &\approx 
    \frac{\Gamma_{a \rightarrow \gamma \gamma}}{m_a^3}
    \sum_{i = e, \mu, \pi_{\pm}}n_i e_i^2 \log\left[1 + \frac{16 E_a^2(m_i + 3T)^2}{m_\gamma^2(m_i^2 + (m_i + 3T)^2)}\right].
\end{align}
Here, $\Gamma_{a \rightarrow \gamma \gamma}$ is the axion-to-two photon decay rate, \eqref{eq:agammagamma}, $m_\gamma$ is the photon plasma mass, and $n_i$ is the number density of the $i$th electromagnetically charged particle  of charge $e_i$ in the thermal bath. When the electron is relativistic, $m_\gamma \simeq e T/3$ \cite{weldon1982covariant}, but when the electron becomes non-relativistic at  $T \lesssim m_e$, $m_\gamma$ reduces to the classical plasma frequency of $m_\gamma \simeq e\sqrt{n_e/m_e}$, which is exponentially suppressed. For simplicity, we piecewise-connect the two regimes for $m_\gamma$ when they intersect, which occurs roughly at $T \simeq m_e/2$. 

The axion-pion scattering collision term for a massive axion has not been computed in the literature and we do so for the first time in Appendix \ref{app:axionPionCollisionTerm}. $C_\pi$ takes the form
\begin{align}
    C_\pi = \left(\frac{A}{f_a f_\pi}\frac{1}{1-r^2}\right)^2 \frac{T^6}{2 E_a} \times \mathcal{F}_{\rm PS}\left(\frac{m_a}{T}, \frac{p_a}{T}\right) \, ,
\end{align}
where  $A = \frac{1}{3}(1-z)/(1+z) \simeq 0.12$ and $r \equiv m_a/m_\pi$ as before.
The function $\mathcal{F}_{\rm PS}$ contains the phase space integration over the axion-pion scattering matrix. In Appendix \ref{app:axionPionCollisionTerm}, we determine this axion-pion scattering matrix, numerically perform the phase integration, and show how $C_\pi$ agrees with the massless axion result found in literature \cite{Hannestad_2005,DiLuzio:2021vjd}. Note that integration of $C_\pi$ over the axion phase space defines $\Gamma_{a \pi \leftrightarrow a \pi}$ given in Eq.~\eqref{eq:axionPionScatteringRate}.
\begin{figure}[tb]
    \centering
    \includegraphics[width=.46\textwidth]{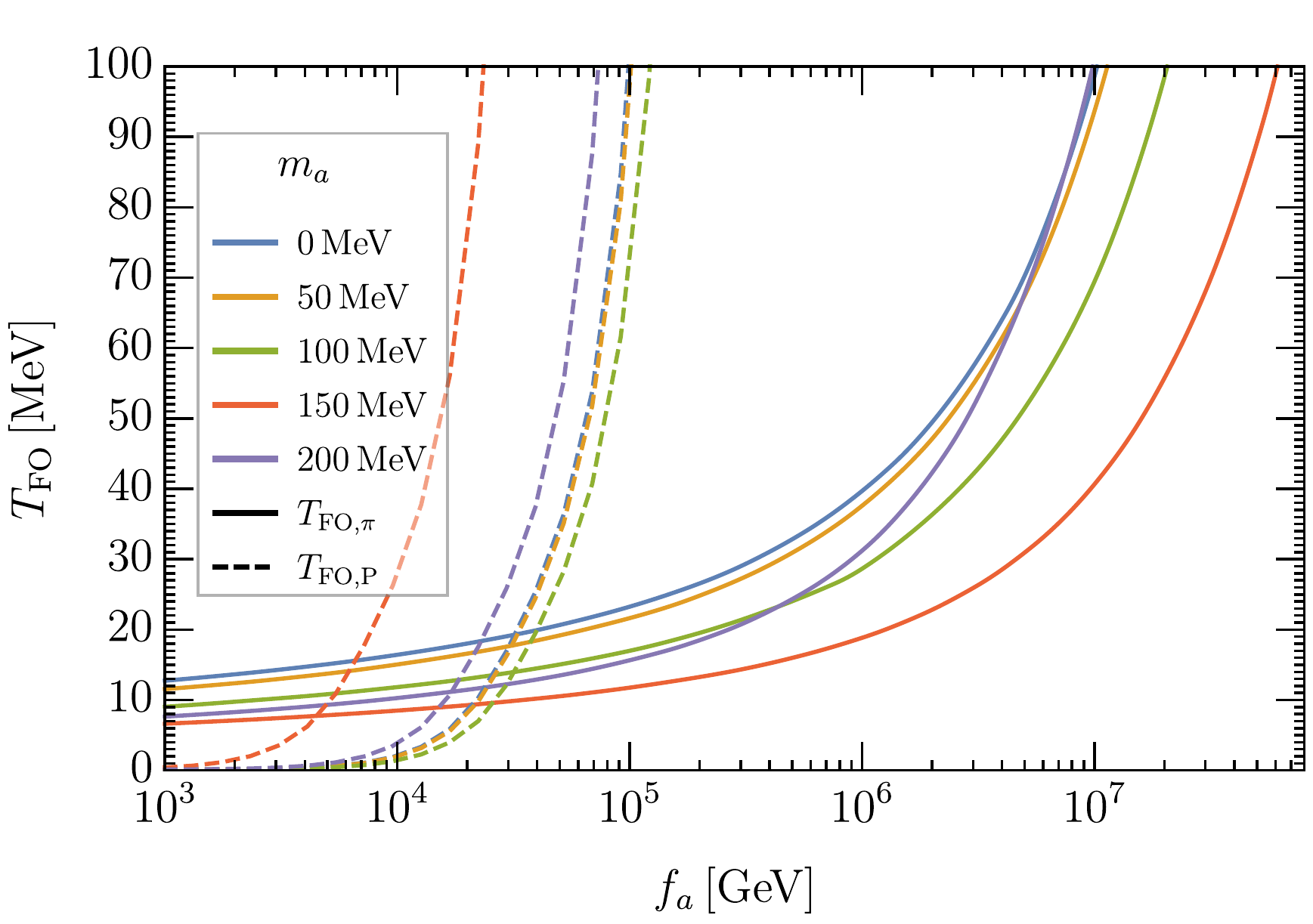} 
    \includegraphics[width=.475\textwidth]{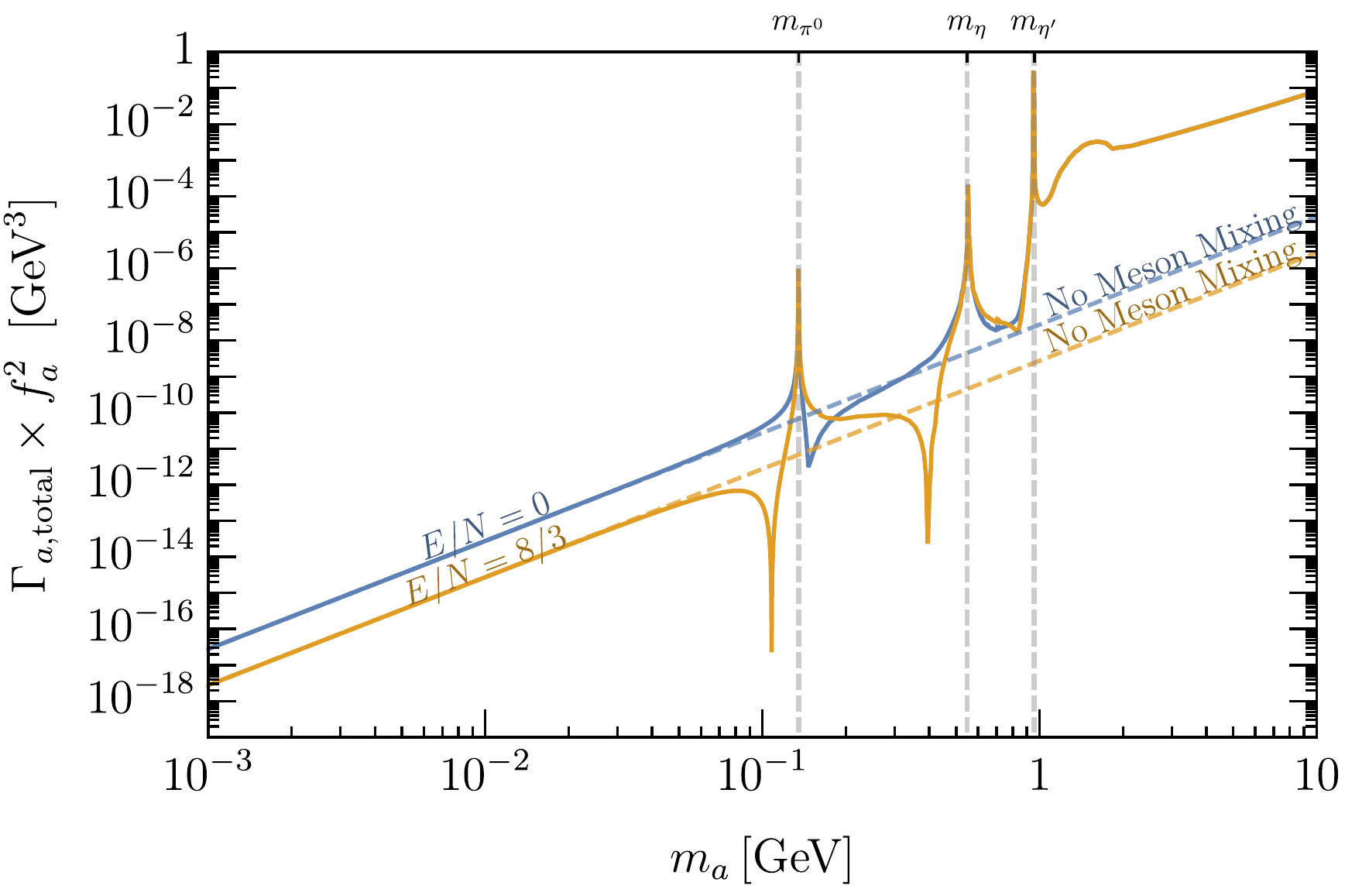} 
    \caption{\small (Left) Axion freeze-out temperature from pion (solid) and Primakoff (dashed) scatterings as a function of $f_a$ for a variety of $m_a$ between $0$ and $200$ MeV. For $f_a \gtrsim 3 \times 10^4$ GeV, the axion-pion scattering dominates over Primakoff scattering, keeping the axions in thermal equilbrium until $T \sim 10$ MeV for $f_a \lesssim 10^6$ GeV.}
    \label{fig:Tdecoupling}
     \caption{\small (Right) Total axion decay rate ($\tau^{-1}$) as a function of $m_a$ for $E/N = 0$ (blue) and $E/N = 8/3$ (orange). Resonant peaks in the decay rate arise from axion-meson mixing when $m_a$ is near $m_{\pi^0}$, $m_\eta$, and $m_{\eta}'$. Troughs arise from cancellations in $g_\gamma$ between the anomaly, axial rotation, and meson-mixing contributions. For $m_a \gtrsim 2$ GeV, the decay rate is set by gluons. The dashed contours show the axion-photon decay rate without meson mixing.}
    \label{fig:totalAxionDecay}
\end{figure}

The axion-meson decay collision term has also never been considered in the literature. Under the Boltzmann approximation in the kinetic equilbrium limit, $C_\Gamma$ is simply
\begin{align}
    \label{eq:collisionQCD}
    C_\Gamma \approx \frac{m_a}{E_a} \Gamma_{\rm QCD} \, ,
\end{align}
where $\Gamma_{\rm QCD}$ is the total axion decay rate to all final states containing mesons \eqref{eq:aMesonDecayRate}. In \eqref{eq:collisionQCD}, we use the Boltzmann kinetic equilibrium approximation.

It is fruitful to estimate the impact of the axion-pion scattering and axion-meson decays compared with the standard Boltzmann calculations in literature which only include Primakoff scattering and axion-photon decays. 
For example, the solid contours in  Fig.~\ref{fig:Tdecoupling} show the axion-pion decoupling temperature, $T_{\rm FO,\pi}$ vs $f_a$ over a range of axion masses. For $f_a \lesssim 10^{6} \,{\rm GeV}$, the axion remains in thermal equilibrium from pionic interactions until a few $10$s of MeV which is typically far lower than the Primakoff decoupling temperature  \cite{Cadamuro:2011fd},
\begin{align}
    T_{\rm FO, P} \approx 91\,  \frac{\sqrt{g_*(T_{\rm FO, P})}}{g_c(T_{\rm FO, P})} \left(\frac{f_a}{10^{6} \, \rm GeV} \right)^{2} \left(\frac{g_\gamma}{\alpha/2\pi} \right)^{-2} \, {\rm GeV}\, ,
\end{align}
where $g_c$ is the the sum of the charged relativistic degrees of freedom in the bath. The Primakoff decoupling temperature is shown by the dashed contours in the left panel of Fig.~\ref{fig:Tdecoupling}. For $f_a \gtrsim 5 \times 10^4 \, {\rm GeV}$, $T_{\rm FO,P} \gg T_{\rm FO, \pi}$, demonstrating the importance of the pions at maintaining thermal equilibrium with the axion all the way to $O(10)$ MeV temperatures. The lower pionic decoupling temperature compared to standard Primakoff decoupling leads to two important effects: (1) it can reduce the abundance of axions with masses above $T_{\rm FO,\pi}$, as they now follow an exponentially suppressed distribution relative to the standard,  non-Boltzmann suppressed distribution; and (2) it can enhance the initial abundance of axions with masses below $T_{\chi \rm PT}$ as they are not diluted by $g_{*S}$ like axions that decouple from Primakoff interactions in the early universe.

Similarly, axion-meson decays and inverse decays can keep the axion in thermal equilbrium at low temperatures. Most importantly, for a fixed $f_a$, the additional QCD decay channels can significantly decrease the axion lifetime relative to $\Gamma_{a \rightarrow \gamma \gamma}^{-1}$, which is the standard axion lifetime taken in previous Boltzmann calculations. For example, the solid blue and orange contours of Fig.~\ref{fig:totalAxionDecay} show the total axion decay rate including QCD decay channels for $E/N = 0$ and $8/3$, respectively. Compared to the dashed contours used in past literature, which show  $\Gamma_{a \rightarrow \gamma \gamma}$ when meson mixing and QCD channels are absent, the realistic total axion decay rate can be significantly different. Moreover, because the axion lifetime relative to neutrino decoupling dominantly sets the $N_{\rm eff}$ signal, we expect that incorporating axion-to-QCD channels will significantly alter the allowed parameter space in the $(m_a, f_a)$ plane.

To precisely quantify these new effects, the phase space evolution of the axion,     \eqref{eq:axionPhaseSpace}, together with the evolution of the Standard Model particles in the thermal bath must be computed to determine the effect of axion decoupling and decay on the relative temperature differences between photons and neutrinos in our present Universe, as typically characterized by the effective number of neutrino species
\begin{align}
    \label{eq:neff}
    N_{\rm eff} = \frac{8}{7}\left(\frac{11}{4} \right)^{4/3} \frac{\rho_\nu}{\rho_\gamma} \, .
\end{align}
Here, $\rho_\nu$ and $\rho_\gamma$ are the neutrino and photon energy densities. Note that $N_{\rm eff}$ is most sensitive to heavy axions that decay at temperatures near neutrino decoupling, which occurs around the MeV scale. The particles in the thermal bath from $T_{\rm \chi PT}$ through neutrino decoupling are photons, neutrinos, electrons, muons, pions and a small density of heavier mesons. The contribution to the energy density from these heavier mesons and from deviations of pions from the ideal gas law from self-interactions, $\rho_{\delta_{\rm QCD}}$, is  approximately $5\%$ of the total energy density at $T_{\rm \chi PT}$ and quickly drops far below $1\%$ by $T = 60$ MeV (see Fig.~\ref{fig:gstarComparison}). 

The evolution of the energy density of species in the thermal bath
that are tightly thermally coupled follows the energy density Boltzmann equation
\begin{align}
    \label{eq:boltzmannSM}
    \sum_{i = \gamma, e, \mu, \pi, \delta_{\rm QCD}} \frac{\partial \rho_i}{\partial t} + 3H( \rho_{i} + P_i) &= 
    \frac{\Gamma_{\nu e}}{T^4}(\rho_{\nu_e}^2 - \rho_{\nu_e, \rm eq}^2)
    + \frac{\Gamma_{\nu_{\mu \tau}}}{T^4}(\rho_{\nu_{\mu \tau}}^2 - \rho_{\nu_{\mu \tau}, \rm eq}^2)
     \\
    &+ \int \frac{d^3 p_a}{(2\pi)^3} \sqrt{p_a^2 + m_a^2}( C_\gamma + C_{P} + C_\pi  + C_{\Gamma})(f_a -f_{a,\rm eq}), \nonumber
\end{align}
where 
\begin{align}
\label{eq:energyDensities}
    \rho_i = \int \frac{d^3 p_i}{(2\pi)^3} \sqrt{p_i^2 + m_i^2} \, f_i(p_i) ,
  \hspace{0.3in}  P_i = \int \frac{d^3 p_i}{(2\pi)^3} \frac{1}{3}\frac{p_i^2}{\sqrt{p_i^2 + m_i^2}} \, f_i(p_i) 
\end{align}
are the energy densities and pressures of the $i$th  tightly coupled species in the bath. We assume that the photons and pions follow Bose-Einstein distributions and the electrons and muons follow Fermi-Dirac distributions. As discussed more in Appendix C, we infer $\rho_{\delta_{\rm QCD}}$ and $P_{\delta_{\rm QCD}}$ from the work of \cite{Saikawa:2018rcs}, which computes the Standard Model equation of state across the QCD phase transition and takes into account the deviations from the ideal gas law arising from the strongly coupled QCD bath. 

The energy densities of particles not strongly thermally coupled electromagnetically, namely the axion and neutrinos, must be solved for numerically. 
Specifically, the rate of change of the neutrino energy densities on the right side of \eqref{eq:boltzmannSM} follow the Boltzmann equations
\begin{align}
    \label{eq:boltzmannNu}
    \frac{\partial \rho_{\nu_e}}{\partial t} + 4H\rho_{\nu_e}  &= -\frac{\Gamma_{\nu e}}{T^4}(\rho_{\nu_e}^2 - \rho_{\nu_e, \rm eq}^2)
    \\
    \frac{\partial \rho_{\nu_{\mu \tau}}}{\partial t} + 4H\rho_{\nu_{\mu \tau}} &= -\frac{\Gamma_{\nu_{\mu \tau}}}{T^4}(\rho_{\nu_{\mu \tau}}^2 - \rho_{\nu_{\mu \tau}, \rm eq}^2).
\end{align}
Here, $\Gamma_{\nu_e} \simeq 0.68 \, G_F^2 T^5$ and $\Gamma_{\nu_{\mu \tau}} \simeq 0.15 \, G_F^2 T^5$ \cite{Cadamuro:2011fd} are thermally averaged neutrino interaction rates with the thermal bath for electron neutrinos and for muon and tau neutrinos, respectively.

Last, the Hubble rate, $H$, quantifies the expansion rate of the universe and sets the decoupling time of all interactions. The squared Hubble rate is set by the sum of all energy densities,
\begin{align}
    \label{eq:hubble}
     H^2 = \left(\frac{\dot{R}}{R}\right)^2 &= \frac{8 \pi G}{3}\left(\rho_\gamma + \rho_e + \rho_\mu + \rho_\pi + \rho_{\delta_{\rm QCD}} +
    \rho_{\nu_e} +
    \rho_{\nu_{\mu\tau}} +
    \rho_a\right).
\end{align}

We numerically solve the system of equations \eqref{eq:axionPhaseSpace}-\eqref{eq:hubble} using the method of lines \cite{methodOfLines}.
The method of lines is a numerical technique for solving a system of partial differential equations by discretizing one independent variable direction (comoving momentum in our case) while keeping the other independent variable continuous (logarithmic time in our case). The main advantage of the method of lines technique is the conversion of the Boltzmann system of \textit{partial} differential equations in $\{|\mathbf{p}_a|,t\}$ into a system of many \textit{ordinary} differential equations in $\{t\}$ which is computationally easier to solve. Moreover, by keeping the time-like variable continuous, useful techniques such as dynamical step-sizes can be employed to speed up the computation by automatically taking large temporal time-steps when changes in the interactions are small (such as at thermal equilibrium) while taking small temporal time-steps when changes are sudden (such as decays, decouplings, or re-thermalizations). See Appendix C for more details of our numerical setup.

\subsection{\texorpdfstring{$\Delta N_{\rm eff}$}{DNeff}}
\label{sec:DNeff}
In this section, we present the numerical results of $N_{\rm eff}$ as determined from the Boltzmann equations of Sec.~\ref{sec:boltzmannEqns} describing the cosmological evolution of the heavy QCD axion below $T_{\chi \rm PT}$. 

\begin{figure}[b]
    \centering
    \includegraphics[width=.45\textwidth]{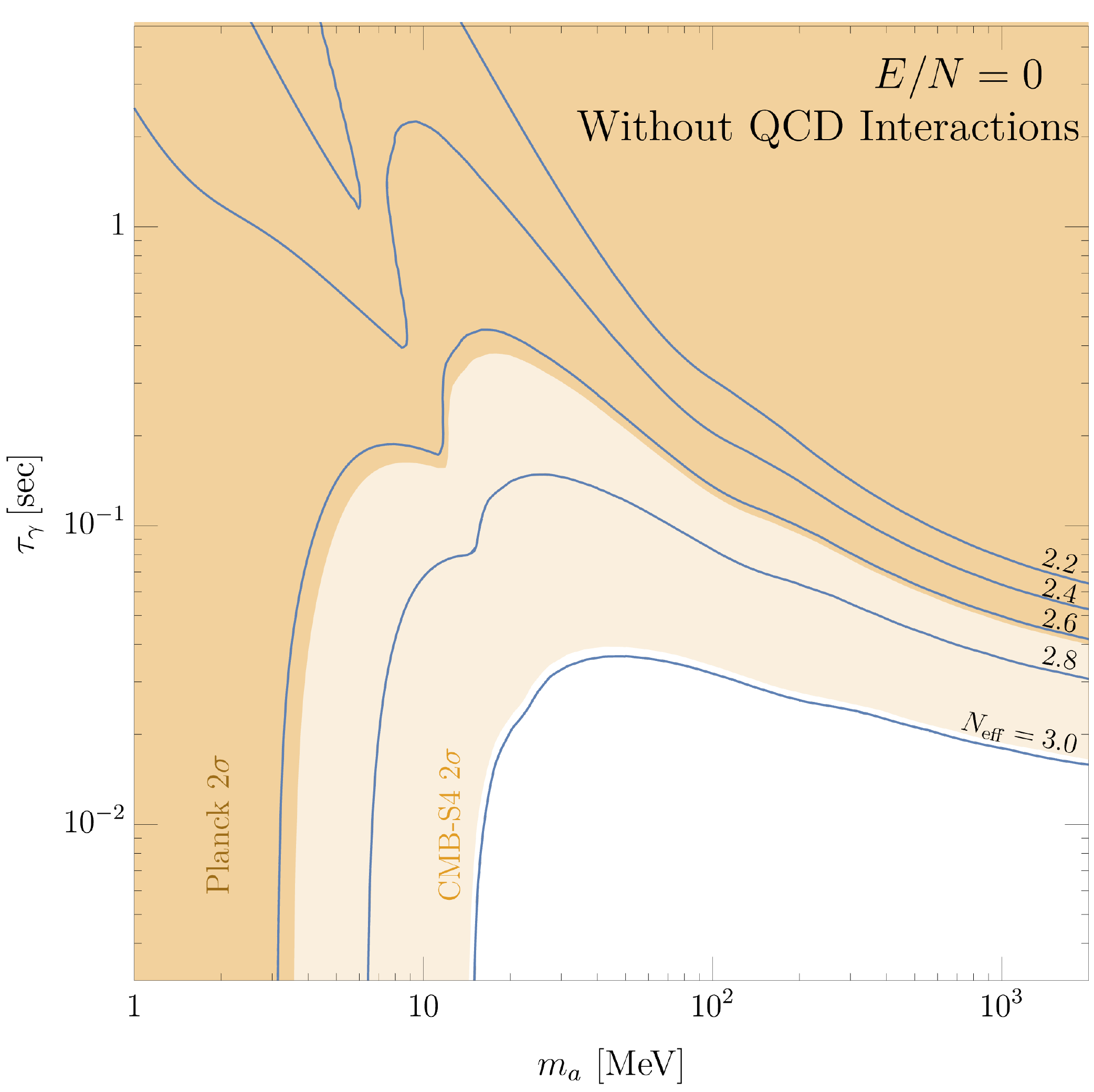} 
    \includegraphics[width=.45\textwidth]{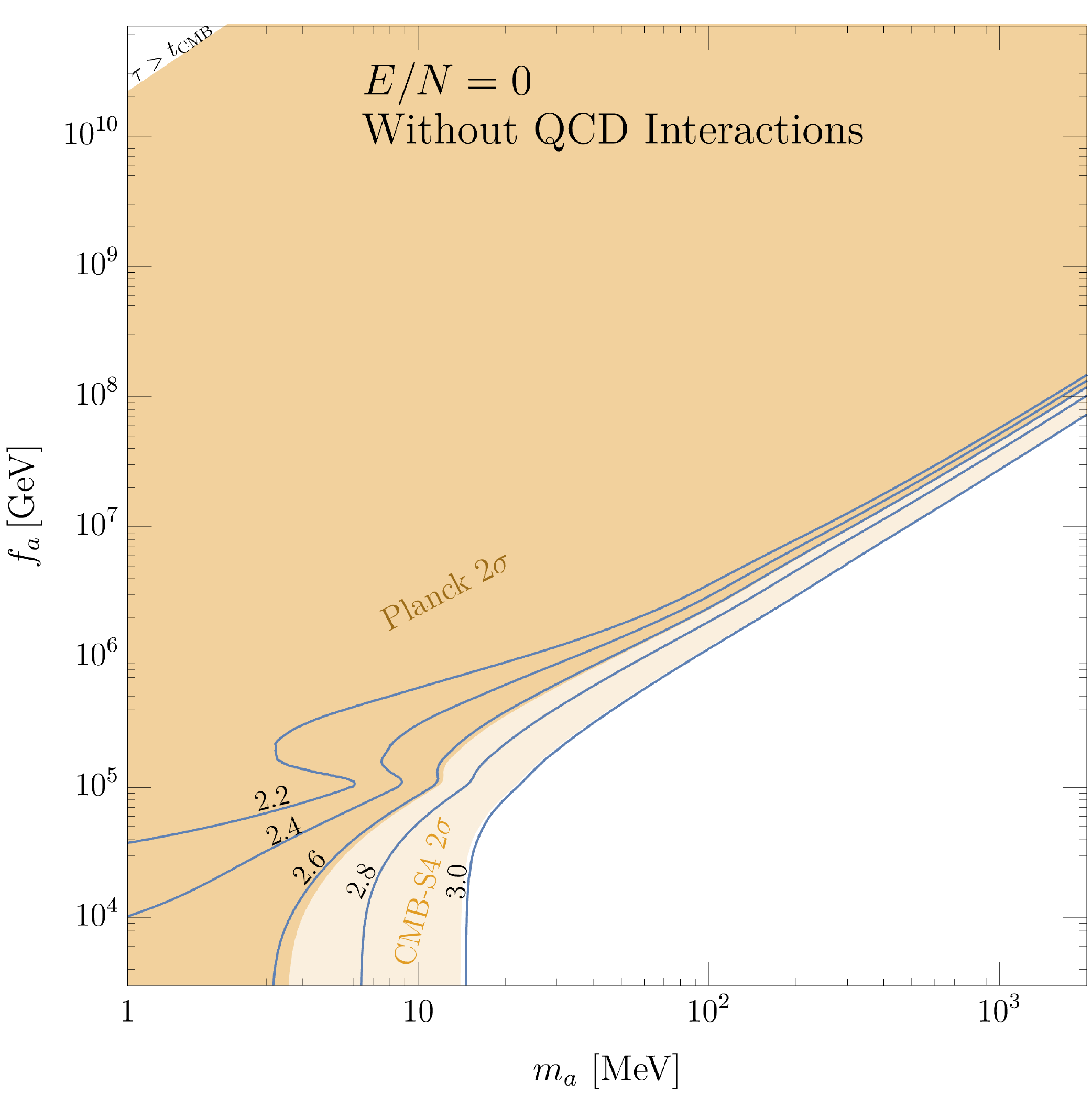} 
    \caption{
    Contours of $N_{\rm eff}$ in the $m_a - \tau_\gamma$ (left) and $m_a - f_a$ (right) planes \textit{neglecting} the following crucial effects that we take into account in future figures: axion-pion scattering, axion-meson decays, axion-meson mixing in $g_\gamma$, the proper frozen-out initial axion abundance, and the QCD contributions to the background evolution. Note that by neglecting these effects, as done in past works, the axion-lifetime is incorrectly set by the axion-photon decay rate with no axion-meson mixing, as shown by the y-axis of the left panel. The dark orange region is excluded at $95 \%$ confidence by Planck, while the light orange shows the future reach of CMB-S4 experiment at $95 \%$ confidence.
    }
    \label{fig:DneffPlotNoPion}
\end{figure}

First, to compare with past literature and highlight the importance of the new effects discussed in this work, we show $N_{\rm eff}$ for heavy axion cosmologies \textit{without} incorporating the following crucial elements in the Boltzmann code: axion-pion scattering, axion-meson decay, axion-meson mixing in $g_{\gamma}$, the proper frozen-out initial axion abundance, and the QCD contributions to the background energy density as described by $\rho_{\pi}$ and $\delta_{\rm QCD}$. Fig.~\ref{fig:DneffPlotNoPion} shows the numerical results of $N_{\rm eff}$ when these terms are neglected. That is, including only the photon ($C_\gamma$) and Primakoff ($C_P$) collision terms in Eq.~\eqref{eq:axionPhaseSpace} and Eq.~\eqref{eq:boltzmannSM}; including only  $\gamma, e$ and $\mu$ in the sum of the thermally coupled species of Eq.~\eqref{eq:boltzmannSM}; including only $\rho_\gamma, \rho_e, \rho_\mu, \rho_{\nu_e}, \rho_{\nu_{\mu \tau}}$, and $\rho_a$ in Hubble \eqref{eq:hubble}; setting the initial abundance of axions at the start of the Boltzmann code to that of a frozen-out abundance set by Primakoff scatterings such that $T_* = {\rm Max}(T_{\rm FO,P}, T_{\chi \rm PT})$; and finally, in Eq.~\eqref{eq:g_gamma}, setting $E/N =0$ and $\mathcal{F_\theta} = (1-z)/(1+z)$, which is the axion-meson mixing contribution in the inapplicable $m_a \ll m_\pi$ limit. The left panel of Fig.~\ref{fig:DneffPlotNoPion} shows contours of $\Delta N_{\rm eff}$ in the $m_a-\tau_\gamma$ plane, where $\tau_{\gamma}$ is the axion to photon lifetime when axion-meson mixing is neglected.
Note that $\tau_\gamma$ is the total lifetime of the axion since QCD decay channels are neglected in this particular case.

For $\tau_\gamma \gtrsim 10^{-1}$ s the axion decays after neutrino decoupling, heating up the photons relative to the neutrinos and giving rise to $\Delta N_{\rm eff} < 0$ as can be seen from the enhanced denominator of Eq.~\eqref{eq:neff}. For $\tau_\gamma \lesssim 10^{-1}$ s and $m_a \gtrsim 10$ MeV, the axion decays sufficiently early that the photons and neutrinos rethermalize before the neutrino decouples. In this scenario, the  $N_{\rm eff}$ signal of the heavy axion is absent and $\Delta N_{\rm eff} \simeq 0$. For $\tau_\gamma \lesssim 10^{-1}$ s and $m_a \lesssim 10$ MeV, the axion remains in thermal equilibrium past neutrino decoupling, heating up the photons and again giving rise to negative $\Delta N_{\rm eff}$. The right panel of Fig.~\ref{fig:DneffPlotNoPion} shows the same $\Delta N_{\rm eff}$ contours as the left panel but in the $m_a-f_a$ plane. Both panels assume the usual hadronic axion with $E/N = 0$. Taking the GUT motivated value of $E/N = 8/3$ only slightly shifts the contours in the right panel vertically.

\begin{figure}[t]
    \centering
    \includegraphics[width=.49\textwidth]{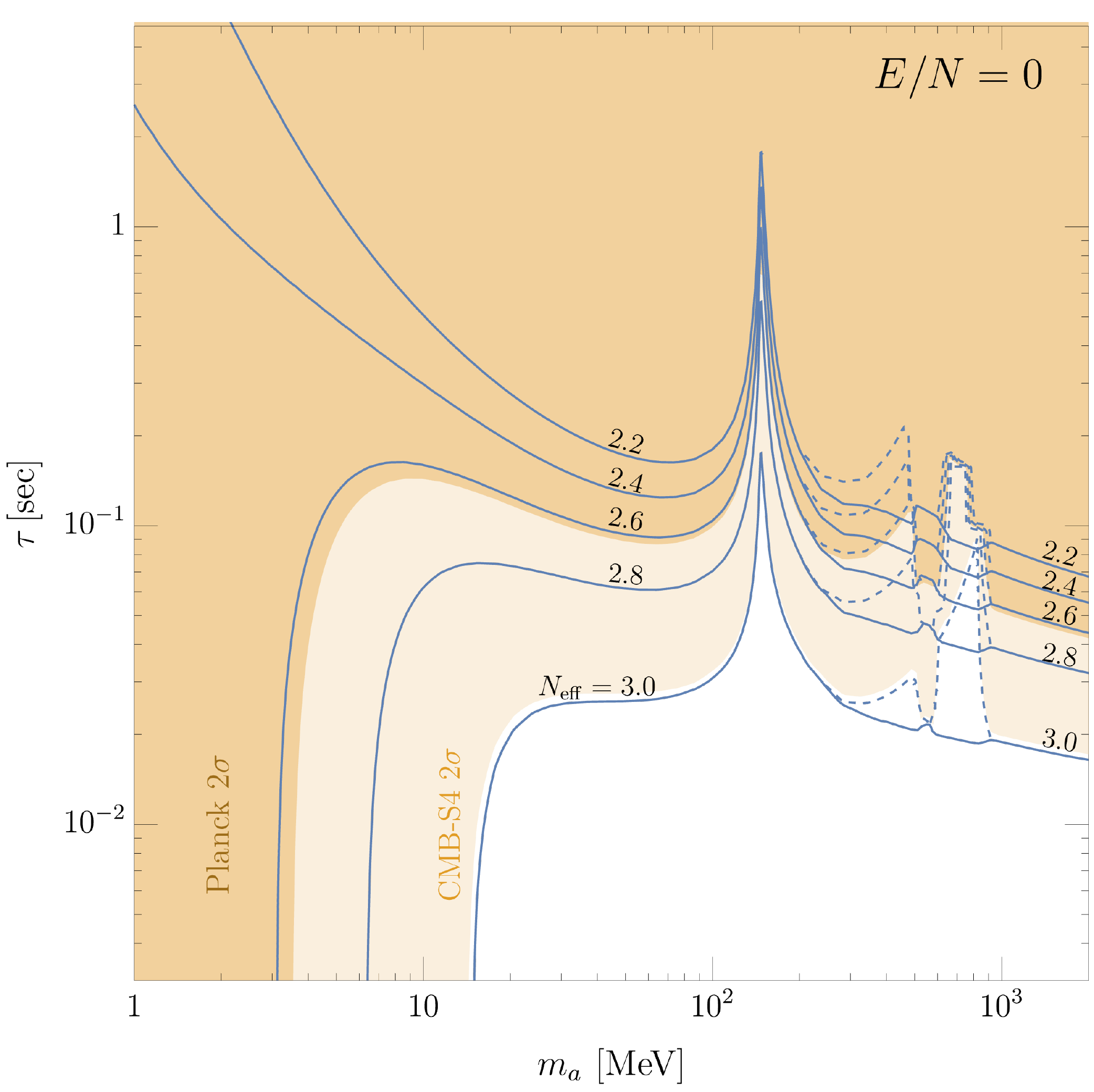} 
    \includegraphics[width=.49\textwidth]{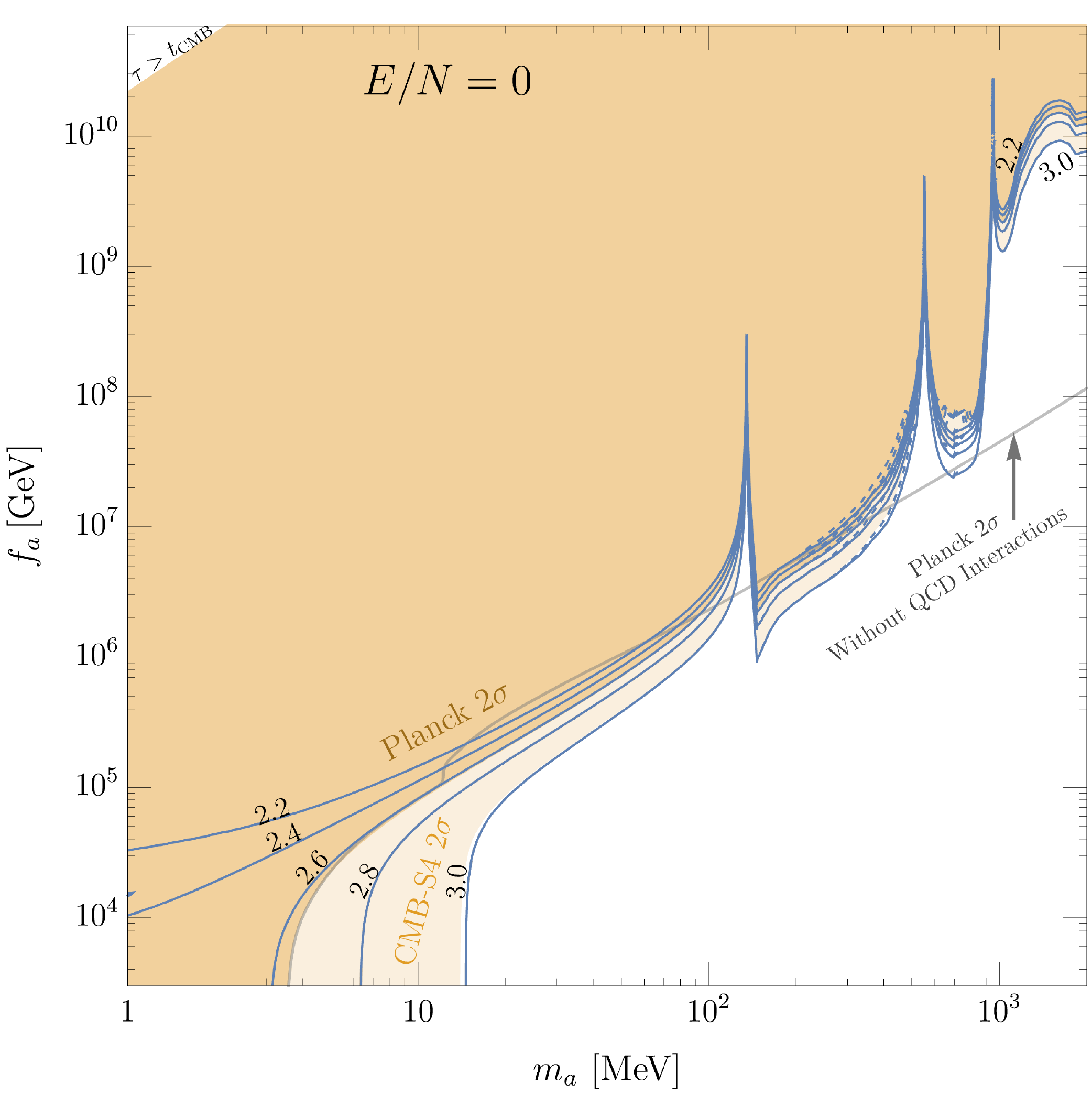} 
    \caption{Contours of $N_{\rm eff}$ in the $m_a - \tau$ (left) and $m_a - f_a$ (right) planes for $E/N = 0$. The dark orange region is excluded at $95 \%$ confidence by Planck, while the light orange shows the future reach of CMB-S4 experiment at $95 \%$ confidence. The dashed contours indicate where the initial axion yield is uncertain because $T_{\rm FO}$ lies between $T_{\chi \rm PT}$ and $T_{\rm g PT}$.  The dashed contours bound this uncertainty by showing the value of $N_{\rm eff}$ taking the minimum initial axion yield while the solid contours show the value of $N_{\rm eff}$ taking the maximum initial axion yield as given in Fig.~\ref{fig:TFOUncertainty}.}
    \label{fig:DneffPlotEN0}
    \includegraphics[width=.49\textwidth]{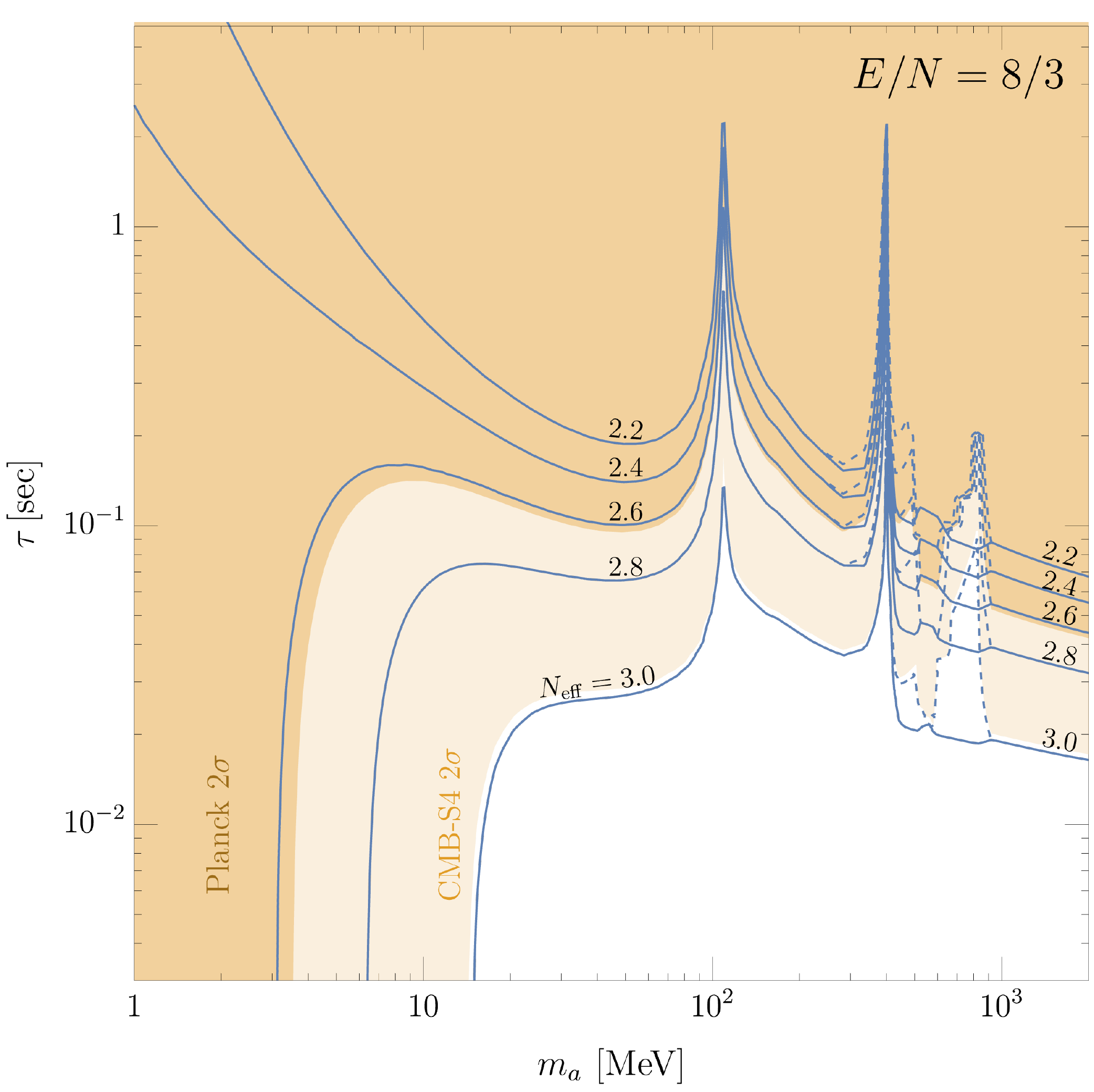} 
    \includegraphics[width=.49\textwidth]{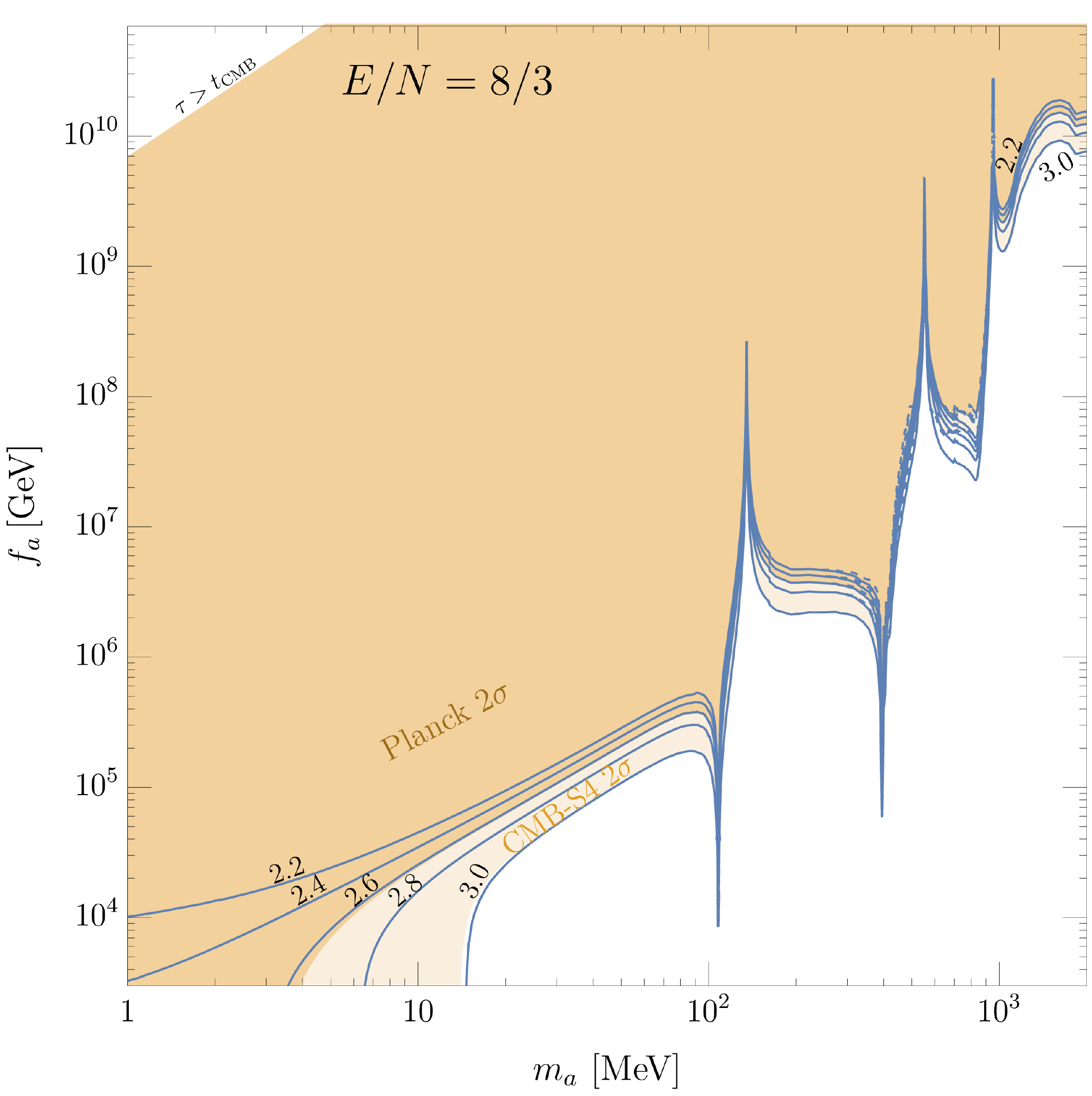} 
    \caption{
    Same as Fig.~\ref{fig:DneffPlotEN0}, but for E/N = 8/3.}
        \label{fig:DneffPlotEN83}
\end{figure}

We now consider $N_{\rm eff}$ for heavy axion cosmologies incorporating the new effects included in this work: axion-pion scattering, axion-meson decay, axion-meson mixing in $g_{\gamma}$, the proper (and occasionally uncertain) frozen-out initial axion abundance, the QCD contributions to the background energy density as described by $\rho_{\pi}$ and $\delta_{\rm QCD}$, as well results for the KSVZ $E/N = 0$ and the GUT motivated $E/N = 8/3$. Figs.~\ref{fig:DneffPlotEN0} and \ref{fig:DneffPlotEN83} show the contours of $\Delta N_{\rm eff}$ for heavy axions with these additional contributions for $E/N = 0$ (KSVZ) and $E/N = 8/3$ (GUT), respectively. In both figures, the left and right panels show the parameter space in the $(m_a, \tau)$ and $(m_a, f_a)$ planes, respectively. Note that here, $\tau$ is the \textit{total} lifetime of the axion, which 
begins differing from $\tau_\gamma$ for $m_a \gtrsim m_\pi$ where axion-meson mixing becomes important and then becomes even more disparate when axion-meson channels open for $m_a \gtrsim 3 m_\pi$, as shown in Fig.~\ref{fig:totalAxionDecay}. The solid and dashed blue contours in each panel correspond to taking the maximum possible $Y_a$ (blue) and minimum possible $Y_a$ (dashed) when $T_{\rm FO}$ lies in the uncertain region between $T_{\chi \rm PT}$ and $T_{g \rm PT}$, as indicated in Fig.~\ref{fig:TFOUncertainty}. The separation between the solid and dashed blue contours indicates the uncertainty in $N_{\rm eff}$ arising from the uncertainty in $T_{\rm FO}$ in this region. As can be seen, this region is localized roughly between $250 \,{\rm MeV} \lesssim m_a \lesssim 800 \,{\rm MeV}$ and, for any value of $m_a$ in this region, the uncertainty in the value of $f_a$ for any $N_{\rm eff}$ contour is typically only a several 10s of percent and always less than a factor 3.

Figs.~\ref{fig:DneffPlotEN0} and \ref{fig:DneffPlotEN83} demonstrate three important differences in the $N_{\rm eff}$ signal from heavy axions as currently considered in literature (Fig.~\ref{fig:DneffPlotNoPion}): First, the effect of axion-pion resonance on the mixing angle can be seen by the triangular shaped peaks near $m_a = m_\pi^0$.  In this regions, the axion is tightly coupled thermally to pions so that when the axion decays, its abundance is sufficiently exponentially suppressed that it does not heat up the photons even when decaying past neutrino decoupling. Second, for $m_a \gtrsim 3 m_\pi$, $|\Delta N_{\rm eff}|$ is reduced for fixed $(m_a, f_a)$ due to the meson-decay channels now open which cause the axion to decay earlier, especially near resonances in the mixing angle at $m_a = m_\pi, \, m_\eta$, and $m_{\eta'}$. This can be seen more clearly in the $(m_a, f_a)$ planes. Note that for some $m_a$, incorporating axion-meson mixing can \textit{increase} the axion lifetime due to cancellations between contributions of $g_\gamma$ \eqref{eq:agammagamma}. The increased  $|\Delta N_{\rm eff}|$ for fixed $(m_a, f_a)$ due to the increased axion lifetime is important for $m_a < 3 m_\pi$ when only the axion-photon decay channel is open. 
Third, for $m_a \lesssim m_\pi$, $|\Delta N_{\rm eff}|$ is reduced for fixed $(m_a, f_a)$ because the axion is kept in thermal equilibrium by axion-pion scattering to lower temperatures compared to when the pions are absent. This leads to a reduced axion abundance at neutrino decoupling which reduces the $|\Delta N_{\rm eff}|$ contribution from the axion.

\begin{figure}[tb]
    \centering
    \includegraphics[width=.45\textwidth]{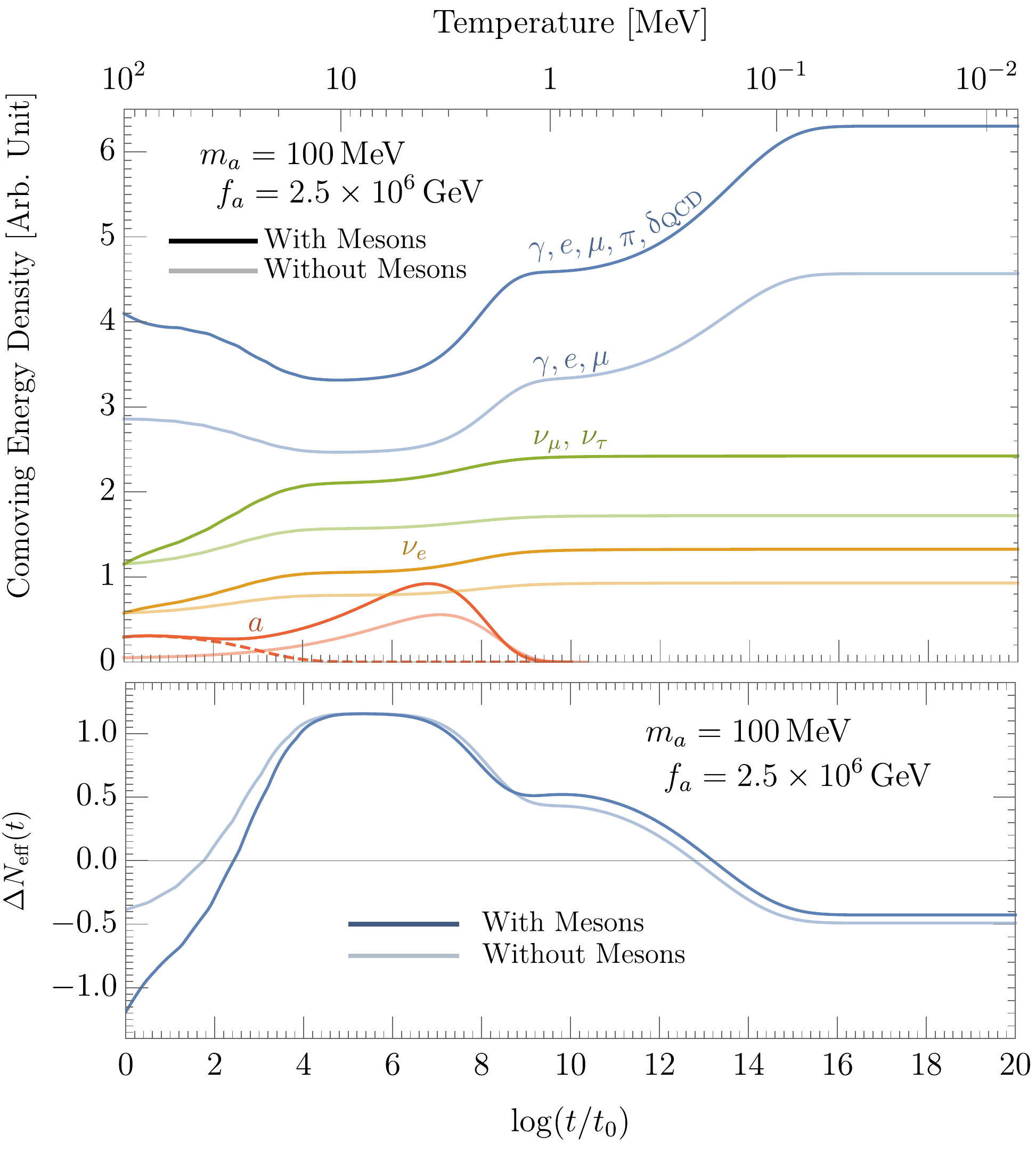} 
    \includegraphics[width=.45\textwidth]{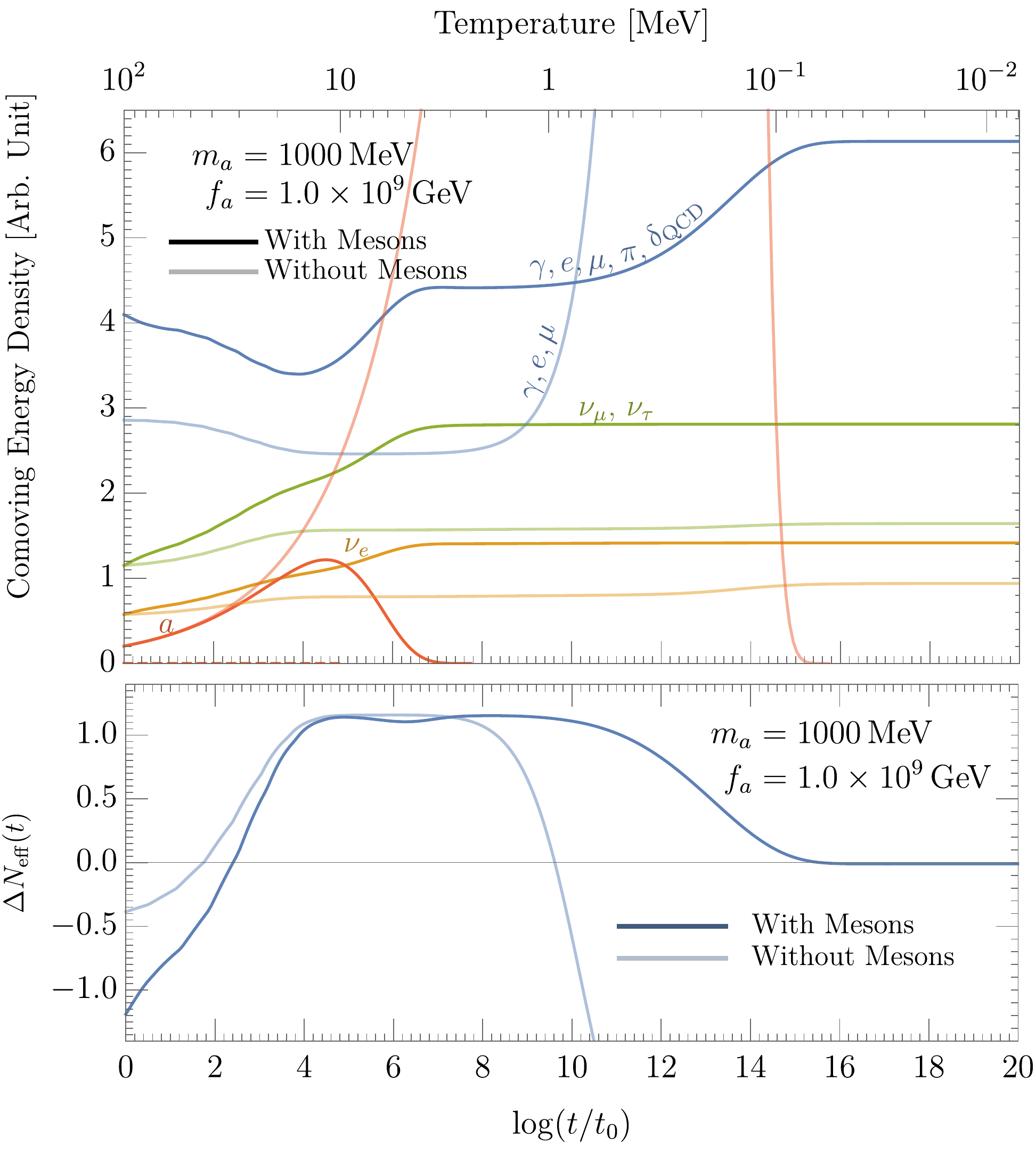}
    \caption{Top panels show the comoving energy density evolution of the axion (dark red), electron neutrinos (dark orange), muon and tau neutrinos (dark green), and strongly coupled species in the Standard Model thermal bath (dark blue) for $(m_a, \tau) = (100 \, {\rm MeV}$, $0.1 {\rm s})$, left, $(m_a, \tau) = (1000 \, {\rm MeV}$, $0.1 {\rm s})$, right. The correspondingly lighter shaded contours show the same evolution but without pion or meson interactions. The dashed red contour shows the axion comoving energy density if it were to maintain a thermal density at all times. The bottom panels show the instaneous value of Eq. \eqref{eq:neff} minus the Standard Model result, $N_{\rm eff}^{\rm SM} = 3.044$. The left and right panels respectively highlight how including QCD interactions can keep the axion thermally coupled to the Standard Model bath for a longer time or enhance the axion decay rate at a fixed $f_a$. The left panel demonstrates that for low $m_a$, the axion can be in thermal equilbrium at $T_{\chi \rm PT}$ when including pions-interactions but be out of equilbrium when including only Primakoff processes which freeze-out at much higher temperatures that lead the axion to possess a $g_{*S}$ diluted abundance at $T_{\chi \rm PT}$. In the left panel, the axion lifetime with or without mesons is approximately the same ($\tau \approx 0.1$ s) because the meson decay channels are forbidden and axion-pion mixing is not too appreciable yet. In contrast, the right panel demonstrates that for high $m_a$, the axion can decay much earlier due to the kinematic availability of meson decay channels ($\tau \approx 0.01$ s with meson interactions and $24$ s without). This leads to a reduced $\Delta N_{\rm eff}$, as shown on the bottom right panel. In both panels, we take $E/N = 0$.} 
    \label{fig:pionNoPionPlots}
\end{figure}

Each of these effects can be seen more clearly in the top panels of Fig.~\ref{fig:pionNoPionPlots} which show the evolution of the energy densities of the axion and other species in the thermal bath as a function of time and temperature (top horizontal axis) for fixed ($m_a$, $f_a$) = ($100$ MeV, $2.5 \times 10^6$ GeV) and ($1000$ MeV, $1.0 \times 10^9$ GeV), in the top left and top right panels, respectively. The dark colored contours show the evolution of the comoving energy density, $X_i = \rho_i R^4$ of the $i$th species when including the axion-meson interactions while the light colored contours show the same evolution when the axion-meson interactions are absent. The electromagnetic component of the thermal bath, $\gamma, e, \mu, \pi, \delta_{\rm QCD}$ is shown in blue, $\nu_{\mu, \tau}$ in green, $\nu_{e}$ in orange, and the axion in red. The dashed red contour shows the comoving energy density of the axion if it were to maintain a thermal distribution for all times. 
As can be seen from the $m_a = 100$ MeV panel, the axion starts off in thermal equilbrium compared to the case without meson interactions in which the axion possesses a $g_{*S}$ suppressed abundance from earlier Primakoff freeze-out. Moreover, the axion abundance with meson interactions follows the dashed thermal distribution to lower temperatures than without meson interactions. This leads to a relative suppression in the non-relativistic abundance of the axion prior to decaying around $2$ MeV. Consequently, including  axion-pion scattering $\Delta N_{\rm eff}$ is not as negative as previous results in the literature. The temporal evolution of $\Delta N_{\rm eff}(t) = N_{\rm eff}(t) - 3.044$, as shown by the dark blue (with mesons) and light blue (without mesons) contours in the lower left panel of Fig.~\ref{fig:pionNoPionPlots}, demonstrates this difference explicitly. Note that for $m_a = 100$ MeV, the axion mass is on the cusp of the axion-pion resonance for $E/N = 0$. For $m_a$ closer to $m_\pi$, the axion follows the dashed thermal abundance for a longer duration which generates the large triangular peak in the allowed $N_{\rm eff}$ plot of Fig.~\ref{fig:DneffPlotEN0}.

For $m_a = 1000$ MeV, the axion with meson interactions again follows the dashed thermal distribution to slightly lower temperatures than the axion without meson interactions. More important though is the difference in decay time between the two cases. In particular, the axion with meson interactions (dark red) decays earlier than the axion without (light red) due to the $a \rightarrow \eta \pi \pi, \pi \pi \gamma$ decay channels that are now kinematically open to the $1000$ MeV axion that are absent from the $100$ MeV axion. These extra decay channels lead to a much smaller $\Delta N_{\rm eff}$ as shown explicitly by the evolution of $\Delta N_{\rm eff}$ in the bottom right panel of Fig.~\ref{fig:pionNoPionPlots}.

We note that for $m_a > 2$ GeV, the axion decays dominantly into gluons. Here, the $2\sigma$ limit on $N_{\rm eff}$ as constrained by Planck approximately follows the contour $\tau \approx 0.05$ s around $m_a \sim$ GeV and slowly drops with increasing axion mass. The slight decrease in the maximum allowed $\tau$ in this region originates from the increase in the axion energy density at decay with axion mass: For such heavy axions with long lifetimes, $f_a$ is large and the axion freezes-out early, leading to freeze-out yield roughly independent of $m_a$. Consequently, the heavier the axion, the earlier it must decay so that its energy density at neutrino decoupling is further exponentially suppressed to counter its larger frozen-out energy density. As shown in Appendix \ref{app:AsymptoticNeff}, the exponentially decaying energy density of axions, $\rho_a \propto e^{-t/\tau}$, leads to a logarithmic decrease in the maximum allowed $\tau$ given by the semi-analytic function
\begin{align}
        \label{eq:tauAsymptotic}
        \tau_{\rm max}(m_a) < \frac{4.3 \times 10^{-2} \, \rm s}{1 + 0.25 \ln{\frac{m_a}{2 \rm \, GeV}}} \qquad (m_a  \geq 2 \, \rm GeV).
\end{align}
Eq.~\eqref{eq:tauAsymptotic} can also be written in terms of $f_a$ by equating $\tau_{\rm max}^{-1}$ with the analytic decay rate into gluons given by Eqns. \eqref{eq:aMesonDecayRate} and \eqref{eq:agg},
\begin{align}
\label{eq:gluonConstraint}
f_a \lesssim 1.2 \times 10^{10} \, \GeV  \left(\frac{m_a}{2 \, \rm GeV}\right)^{ \scalebox{1.01}{$\frac{3}{2}$}} \left(\frac{\alpha_3}{0.3}\right)\left(\frac{1 + \frac{83}{4\pi}  \alpha_3}{1 + \frac{83}{4\pi} 0.3}\right)^{ \scalebox{1.01}{$\frac{1}{2}$}}  \left(1 + 0.25 \ln \frac{m_a}{2 \, \rm GeV} \right)^{ \scalebox{1.01}{$-\frac{1}{2}$}}\quad (m_a  \geq 2 \, \rm GeV).
\end{align}
Thus, for $m_a \gg 2 $ GeV and $f_a$ small enough to be probed by accelerator experiments \cite{Ertas:2020xcc,Dobrich:2015jyk,Dolan:2017osp,NA64:2020qwq,FASER:2018eoc,Gori:2020xvq,Aloni:2018vki,Mariotti:2017vtv,Kelly:2020dda,Chakraborty:2021wda,Bertholet:2021hjl}, the dark radiation constraint is absent.

\section{Including a Mirror Photon}
\label{sec:mirrorPhoton}
In this section we add a mirror photon $\gamma'$ to the theory, with a mass sufficiently small that it can be ignored in our analysis. A mirror photon is natural in theories with a $Z_2$ symmetry that not only doubles the $SU(3)_c$ sector of the Standard Model to achieve a heavy QCD axion, but also doubles the $SU(2)_L \times U(1)_Y$ sector.
This complete mirroring of the Standard Model gauge group introduces another axion coupling relevant in computing the amount of dark radiation 
\begin{align}
    {\cal L}_{a,\gamma'} &=
    \frac{g_{\gamma'}}{4}\frac{a}{f_a} F'_{\mu \nu}\tilde{F'}^{\mu \nu}. 
\end{align}
Above the mirror QCD scale, $\Lambda_{\rm QCD}^{'}$, the $Z_2$ symmetry ensures that
\begin{align}
    g_{\gamma'} =
    \frac{e'^2}{8\pi^2}\left( \frac{E}{N} \right) \, , 
    \label{eq:g_gamma'}
\end{align}
with $e'$ differing from $e$ only by renormalization group scaling, which we ignore. 

Note that unlike $g_\gamma$ in \eqref{eq:agammagamma}, $g_{\gamma'}$ does not include contributions from axion-mirror meson mixing nor from the axial rotation onto mirror quarks because the masses of the lightest mirror quarks are typically much heavier than  $\Lambda_{\rm QCD}^{'}$ and thus irrelevant to the theory below $\Lambda_{\rm QCD}^{'}$. If, however, any mirror quark is lighter than the mirror QCD scale, then there is an additional contribution to $g_{\gamma'}$ analogous to the second and third terms in (\ref{eq:g_gamma}) for $g_\gamma$. In the minimal theory, where the $Z_2$ symmetry exchanges the Standard Model with its mirror and is spontaneously broken by a difference between the electroweak vevs with  $v' \gg v$, all mirror quarks are heavier than $\Lambda_{\rm QCD}^{'}$ for $f_a m_a \gtrsim 25 \; \GEV^2$ \cite{Dunsky:2019upk}. This relation is satisfied for nearly the entire parameter region of interest to us,  so that all mirror quarks are well above the QCD scale; $g_{\gamma'}$ is thus uncorrected and given by (\ref{eq:g_gamma'}).

Numerical results in this section are calculated taking $e'=e$ (that is, neglecting the small running of $e'$ below $v'$) and with a non-zero $E/N$ so that $g_\gamma'$ is non-zero. As a result, the axion-mirror photon decay rate is 
\begin{align}
    \label{eq:axionGammaPrimeDecayRate}
    \Gamma_{a \rightarrow \gamma' \gamma'} = \frac{g_{\gamma'}^2}{64\pi} \frac{m_a^3}{f_a^2} \simeq \frac{1}{64\pi}\left(\frac{e^2}{8 \pi^2} \frac{E}{N}\right)^2 \frac{m_a^3}{f_a^2} \, .
\end{align}
\begin{figure}[tb]
    \centering
    \includegraphics[width=.49\textwidth]{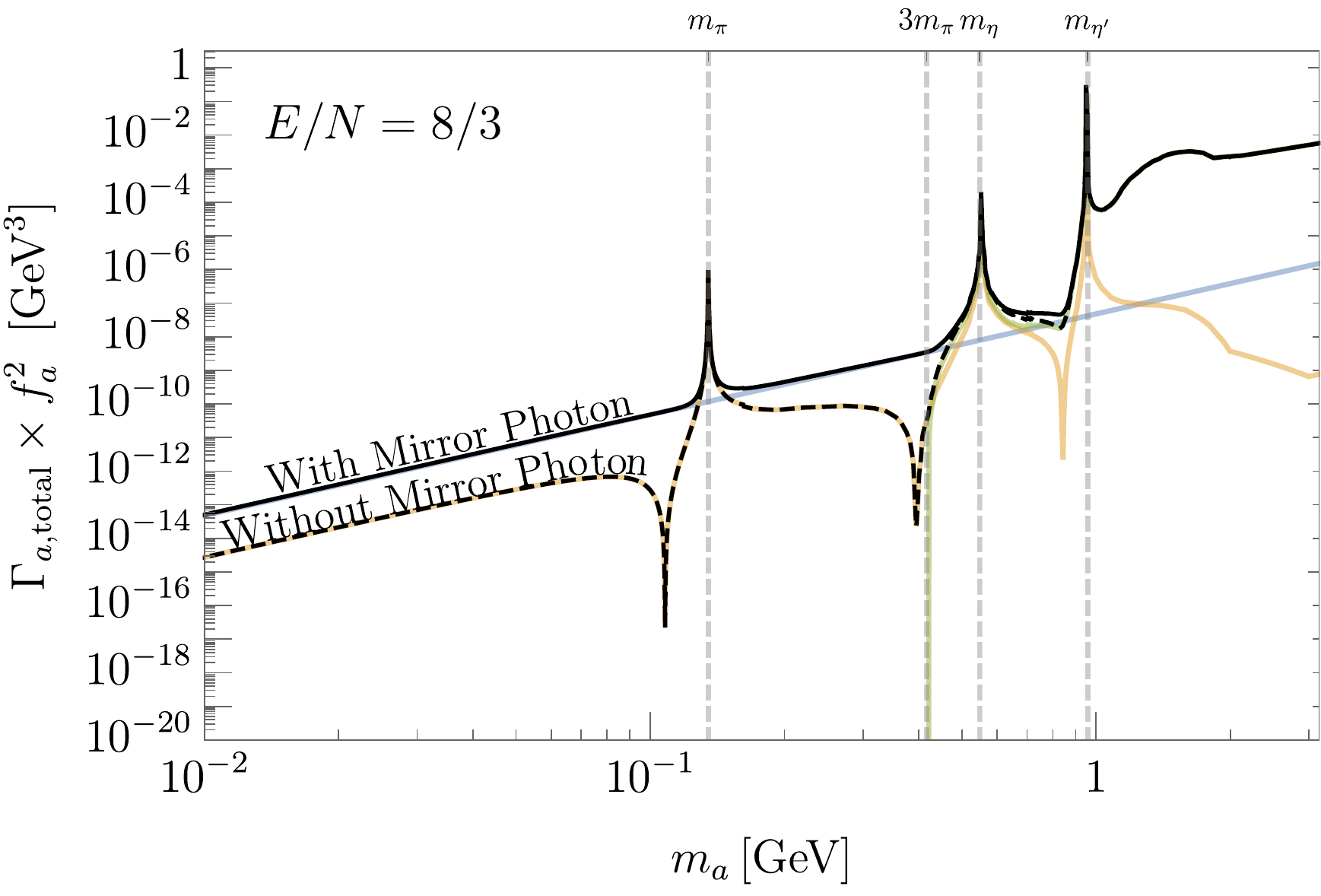} 
    \includegraphics[width=.48\textwidth]{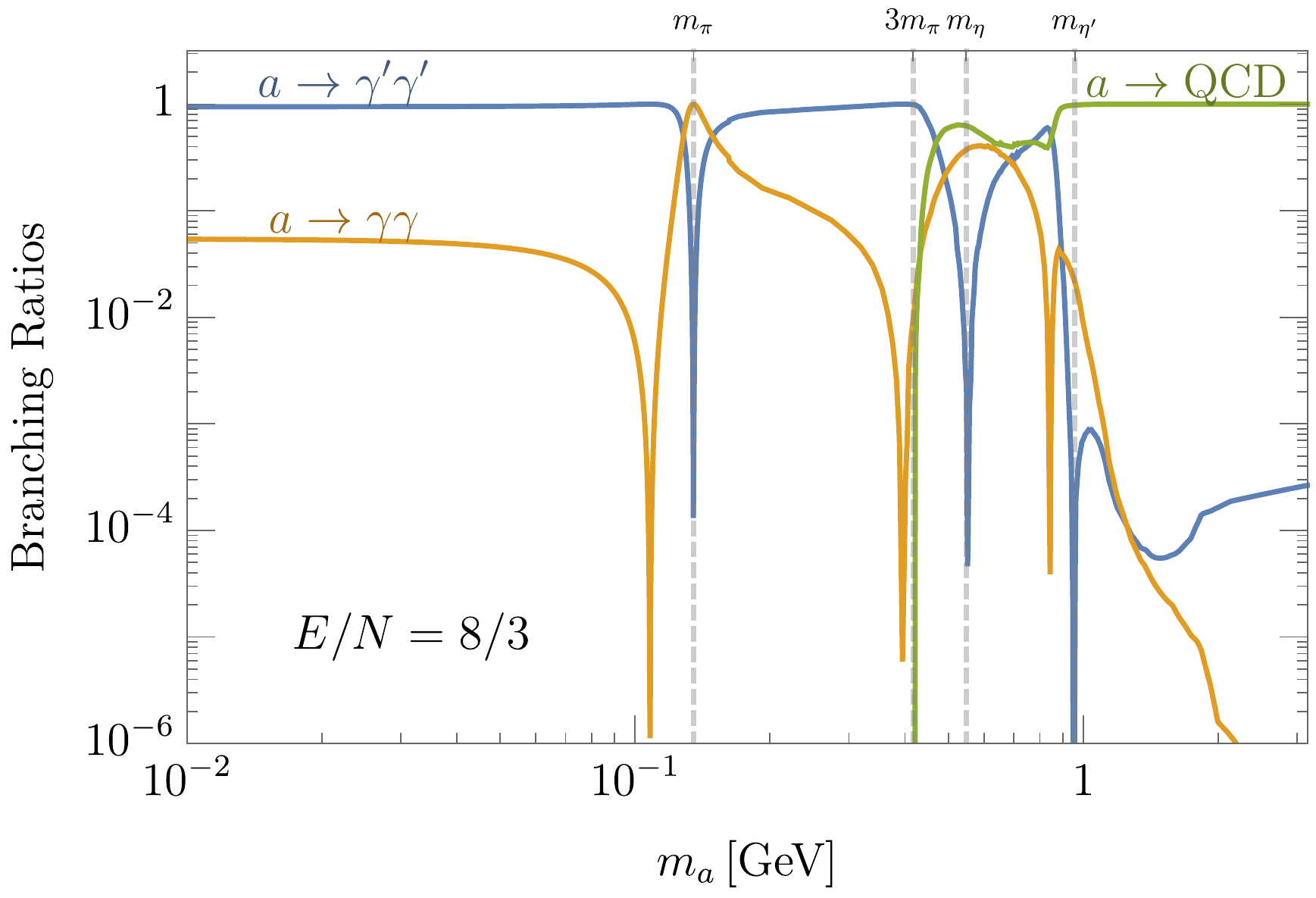} 
    \caption{The axion total decay width (left) and branching ratios (right) as a function of axion mass, for $E/N = 8/3$. The orange (blue) contours refer to decays to photons (mirror photons); the green contour is for decays to states that include hadrons. In the left panel, the solid (dashed) black contours give the axion decay width in the theory with (without) mirror photons.}
    \label{fig:darkPhotonDecayComparisonEN83}
\end{figure}
\begin{figure}[tb]
    \centering
    \includegraphics[width=.49\textwidth]{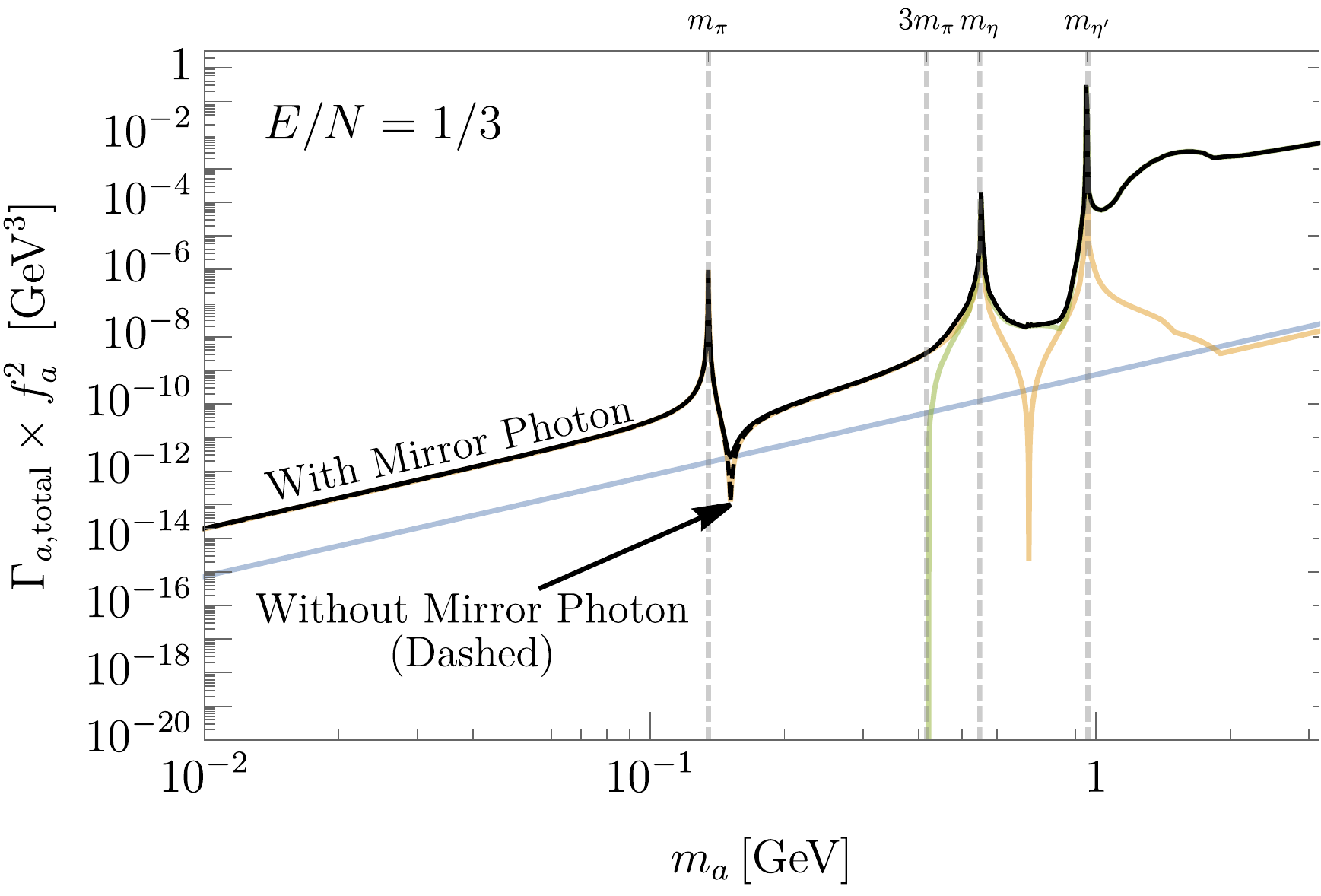} 
    \includegraphics[width=.48\textwidth]{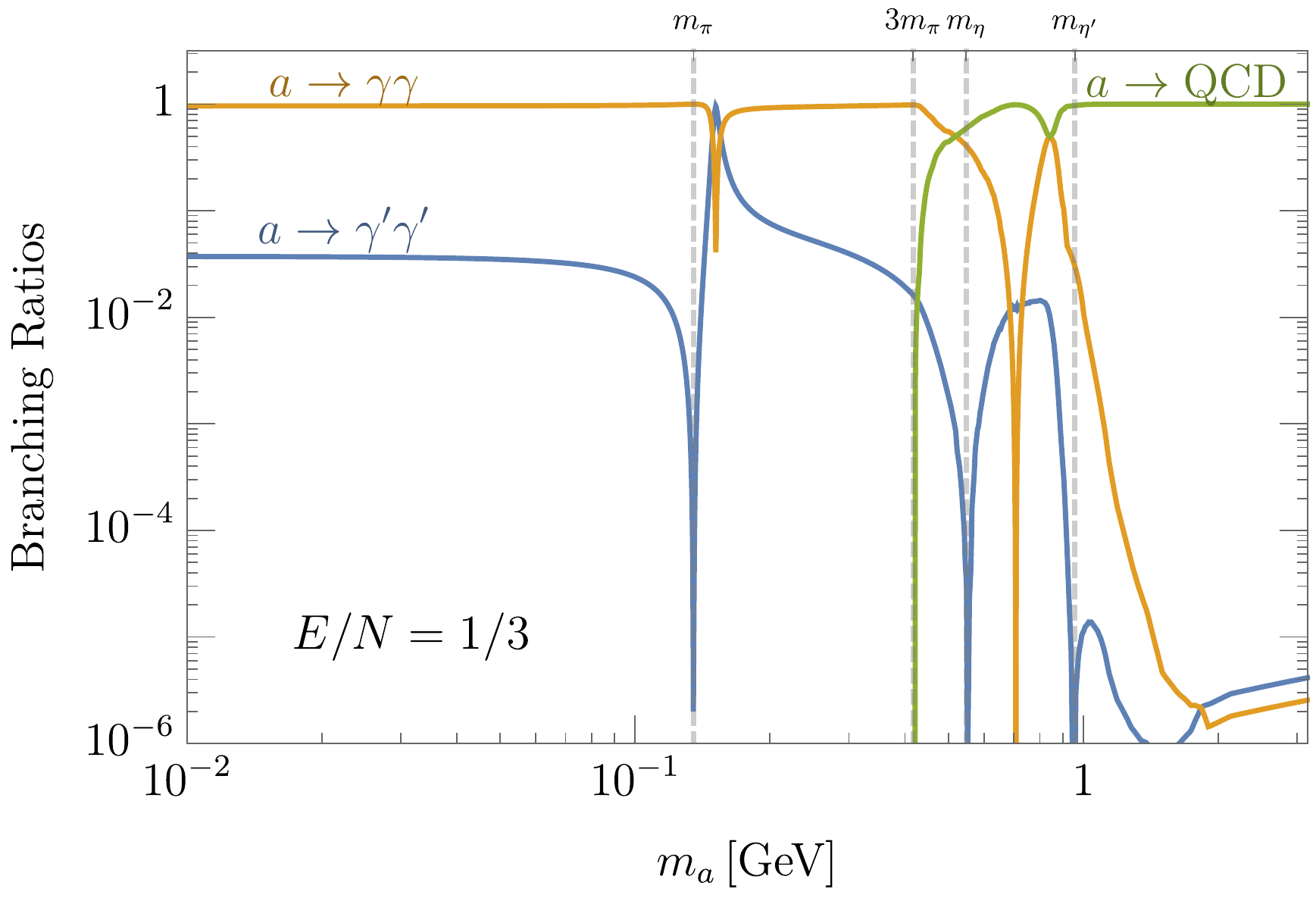} 
    \caption{As in Fig.~\ref{fig:darkPhotonDecayComparisonEN83}, but for $E/N = 1/3$.  In the left panel, the solid and dashed black contours, giving the axion decay width with and without mirror photons, are almost conincident, as the partial width to mirror photons is sub-dominant, as shown by the straight blue line.}
    \label{fig:darkPhotonDecayComparisonEN13}
\end{figure}
In particular, we consider two values for $E/N$: 8/3, motivated by grand unification, and $1/3$, which can be achieved by
an appropriate choice of KSVZ fermions.
For $E/N = 8/3$, the total axion decay rate with and without mirror photons is shown by the solid and dashed black contours in the left panel of Fig.~\ref{fig:darkPhotonDecayComparisonEN83}. The light blue, orange, and green contours indicate the axion decay rates into $\gamma'$, $\gamma$, and QCD degrees of freedom, respectively. Because there is no cancellation of terms in $g_{\gamma'}$ as compared to $g_\gamma$, the $a \rightarrow \gamma' \gamma'$ decay rate (blue) is roughly an order of magnitude \textit{greater} than the $a \rightarrow \gamma \gamma$ decay rate (orange) until $m_a \gtrsim 3 m_\pi$: thus, axions below this mass dominantly decay into dark photons. This can be seen more clearly in the right panel of Fig.~\ref{fig:darkPhotonDecayComparisonEN83}, which shows the branching ratios for the same three axion decay channels. The case for $E/N = 1/3$ is significantly different as demonstrated in Fig.~\ref{fig:darkPhotonDecayComparisonEN13}. In particular, the cancellation between terms in $g_{\gamma}$ is negligible which leaves the now smaller $a \rightarrow \gamma' \gamma'$ decay rate roughly an order of magnitude \textit{weaker} than the $a \rightarrow \gamma \gamma$ decay rate. 

We highlight these two representative values of $E/N$ since they generate substantially different decay branching ratios into mirror photons.\footnote{The case for $E/N = 2/3$, which can be achieved by a KSVZ quark that possesses the same gauge quantum numbers as down quarks, is intermediate between these two cases and like the case $E/N = 1/3$, also yields a region of $\Delta N_{\rm eff} \approx 0$ at low $m_a$.}
This disparity is important since the parameter space where the axion branching ratio into dark photons is $\mathcal{O}(1)$ can be cosmologically dangerous as the mirror photon decay mode \eqref{eq:axionGammaPrimeDecayRate} increases $N_{\rm eff}$, by directly generating dark radiation in the form of $\gamma'$, and reduces the heating of the Standard Model bath as fewer axions decay into $\gamma$.

Quantitatively, $N_{\rm eff}$ with a mirror photon is
\begin{align}
    \label{eq:NeffDarkPhoton}
    N_{\rm eff} = \frac{8}{7}\left(\frac{11}{4} \right)^{4/3} \frac{\rho_\nu + \rho_{\gamma'}}{\rho_\gamma} \, ,
\end{align}
where $\rho_\nu$,  $\rho_{\gamma'}$,  and  $\rho_\gamma$, are the relic energy densities of neutrinos, mirror photons, and photons, respectively.
According to Eq.~\eqref{eq:NeffDarkPhoton}, an additional mirror photon in thermal equilibrium significantly increases $N_{\rm eff}$ and is generally excluded by current $\Delta N_{\rm eff}$ limits \cite{Planck:2018vyg}. Nevertheless, a mirror photon can be allowed if it does not achieve a thermal abundance.

This leads us to consider the Boltzmann equation for $f_a(\mathbf{p})$ in the freeze-in picture
\begin{align}
    \label{eq:axionPhaseSpaceDarkPhoton}
    \frac{\partial f_a}{\partial t} - p_a H \frac{\partial f_a}{\partial p_a} &= ( C_\gamma + C_{P} + C_\pi  + C_{\Gamma})(f_{a,\rm eq}-f_a) - C_{\gamma'} f_a \, ,
\end{align}
where the mirror photon collision term, $C_{\gamma'}$, is given by
\begin{align}
    C_{\gamma'} \simeq \frac{m_a}{E_a}\Gamma_{a \rightarrow \gamma' \gamma'} \, .
\end{align}
Eq.~\eqref{eq:axionPhaseSpaceDarkPhoton} replaces Eq.~\eqref{eq:axionPhaseSpace} when including mirror photons.
In addition, while the energy density evolution of the Standard Model bath remains as given in Eqns. \eqref{eq:boltzmannSM} and \eqref{eq:boltzmannNu}, \footnote{The axion-mediated interaction $\gamma' + \gamma' \leftrightarrow \gamma + \gamma$ can contribute to the energy transfer to the Standard Model thermal bath and hence to additional terms on the right-hand-side of Eq .\eqref{eq:boltzmannSM}. However, this interaction is $\mathcal{O}(1/f_a^{4})$ and generally negligible. Similarly, the decay and inverse decay $a \leftrightarrow \gamma + \gamma'$ can also modify \eqref{eq:boltzmannSM}, but this requires $O(1)$ kinetic mixing or $a F' \tilde{F}$. The former is constrained by searches for relic mirror charged particle~\cite{Dunsky:2018mqs}, and the latter requires $U(1)\times U(1)'$ charged particles around the mass scale $f_a$, whose relic is also constrained.} the energy density evolution of ${\gamma'}$ is described by  
\begin{align}
    \label{eq:rhoGammaPrime}
    \frac{\partial\rho_{\gamma'}}{\partial t} + 4 H \rho_\gamma' = \int \frac{d^3p}{(2\pi)^3} m_a \Gamma_{a \rightarrow \gamma' \gamma'} f_a = m_a \Gamma_{a \rightarrow \gamma' \gamma'} \, n_a \, .
\end{align}
\begin{figure}[tb]
    \centering
    \includegraphics[width=.49\textwidth]{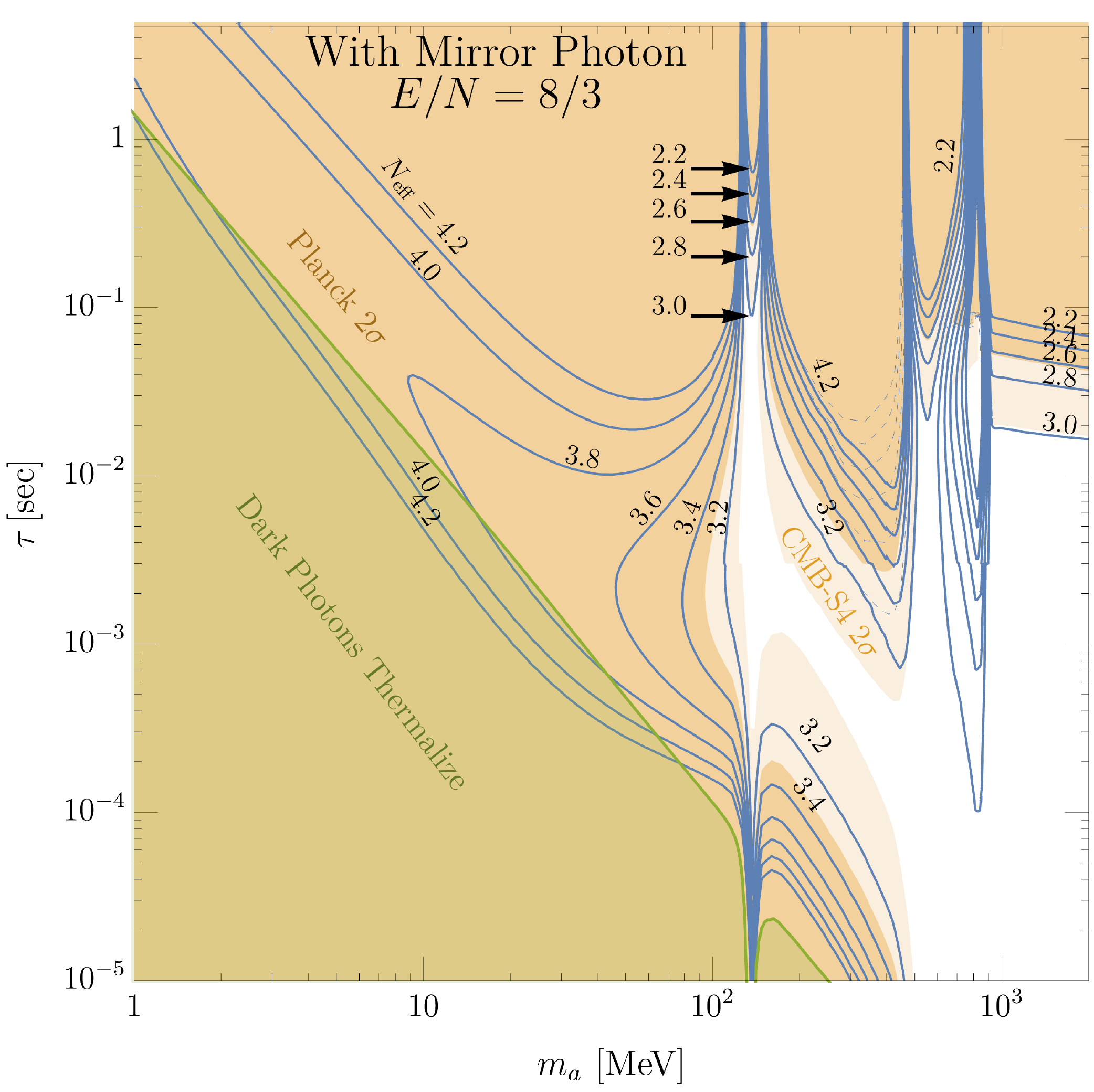} 
    \includegraphics[width=.49\textwidth]{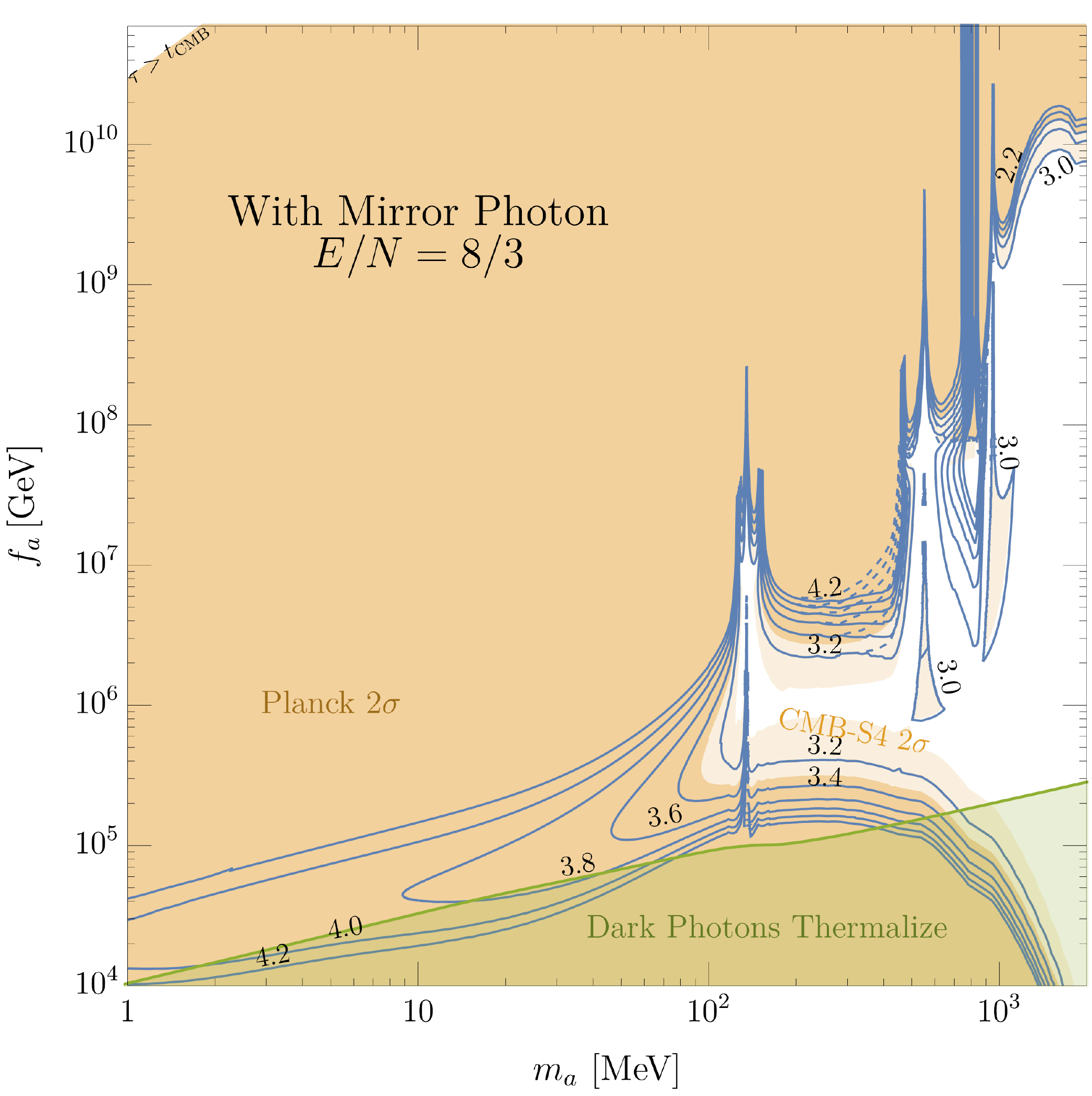} 
    \caption{Contours of $N_{\rm eff}$ in the $m_a - \tau$ (left) and $m_a - f_a$ (right) planes for $E/N = 8/3$ when including a mirror photon. The dashed contours indicate where the initial axion yield is uncertain because $T_{\rm FO}$ lies between $T_{\chi \rm PT}$ and $T_{\rm g PT}$.  The dashed contours bound this uncertainty by showing the value of $N_{\rm eff}$ taking the minimum initial axion yield while the solid contours show the value of $N_{\rm eff}$ taking the maximum initial axion yield as given in Fig.~\ref{fig:TFOUncertainty}. The green region shows where $f_a$ is sufficiently small that the dark photon reaches a thermal abundance and the freeze-in picture we employ breaks down. The dark orange region is excluded at $95 \%$ confidence by Planck, while the light orange shows the future reach of CMB-S4 experiment at $95 \%$ confidence.}
    \label{fig:darkPhotonNeffEN83}
    \includegraphics[width=.49\textwidth]{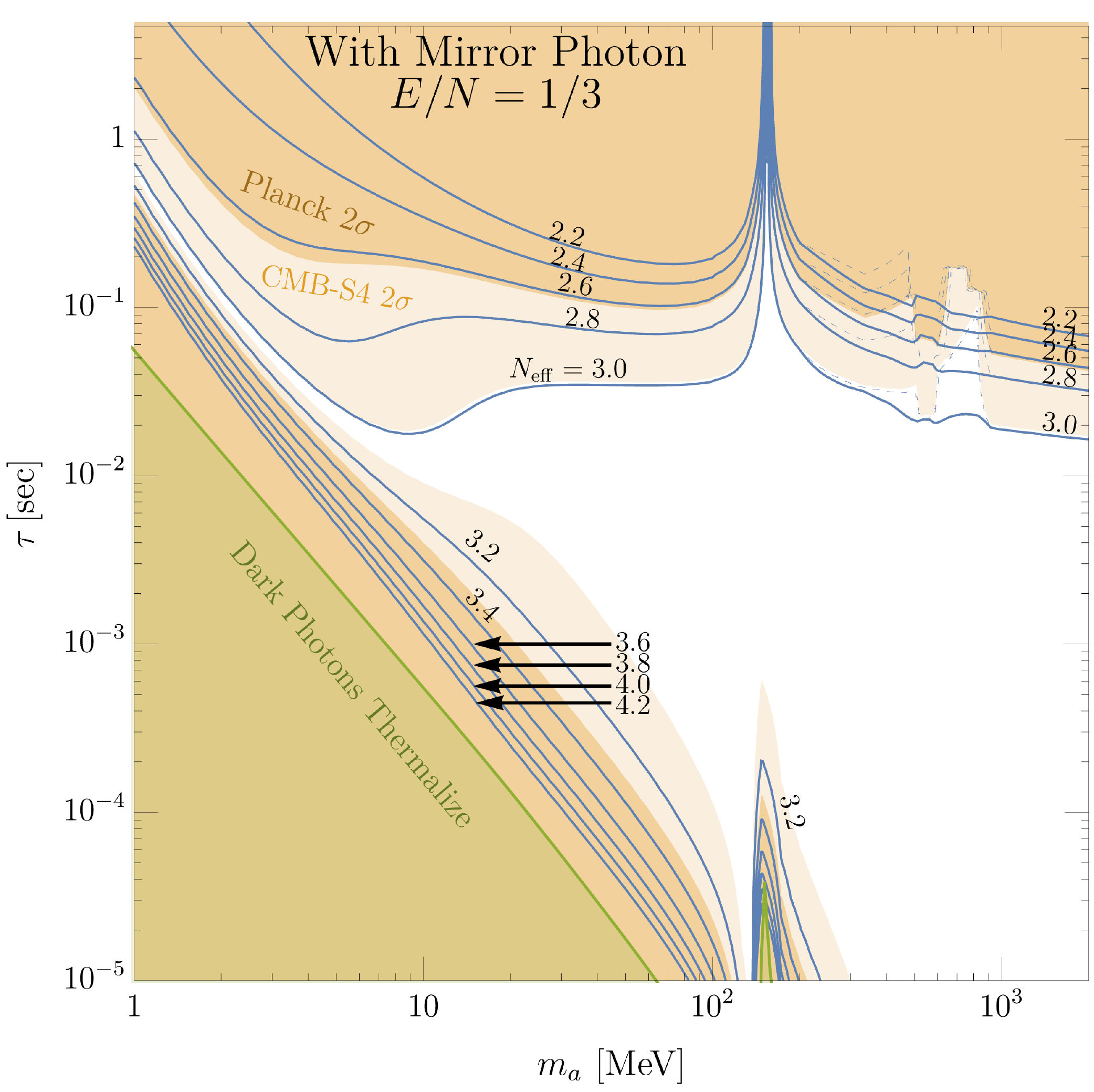} 
    \includegraphics[width=.49\textwidth]{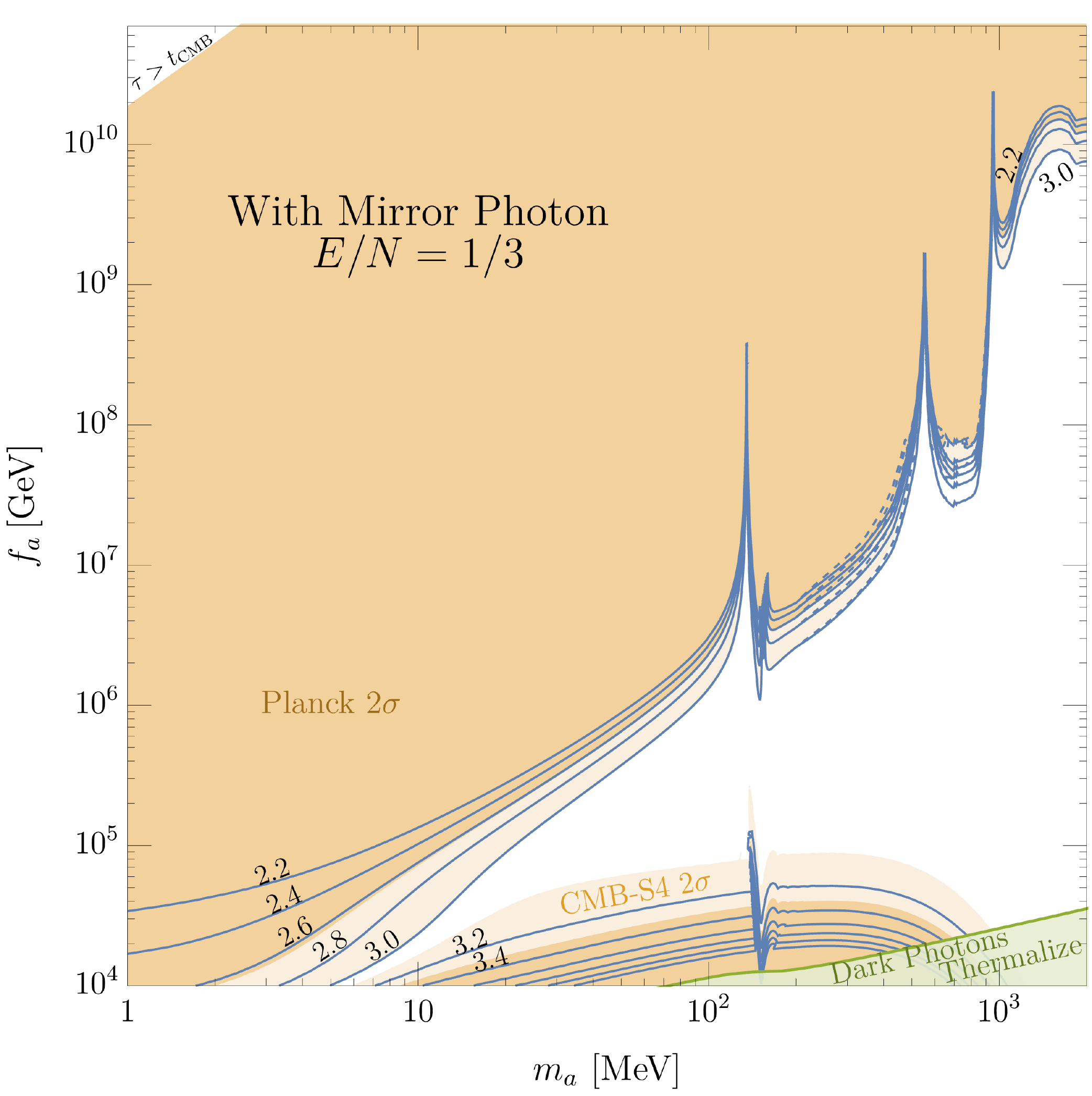} 
    \caption{
    Same as Fig.~\ref{fig:darkPhotonNeffEN83}, but for $E/N = 1/3$.}
     \label{fig:darkPhotonNeffEN13}
\end{figure}
Conservatively, we take the initial $\gamma'$ density at $T_{\chi \rm PT}$ to be zero. Due to the substantial change in Standard Model degrees of freedom across $T_{\rm QCD}$, freeze-in production of $\gamma'$ much earlier than $T_{\chi \rm PT}$ is diluted and this conservative estimate is a fairly good approximation to the true initial abundance of $\gamma'$. Last, the Hubble expansion rate, \eqref{eq:hubble},  is modified to include the additional mirror photon energy density,
\begin{align}
    \label{eq:hubbleDarkPhoton}
     H^2 = \left(\frac{\dot{R}}{R}\right)^2 &= \frac{8 \pi G}{3}\left(\rho_\gamma + \rho_e + \rho_\mu + \rho_\pi + \rho_{\delta_{\rm QCD}} +
    \rho_{\nu_e} +
    \rho_{\nu_{\mu\tau}} +
    \rho_a + \rho_{\gamma'}\right).
\end{align}

Figures \ref{fig:darkPhotonNeffEN83} and \ref{fig:darkPhotonNeffEN13} show the contours of $N_{\rm eff}$ when including a massless mirror photon in heavy axion cosmologies for $E/N = 8/3$ and $1/3$, respectively. As before, the blue region indicates where $N_{\rm eff}$ is excluded by current CMB measurements at the 2$\sigma$ level. Note that whereas $\Delta N_{\rm eff}$ is strictly negative in the case without the mirror photon (Figs.~\ref{fig:DneffPlotEN0} and \ref{fig:DneffPlotEN83}), the case with the mirror photon gives positive $\Delta N_{\rm eff}$ for most of the parameter space where the mirror photon dominates the branching ratio.

The green region indicates where the mirror photon reaches equilibrium and the freeze-in picture breaks down. This occurs when $\Gamma_{a \gamma' \gamma'} \gtrsim H(T = m_a)$, or equivalently, roughly when
\begin{align}
f_a \lesssim 10^5 {\, \rm GeV} \left(\frac{m_a}{125 \, {\rm MeV}} \right)^{ \scalebox{1.01}{$\frac{1}{2}$}} \left(\frac{E/N}{8/3}\right). \qquad (\text{Mirror Photon reaches thermal equil.})
\end{align}
Within this green region, the contour values for $N_{\rm eff}$ are artificially high because the mirror photon acquires a greater than thermal abundance due to the lack of a back reaction in Eq.~\eqref{eq:axionPhaseSpaceDarkPhoton}. In principle, capping the mirror photon abundance at a thermal abundance suggests that realistic $N_{\rm eff}$ contours within the green region are roughly fixed at the value of $N_{\rm eff}$ on the boundary of the green region; that is, the value of $N_{\rm eff}$ when the mirror photon \textit{just} acquires a thermal abundance from the freeze-in picture. For axions decaying prior to neutrino decoupling, this argument suggests $N_{\rm eff} \gtrsim 3.8$ in the green region when $E/N = 8/3$. Such a large $\Delta N_{\rm eff}$ is already excluded by experiments and thus the freeze-in picture is generally valid within the experimentally allowed region. However, for axions decaying after neutrino decoupling, it is possible that a tuned cancellation between the $\rho_{\gamma'}$ energy deposit (positive $\Delta N_{\rm eff}$ contribution) and the heating of $\rho_\gamma$ relative to neutrinos (negative $\Delta N_{\rm eff}$ contribution) can occur in the green region. We leave the calculation of such a tuned cancellation to future work, but we expect that the parameter region with $\tau \gtrsim 1$ sec is excluded by BBN. This is because  to cancel the positive $\Delta N_{\rm eff}$, the axion decays before the proton-neutron conversion completes, and the Helium abundance will be affected.   

\section{Light Axion and Dark Matter Over-Production}
\label{sec:lightaxion}

In this section, we discuss constraints on the heavy QCD axion for $m_a < 1$ MeV. As its mass is decreased the axion remains excluded by $\Delta N_{\rm eff}$ until it decays after the CMB era.  However, at this point, the axions remains excluded from its contribution to dark matter at the CMB era, until a significant further reduction in its mass.  Before entering the allowed light axion region, there is an excluded region from free-streaming effects on large scale structure.

\begin{figure}[tb]
    \centering
    \includegraphics[width=.85\textwidth]{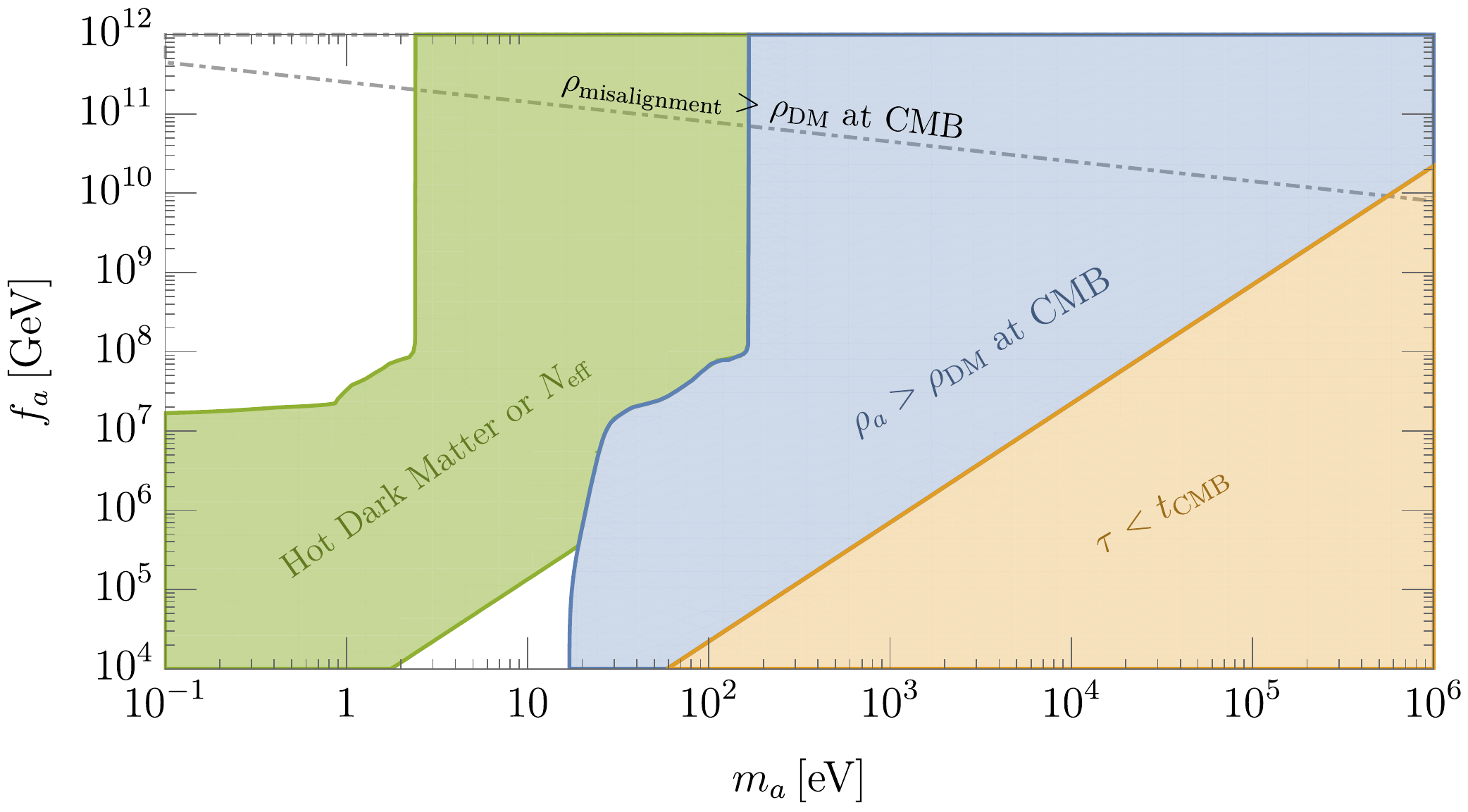} 
    \caption{Cosmological constraints on the low mass axion region, $m_a < 1$ MeV. The axion decays prior to recombination ($t_{\rm CMB} \approx 370,000$ yr) in the orange region, which is excluded by the $N_{\rm eff}$ constraint. The boundary denoting $\tau = t_{\rm CMB}$ is given for $E/N = 0$ (solid) and $8/3$ (dashed) without a mirror photon; and for $1/3$ (dotted) and $8/3$ (dot-dashed) with a mirror photon decay channel. In the blue region, the axion decays after recombination, but possesses a matter energy density during the CMB era exceeding that measured by Planck~\cite{Planck:2018vyg}, assuming the axion froze-out in the early Universe. For $f_a \lesssim 10^8$ GeV, $T_{\rm FO}$ occurs below the electroweak scale, and hence the axion energy density at $t_{\rm CMB}$ grows because of the reduction in  $g_*$.
    In the green region, the axion suppresses the structure formation as hot dark matter or contributes to dark radiation.
    }
    \label{fig:lowMassConstraintPlot}
\end{figure}

As can be seen from Fig.~\ref{fig:DneffPlotEN0} and \ref{fig:DneffPlotEN83}, the CMB limit on $\Delta N_{\rm eff}$ excludes 1 MeV $\lesssim m_a \lesssim 3$ MeV for any $f_a$. This exclusion from $\Delta N_{\rm eff}$ continues for $m_a < 1$ MeV, until the axions decay after recombination.%
\footnote{The parameter space where the axions decay before recombination but are in thermal equilibrium at BBN and generate a positive $\Delta N_{\rm eff}$ is discussed in \cite{Salvio:2013iaa,DEramo:2021lgb,DEramo:2021psx}, but is weaker than the bounds in this work.} 
Hence, the blue region of Fig.~\ref{fig:lowMassConstraintPlot}, where $\tau < t_{\rm CMB} \equiv 370,000$ yrs ($z \approx 1100$) \cite{ParticleDataGroup:2020ssz}, is excluded by $\Delta N_{\rm eff}$.  
For $\tau \gtrsim t_{\rm CMB}$ and $m_a  > 0.13 (\Omega_{\rm DM} h^2/0.12)Y_a(T_{\rm FO,g})^{-1}$ eV, the axion is sufficiently heavy and long-lived to exceed the observed dark matter density at the CMB era, as shown by the excluded orange region of Fig.~\ref{fig:lowMassConstraintPlot}. Here, we assume the reheat temperature is sufficiently high that axions undergo freeze-out, as shown in Fig.~\ref{fig:overviewPlot}, giving an axion freeze-out yield, $Y_a(T_{\rm FO,g})$, typically between $2 \times 10^{-3}$ to $2 \times 10^{-2}$. Relaxing this assumption, by taking $T_{\rm RH}$ below $T_{\rm FO,g}$ or by introducing dilution between axion freeze-out and BBN, reduces the orange excluded region.

At lower axion masses, the free-streaming of axions suppresses the matter spectrum (i.e., the axion is hot dark matter) and for even smaller masses, the axion works as dark radiation. We reinterpret the bound derived in \cite{Xu:2021rwg} for our framework and exclude the green-shaded region in Fig.~\ref{fig:lowMassConstraintPlot}. Here we conservatively impose the bound only for $\tau > t_0$, where $t_0$ is the present age of the Universe, but we expect that the bound is also applicable as long as $\tau >t_{\rm eq}$, since the suppression of the matter spectrum is dominated by the free-streaming of axions before the matter-radiation equality.    
A part of orange, blue, or green region is also excluded by other astrophysical constraints (see \cite{ciaran_o_hare_2020_3932430} for an overview), but they are generically weaker.

Finally, the misalignment mechanism~\cite{Preskill:1982cy,Abbott:1982af,Dine:1982ah} overproduces axion dark matter above the dotted-dashed line in~Fig.~\ref{fig:lowMassConstraintPlot}. Here we assume an $O(1)$ misalignment angle.

\section{Conclusions}
\label{sec:conclusions}

The strong CP problem can be addressed in a wide variety of axion models.  The minimal ones, where the QCD axion mass is solely given by strong QCD dynamics, predict $m_a f_a \sim (100 \; \MeV)^2$, but are typically plagued by a quality problem. This quality problem can be ameliorated or solved in a range of ``Heavy QCD Axion" theories, where $m_a f_a$ is orders of magnitude larger than in the minimal models. The constraints and search strategies for these heavy axions are completely different from those for the conventional lighter axion.
An important constraint from CMB data arises if the axion lifetime is in the range of $10^{-1}\,$s -- $10^{-12}\,$s, decaying after neutrino decoupling at the MeV era, but before last scattering of the CMB at the eV era.  In this case, the energy density of neutrinos is diluted, affecting the dark radiation at the CMB era, $N_{\rm eff}$, which has been precisely measured by the Planck Collaboration \cite{Planck:2018vyg} and will be significantly improved by CMB Stage 4 experiments \cite{CMB-S4:2016ple}. Thus, theory and experiment both strongly motivate a detailed study of this cosmological bound on heavy axions. For axion masses above 1 MeV, except for accelerator searches at low values of $f_a$, $N_{\rm eff}$ is the strongest bound on the heavy QCD axion, and is the focus of this work. 

A well-motivated and predictive model involves a mirror copy of the SM with a large axion mass generated by the mirror QCD interaction.  In this case there is a competition between axion decays to photons diluting the neutrino contribution to $N_{\rm eff}$ and axion decays to mirror photons directly enhancing $N_{\rm eff}$. We have also provided a detailed analysis of the $N_{\rm eff}$ bound in theories with a light mirror photon.  

Our analysis of the $N_{\rm eff}$ bound takes into account key pieces missing from previous studies of the heavy QCD axion by developing a Boltzmann code that follows the evolution of the momentum distribution for the axion. The mesons and gluons of QCD play a key role; we include axion-pion scattering, axion decay to final states involving mesons, and axion-meson mixing. In addition we follow a detailed cosmological evolution from the initial axion abundance from freeze-out to the non-trivial QCD contributions in the Friedmann equations.

\begin{figure}[tb]
    \centering
    \includegraphics[width=.95\textwidth]{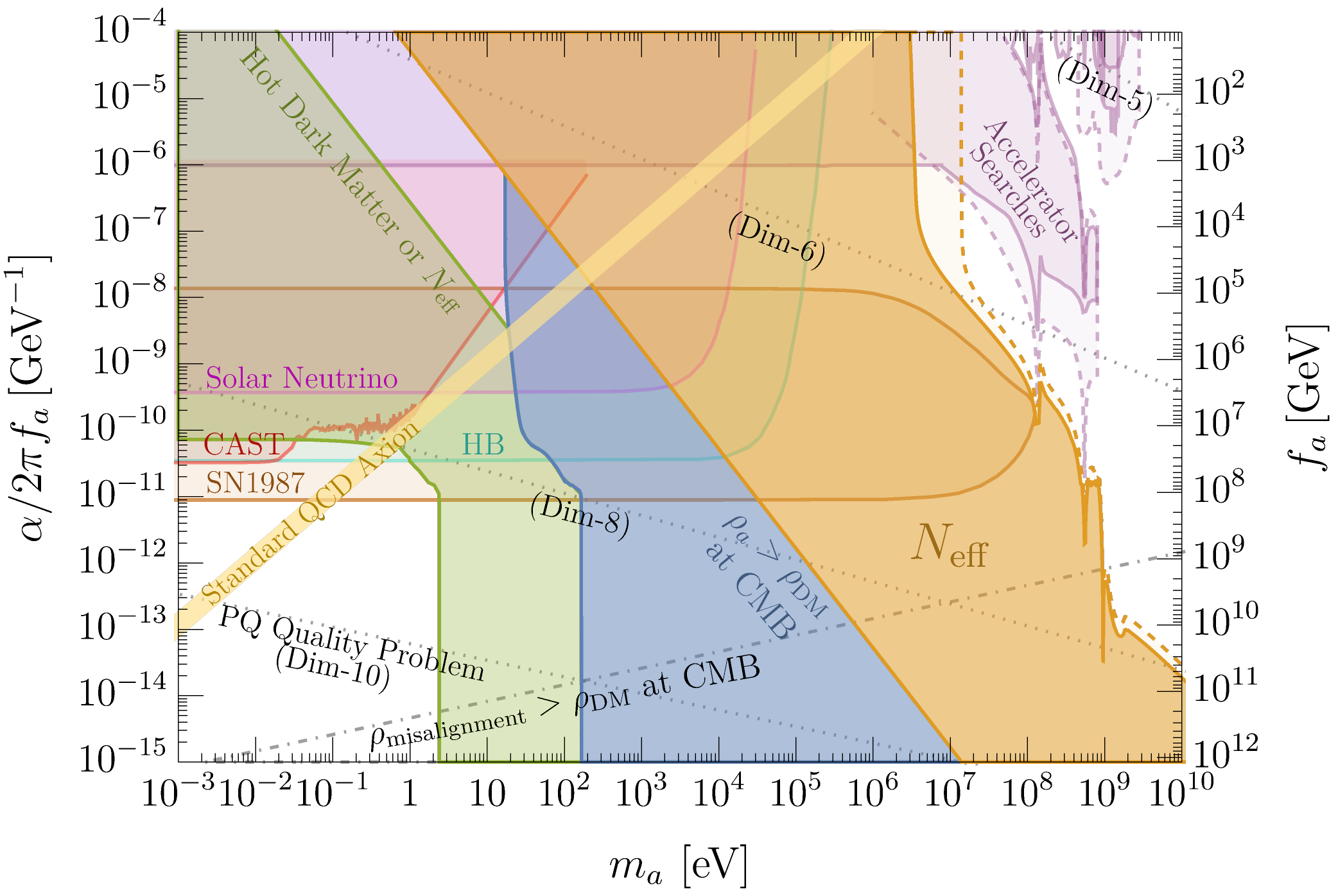} 
    \caption{Overview of the excluded parameter space as determined in this work (blue region) in relationship to other QCD axion bounds. The solid orange region shows the $2\sigma$ exclusion region on $N_{\rm eff}$ as determined by Planck for the heavy QCD axion with $E/N = 0$ (see Fig.~\ref{fig:DneffPlotEN0}). The dashed orange contour shows the future reach of $N_{\rm eff}$ from CMB-S4. The left boundary of the orange region indicates where the axion decays after recombination which marks the parameter space where $N_{\rm eff}$ contstraints become inapplicable (see Fig.~\ref{fig:lowMassConstraintPlot}). The blue region indicates where the axion energy density at recombination is greater than that of dark matter, assuming the reheat temperature of the universe is high enough that the axion froze-out with a thermal abundance. The solid and dashed purple regions shows the present and future bounds from accelerator searches, respectively \cite{Ertas:2020xcc,Dobrich:2015jyk,Dolan:2017osp,NA64:2020qwq,FASER:2018eoc,Gori:2020xvq,Aloni:2018vki,Mariotti:2017vtv,Kelly:2020dda,Chakraborty:2021wda,Bertholet:2021hjl}. The brown region shows the bound from supernova 1987a on axions with hadronic interactions, \cite{Chang:2018rso}, green from horizontal-branch cooling \cite{Ayala:2014pea}, red from CAST \cite{CAST:2007jps,CAST:2017uph}, and pink from solar neutrinos \cite{Vinyoles:2015aba}. Below the dot-dashed contour, the  axion energy density arising from the misalignment mechanism, with a misalignment angle of unity, is greater than the observed dark matter energy density. In the green region, the axion suppresses the structure formation or contributes to $N_{\rm eff}$. Below the dotted contours, the axions suffers a PQ quality problem arising from operators of the labeled dimension. The diagonal yellow strip indicates the standard QCD axion, which highlights the severity of the PQ quality problem for low mass axions.} 
    \label{fig:masterPlot}
\end{figure}

Our results for the CMB $N_{\rm eff}$ constraints on the heavy QCD axion, in the absence of a mirror photon, are shown in Figs.~\ref{fig:DneffPlotEN0} and \ref{fig:DneffPlotEN83}, and are very powerful. Planck excludes large areas of parameter space, especially at large $f_a$ , but large areas remain at low $f_a$, where the quality problem is solved for operators of dimension 6 and larger. The discovery reach of CMB-S4 at larger values of $m_a$ is modest, but improves at lower $m_a$: for example, if $f_a$ is of order $10^4$ GeV, CMB-S4 will see a signal for $m_a$ in the range of (3-10) MeV. 

We find two important differences from standard results, illustrated by comparing Fig.~\ref{fig:DneffPlotNoPion} with our results shown in Figs.~\ref{fig:DneffPlotEN0} and \ref{fig:DneffPlotEN83}. First, resonances occur when the axion mass is around the $\pi_0$, $\eta$, and $\eta'$ masses, greatly affecting $N_{\rm eff}$ for axion masses between $(100-1000)$ MeV. Second, by including mesons and gluons, we correctly take account of the axion lifetime. This is a large effect, especially at large $m_a$, increasing the decay rate by orders of magnitude as $m_a$ rises above $\sim 1$ GeV; this point is apparent in the right panels of Figs. \ref{fig:DneffPlotEN0} and \ref{fig:DneffPlotEN83} where regions with higher $f_a$ open up.

In the presence of a light mirror photon, our results for the CMB $N_{\rm eff}$ constraints are shown in Figs.~\ref{fig:darkPhotonNeffEN83} and \ref{fig:darkPhotonNeffEN13} for $E/N$ = 8/3 and 1/3 respectively. For $E/N=8/3$, $m_a < 100 \; \MeV$ is excluded for all values of $f_a$.  A substantial fraction of the allowed region with $ 100 \; \MeV < m_a < 500 \; \MeV $ will be probed by CMB-S4 via a positive signal for $\Delta N_{\rm eff}$. For $E/N=1/3$, the CMB $N_{\rm eff}$ bound is considerably weaker.  A new allowed region opens up at lower axion masses, $ 1 \; \MeV < m_a < 100 \; \MeV $, where dark radiation from the mirror photon compensates neutrino dilution from axion decays.  A large fraction of this region gives a CMB-S4 signal, with $\Delta N_{\rm eff}$ positive (negative) for smaller (larger) values of $f_a$.
These allowed regions both solve the quality problem for operators of dimension 6 and larger.

The current and future $95 \%$ confidence limit on the axion mass from $N_{\rm eff}$ in this work are shown in comparison to other cosmological and astrophysical constraints in Fig.~\ref{fig:masterPlot}. The limits on $N_{\rm eff}$ in this work provide the strongest constraints on heavy QCD axions for $m_a \gtrsim 1$ MeV and $f_a \gtrsim 10^5$ GeV. Complementary constraints at small $f_a$ arise from direct heavy axion searches at accelerators \cite{Kelly:2020dda, Bertholet:2021hjl} as shown by the purple shaded regions.  Producing axions in a beam dump, such as the DUNE Near Detector, and discovering their subsequent decays,  will allow the region enclosed by the dashed purple contour to be probed \cite{Kelly:2020dda}. Furthermore, the dashed purple contour at higher $m_a$ and low $f_a$ can be probed by observing axions in B meson decays at Belle \cite{Bertholet:2021hjl}. The bound from axion cooling of Supernova 1987A has uncertainties arising from the temperature and density profiles of the supernova, and  has been computed for a variety of such profiles in \cite{Chang:2018rso}; we show a conservative case. 
Constraints on the decay of the axion from extragalactic background light or CMB spectral distortions are derived in~\cite{Cadamuro:2011fd}, but the constraints do not exclude the parameter region that is allowed in Fig.~\ref{fig:masterPlot}.

The bounds shown in Fig.~\ref{fig:masterPlot}, and elsewhere in the paper, are computed assuming that the reheat temperature of the universe $T_{\rm RH}$ is above the axion freeze-out temperature, and that there is no subsequent dilution of the axion abundance, for example from late decaying particles. Removing this assumption relaxes the bounds, since axion production occurs via freeze-in rather than freeze-out, or is diluted after freeze-out. Since the freeze-out temperature decreases as $f_a$ drops, relaxing the bounds becomes harder at lower $f_a$.
For $f_a < 10^4$ GeV, the axion is kept into thermal equilibrium even at $T< 4$ MeV, and the BBN bound $T_{\rm RH}> 4$ MeV~\cite{Kawasaki:1999na,Kawasaki:2000en,Hasegawa:2019jsa} excludes the possibility of relaxing the bound.
It may be plausible that the reheating temperature is below the freeze-out temperature for large $f_a$, but solving the quality problem favors low $f_a$, and it is typically harder to obtain a large enough $m_a$ for large $f_a$; see e.g., Eq.~(\ref{eq:ma}).

The next decade will yield exciting and important answers to axion physics.  Heavy QCD axions provide a highly-motivated solution to the strong CP problem. Unlike the standard QCD axion, which induces a small $\Delta N_{\rm eff}$ signal \cite{DEramo:2021psx}, heavy QCD axions can generate substantial $\Delta N_{\rm eff}$ signals that can be probed by the exquisite sensitivity of current and near future CMB telescopes. Moreover, in theories without a mirror photon, this signal results from a {\it depletion} of the cosmic neutrino abundance, providing a  less common fingerprint of a \textit{negative} contribution to $\Delta N_{\rm eff}$. Such a measurement would determine a correlation between the axion mass and decay constant.

\section*{Acknowledgement}
We thank Jeffrey Anderson, Raymond Co, and Jonathan Wurtele for providing useful computational resources. This work was supported in part by the Director, Office of Science, Office of High Energy and Nuclear Physics, of the US Department of Energy under Contracts DE-AC02-05CH11231 (LJH) and by the National Science Foundation under grant PHY-1915314 (LJH).

\appendix

\section{Calculation of Axion-Pion Collision Term}
\label{app:axionPionCollisionTerm}
\begin{figure}[tb]
    \centering
    \includegraphics[width=.25\textwidth]{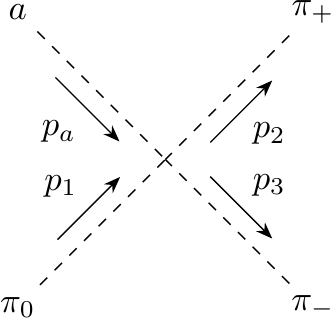}\hfill
    \includegraphics[width=.25\textwidth]{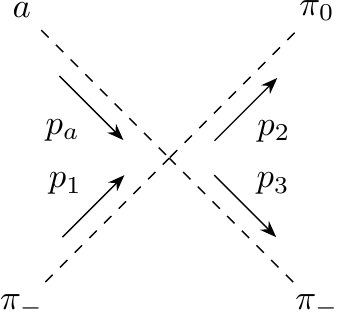}\hfill
    \includegraphics[width=.25\textwidth]{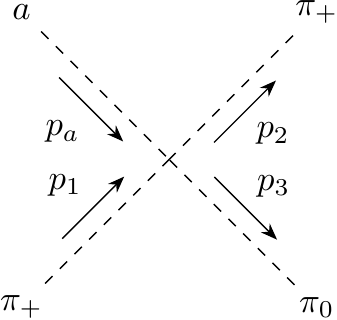}\hfill 
    \caption{Feynman diagrams associated with axion-pion scattering involving charged pions.}
    \label{fig:axionPionFeynmanDiagaom}
\end{figure}

In this section, we compute the axion-pion scattering collision term, $C_\pi$, used in the axion-Boltzmann equation, \eqref{eq:boltzmannSM}. 
The interaction between an axion, neutral pion, and two charged pions includes the following three interactions, $a(p_a) + \pi_0(p_1) \leftrightarrow \pi_+(p_2) + \pi_-(p_3)$, $a(p_a) + \pi_-(p_1) \leftrightarrow  \pi_0(p_2) + \pi_-(p_3)$, and $a(p_a) + \pi_+(p_1) \leftrightarrow \pi_+(p_2) + \pi_0(p_3)$ as shown in Fig.~\ref{fig:axionPionFeynmanDiagaom}. According to the chiral Lagrangian \eqref{eq:chiralLagrangian}, the matrix element for $a+\pi_0 \rightarrow \pi_+ + \pi_-$ is
\begin{align}
    \mathcal{M}_{a + \pi_0 \rightarrow \pi_+ + \pi_-} = \frac{3}{2} \frac{A}{f_a f_\pi}\frac{1}{1-r^2}(s - m_\pi^2),
\end{align}
where $s = (p_a + p_1)^2$. The matrix elements for $a+\pi_- \rightarrow \pi_- + \pi_0$  and $a+\pi_+ \rightarrow \pi_+ + \pi_0$ scattering are obtained by the four momentum mapping $p_1 \rightarrow -p_2$ and $p_1 \rightarrow -p_3$, respectively. As a result, the total squared amplitude for scatterings involving the axion and charged pions is 
\begin{align}
    \label{eq:axionChargedPionScattering}
    |\mathcal{M}|^2_{a + \pi_i \rightarrow \pi_j^\dagger+ \pi_k^\dagger} &=
     |\mathcal{M}|^2_{a + \pi_0 \rightarrow \pi_+ + \pi_-}
     +
     |\mathcal{M}|^2_{a + \pi_- \rightarrow \pi_- + \pi_0}
      +
     |\mathcal{M}|^2_{a + \pi_+ \rightarrow \pi_+ + \pi_0}
     \\
    &= \left(\frac{3}{2}\frac{A}{f_a f_\pi}\frac{1}{1-r^2}\right)^2 \left[(s - m_\pi^2)^2 + (t - m_\pi^2)^2 + (u - m_\pi^2)^2\right] 
    \nonumber \\ 
    &= \left(\frac{3}{2}\frac{A}{f_a f_\pi}\frac{1}{1-r^2}\right)^2 \left[ s^2 + t^2 + u^2 - 3m_{\pi}^4 -2 m_a^2 m_\pi^2 \right] , \nonumber
\end{align}
where $t \equiv (p_a - p_2)^2, \, u \equiv (p_a-p_3)^2$ and $\pi_i, \pi_j, \pi_k$ in the subscript of \eqref{eq:axionChargedPionScattering} refer to three pions of different charge $\pi_0$, $\pi_+$, $\pi_-$. Likewise, the interaction between an axion and three neutral pions generates the squared scattering amplitude
\begin{align}
    \label{eq:axionNeutralPionScattering}
     |\mathcal{M}|^2_{a + \pi_0 \rightarrow \pi_0 + \pi_0} =  \left(\frac{3}{2}\frac{A}{f_a f_\pi}\frac{1}{1-r^2}\right)^2 m_a^4.
\end{align}
Inserting the sum of the squared scattering matrix elements, \eqref{eq:axionChargedPionScattering} and \eqref{eq:axionNeutralPionScattering}, into Eq.~\eqref{eq:collisionTerm} and integrating over the phase spaces of the three pions gives the product of the axion-pion collision term and $f_{a, \rm eq} - f_a$ 
\begin{align}
    \label{eq:apicollisionTerm}
    C_\pi (f_{a, \rm eq} - f_a) =  \frac{1}{2E_a}  \int \Bigl[
    &d\Pi_{1} \, d\Pi_{2} \, d\Pi_{3}  
   (|\mathcal{M}|^2_{a + \pi_i \rightarrow \pi_j^\dagger + \pi_k^\dagger} + \frac{1}{2!}|\mathcal{M}|^2_{a + \pi_0 \rightarrow \pi_0 + \pi_0}) 
   \nonumber \\ 
   &\times \Lambda (2\pi)^4 \delta^4(p_a + p_1 - p_2 - p_3)\Bigr],
\end{align}
where
\begin{align}
       \Lambda &=  
       \label{eq:LambdaExact}
      \left(1+f_a\right)\left(1+f_{1}\right)f_2 f_{3}-
       f_a f_{1}\left(1+f_2\right)\left(1+f_{3}\right)
       \\
       \label{eq:LambdaApprox}
       &\simeq (f_{a, \rm eq} - f_a)f_{1}\left(1+f_2 + f_3\right) \, ,
\end{align}
and $f_i$ is the distribution function of particle $i$ possessing momentum $p_i$ in accordance with Fig.~\ref{fig:axionPionFeynmanDiagaom}. In going from \eqref{eq:LambdaExact} to \eqref{eq:LambdaApprox}, we take the pions to be in thermal equilibrium with the Standard Model thermal bath so that $f_1$, $f_2$, and $f_3$ follow a Bose-Einstein distribution of temperature $T$. The temperature $T$ of the strongly coupled thermal bath is inferred at each numerical time step by solving the following equation for $T$:
\begin{align}
        \label{eq:temperature}
        \sum_{i = \gamma, e, \mu, \pi, \delta_{\rm QCD}}\rho_i =  \rho_\gamma(T) + \rho_e(T) + \rho_\mu(T) + \rho_\pi(T) + \rho_{\delta \rm QCD}(T)
\end{align}
where the left-hand side of \eqref{eq:temperature} is solved from Eq.~\eqref{eq:boltzmannSM} and the right-hand side is calculated from Eq.~\ref{eq:energyDensities} for $\gamma, \mu, e, \pi$ and from \cite{Saikawa:2018rcs}  for $\rho_{\delta_{\rm QCD}}$, as described more in Appendix \ref{app:numerical}.

To calculate $C_\pi$, we first introduce another $\delta$-function in \eqref{eq:apicollisionTerm} by writing $d\Pi_{3}(2\pi)^3 = \frac{d^3 \mathbf{p}_3}{2E_3} =  d^4 p_3 \delta(p_4^2 - m_4^2) \Theta(p_4^0)$. By integrating $p_3$ over the other delta function $\delta^4(p_1 + p_2 - p_3 - p_4)$, Eq.~\eqref{eq:apicollisionTerm} simplifies to
\begin{align}
 C_\pi (f_{a, \rm eq} - f_a) =  \frac{1}{2E_a}  \int \Bigl[
    &d\Pi_{1} \, d\Pi_{2} 
   (|\mathcal{M}|^2_{a + \pi_i \rightarrow \pi_j^\dagger + \pi_k^\dagger} + \frac{1}{2!}|\mathcal{M}|^2_{a + \pi_0 \rightarrow \pi_0 + \pi_0}) 
   \Lambda 2\pi \delta(p_4^2 - m_{\pi4}^2) \Theta(p_4^0)\Bigr] \, ,
\end{align}
with the understanding that $p_3 = p_a + p_1 - p_2$. Note that the argument of the remaining delta-function, $p_3^2 - m_{3}^2$, can be written as 
\begin{align}
        \label{eq:deltaArg}
        Q + 2\left((E_a E_1- |\mathbf{p}_a||\mathbf{p}_1|\cos \alpha) -(E_1 E_2- |\mathbf{p}_1||\mathbf{p}_2| \cos \gamma)-(E_a E_2- |\mathbf{p}_a||\mathbf{p}_2| \cos \theta)\right)
\end{align}
where, in the notation of  \cite{Hannestad:1995rs}, $Q = m_3^2 - m_a^2 + m_1^2 + m_2^2 $, and $\alpha, \theta, \gamma$ are the angles between $\mathbf{p}_a$ and $\mathbf{p}_1$, $\mathbf{p}_a$ and $\mathbf{p}_2$, and $\mathbf{p}_1$ and $\mathbf{p}_2$, respectively. It is convenient to express the latter angle in terms of the former two by $\cos \gamma = \cos \alpha \cos \theta + \sin \alpha \sin \theta \cos \beta$. 

In the massless axion limit, the argument of the remaining delta function can easily be expressed in terms of $E_a$ as done in \cite{DiLuzio:2021vjd}. However, in the massive axion limit, this is impossible and it is thus more useful to move the argument of the delta function onto one of the scattering angles, as done in \cite{Hannestad:1995rs}, which we follow. In particular, the choice of the angle $\beta$ is most convenient as it only occurs once in \eqref{eq:deltaArg}. The resulting integral for $C_\pi$ is 
\begin{align}
        \label{eq:apcollisionTermSimplified}
        C_\pi (f_{a, \rm eq} - f_a) = \frac{1}{2 E_a} \int \Bigl[\left(\frac{d|\mathbf{p}_1|}{(2\pi)^3}\frac{\mathbf{p}_1^2}{2 E_1} d\cos \alpha \, d \beta  \right) \left(\frac{d|\mathbf{p}_2|}{(2\pi)^3}\frac{\mathbf{p}_2^2}{2 E_2} d\cos \theta \, d \phi  \right) \left(\frac{2 \pi \delta(g(\beta))}{\left|\frac{dg}{d\beta} \right|}\right) 
        \nonumber \\  
        \times (|\mathcal{M}|^2_{a + \pi_i \rightarrow \pi_j^\dagger + \pi_k^\dagger} + \frac{1}{2!}|\mathcal{M}|^2_{a + \pi_0 \rightarrow \pi_0 + \pi_0}) 
   \Lambda 2\pi \delta(p_4^2 - m_{\pi4}^2) \Theta(p_4^0)\Bigr]
        \
\end{align}
where $g(\beta) = p_3(\beta)^2 - m_3^2$. The integrals over the azimuthal angles  $\beta$ and $\phi$ can be done analytically due to the delta function and the lack of $\phi$ dependence in the integrand. The remaining integrals over the pion 3-momenta $|\mathbf{p}_1| \in [0,\infty)$ and $|\mathbf{p}_2| \in [0,\infty)$ and the polar angles $\cos \alpha \in [-1,1]$ and $\cos \theta \in [-1,1]$ are done numerically using Monte Carlo integration. To ensure the integration region is performed only in the kinematically allowed region, we include a Heaviside function $\Theta(1 - \cos^2 \beta_i)$ in \eqref{eq:apcollisionTermSimplified}, where $\beta_i$ is the location of the two (equal and opposite) roots of $g(\beta)$ \cite{Hannestad:1995rs}. Note that $\beta_i$ are functions of the other four integration variables. 

We can gain intuition for the cosmological effect of axion-pion scattering by calculating the thermally averaged pion-to-axion scattering rate as introduced in Eq.~\eqref{eq:axionPionScatteringRate},

\begin{align}
    \Gamma_{a \pi \leftrightarrow \pi \pi} = \frac{1}{n_{a, \rm eq}} \int \frac{d^3 \mathbf{p}_a}{(2\pi)^3} \mathcal{C}_{\pi} f_{a, \rm eq} \, \equiv \frac{T^5}{f_a^2 f_\pi^2}\frac{A^2}{(1-r^2)^2} \mathcal{F}_\pi(m_a,T) ,
\end{align}
where $n_{a,\rm eq}$ is the thermal number density of axions with mass $m_a$ at temperature $T$. As before, $A \equiv \frac{1}{3}(1-z)/(1+z)$, $z \equiv m_u/m_d$, and $r \equiv m_a/m_\pi$.
\begin{figure}[tb]
    \centering
    \includegraphics[width=.99\textwidth]{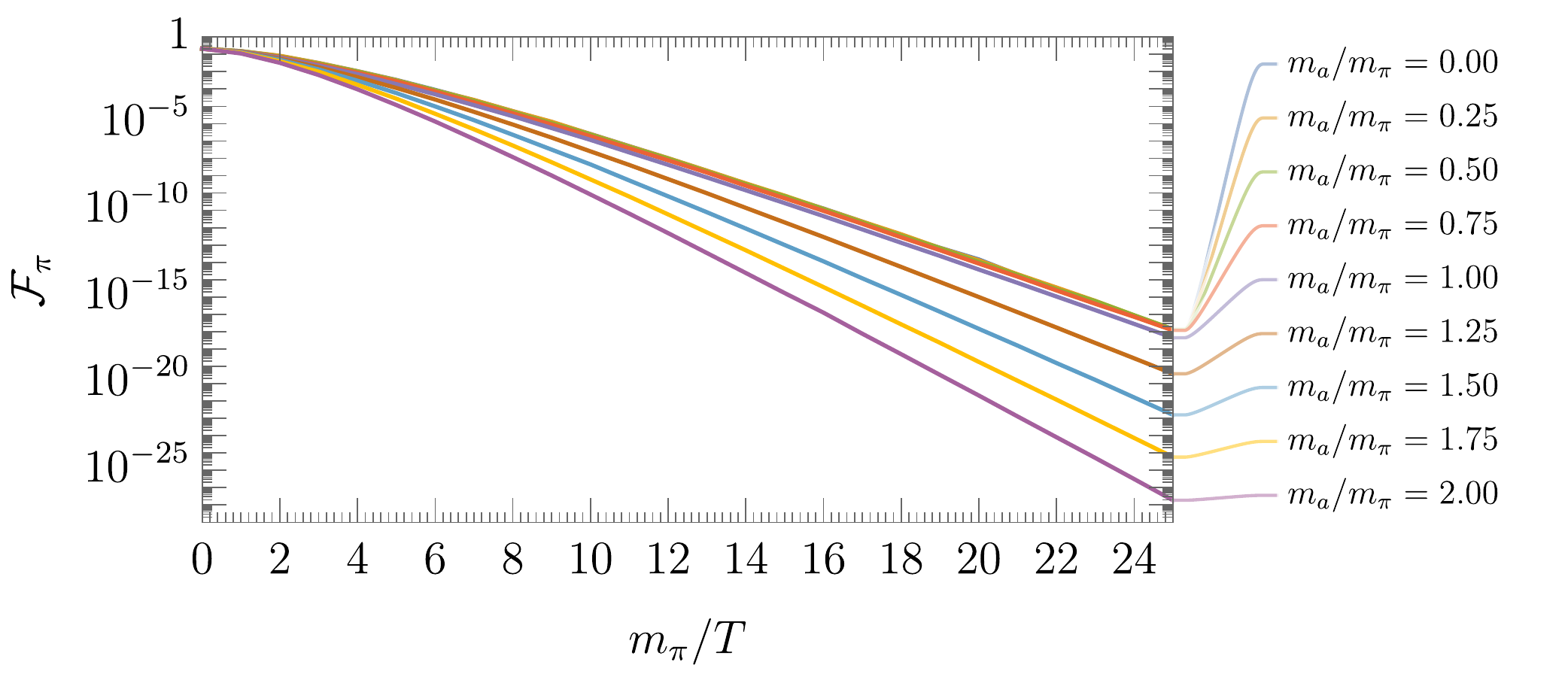} 
    \caption{Numerical evaluation of $\mathcal{F}_\pi$ as a function of $m_\pi/T$ for a variety of axion masses.  $\mathcal{F}_\pi$ becomes suppressed for $m_a > m_\pi$ and $T < m_\pi$ since the production of axions must occur at the Boltzmann tail of the incoming pions. The massless axion limit, which has been computed previously in literature \cite{Hannestad:2005df,DiLuzio:2021vjd}, corresponds to the $m_a/m_\pi = 0$ contour. }
    \label{fig:fPi}
\end{figure}
Fig.~\ref{fig:fPi} shows $\mathcal{F}_\pi$ as a function of $m_\pi/T$ for a variety of axion masses. For $m_a \ll m_\pi$, $\mathcal{F}_\pi$ reduces to previous results in the literature for massless axions \cite{Hannestad_2005,DiLuzio:2021vjd}, with $\mathcal{F}_\pi$ related to the $h_{\rm LO}$ function defined in \cite{Hannestad_2005,DiLuzio:2021vjd} by the mapping $\mathcal{F}_\pi(m_\pi/T, m_a = 0) \equiv 0.212 \, h_{\rm LO}(m_\pi/T)$. According  to Fig.~\ref{fig:fPi}, for $T < m_\pi$, $\mathcal{F}_\pi$ drops as $m_a$ increases. This is because in this regime, only the Boltzmann tail of pions with high energies $m_a$ can kinematically scatter to produce axions.

\section{\texorpdfstring{$N_{\rm eff}$}{Neff} for  Large  Axion Masses}
\label{app:AsymptoticNeff}
In the main text, we show the $N_{\rm eff}$ constraint for $m_a < 2$ GeV.
In this appendix, we derive the constraint for $m_a > 2$ GeV.

In the left panels of Fig.~\ref{fig:DneffPlotEN0}, \ref{fig:DneffPlotEN83}, \ref{fig:darkPhotonNeffEN83} and \ref{fig:darkPhotonNeffEN13}, the $N_{\rm eff}$ contours become approximately horizontal above $m_a \approx 1$ GeV, signifying that $N_{\rm eff}$ is dictated mainly by the lifetime of the axion in this region. However, careful inspection indicates that the slope of the contours is not quite flat, but slightly decreases as $m_a$ grows. The reason is, for fixed $\tau$, the energy density of the axion at decay increases with increasing mass. This follows because axions in this region have such large $f_a$ that they decouple early and decay non-relativistically. Thus, what actually sets the $N_{\rm eff}$ contours in the $m_a > 2$ GeV region is how much energy density they deposit into the thermal bath right at neutrino decoupling.  

For example, let $\rho_{\rm max}$ be the maximum energy density that can be deposited at a certain time $t_*$ so that $N_{\rm eff}$ does not drop below an arbitrary contour, $N_{\rm eff,0}$, which we will take to be $\simeq 2.62$, the $2\sigma$ limit on $N_{\rm eff}$ allowed by Planck. Choose a point $(m_{0}, {\tau_0})$ that lies on this  $N_{\rm eff,0}$ contour in the $m_a \gtrsim 1$ GeV region. Analytically, the energy density of this non-relativistic axion at time $t_*$ is
\begin{align}
        \label{eq:rhoMax}
        \rho(t_*) = \rho_{\rm init} \left(\frac{a_{\rm init}}{a(t_*)}\right)^3 e^{-t_*/\tau_0} = m_{0} Y_{0} s(t_*) e^{-t_*/\tau_0} \equiv \rho_{\rm max}
\end{align}
where $Y_0$ is the axion yield, and $s(t_*)$ the entropy density at time $t_*$. Note Eq.~\eqref{eq:rhoMax} defines $\rho_{\rm max}$. It follows that for axions of different $(m_a, \tau)$ to possess the same energy density as $\rho_{\rm max}$ at time $t_*$, requires 
\begin{align}
        \frac{m_0 Y_0}{m_a Y_a} = \exp t_*\left(\frac{1}{\tau_0} - \frac{1}{\tau}\right) \,
\end{align}
or equivalently,
\begin{align}
        \tau = \frac{\tau_0}{1 + \frac{\tau_0}{t_*} \ln \left(\frac{m_a Y_a}{m_0 Y_0} \right)},
\end{align}
We perform a numerical fit of the $N_{\rm eff,0} = 2.62$ contour with the anchor point $(m_0, \tau_0) = (2.0 \, {\rm GeV}, \, 4.3 \times 10^{-2} \, {\rm s})$ and find $t_* \simeq 0.17 \, {\rm s}$. Other anchor points give similar $t_*$. Eq.~\eqref{eq:tauAsymptotic} follows from this fit.

\section{Numerical Approaches}
\label{app:numerical}
In this section, we discuss the numerical techniques used to solve the Boltzmann equation describing the cosmological evolution of the axion. As mentioned in Sec. \ref{sec:boltzmannEqns}, we employ the method of lines technique to convert the Boltzmann system of partial differential equations into a system of ordinary differential equations. In particular, we discretize the partial differential equation governing the axion phase space density, \eqref{eq:axionPhaseSpace}, into a partition of $N$ ordinary differential equations, $\{f_\mathbf{\tilde{p},i}(t)\}$ of time, with each ODE corresponding to the time evolution of the phase space density at a fixed comoving momentum, $|\tilde{\mathbf{p}}_i| = |\mathbf{p}_i|a(t)$, with $i \in \{1,...,N\}$. In our numerical setup, we split \eqref{eq:axionPhaseSpace} into $N = 24$ ODEs of logarithmically equidistant $|\tilde{\mathbf{p}}_i|$ where $i = 1$ corresponds to the fixed comoving momentum $|\tilde{\mathbf{p}_1}| = e^{-8} T_{\chi \rm PT}$ and $i = N$ corresponds to $|\tilde{\mathbf{p}}_N| = e^{7/2} T_{\chi \rm PT}$. Note that because of entropy conservation --- which holds except for when the axion dominates the energy density of the universe before decaying --- each fixed comoving momentum equals the ratio of the physical momentum to the temperature, $|\mathbf{p}|/T$. We have verified the convergence of our results by checking that $N_{\rm eff}$ changes by less than 1\% when using larger $N$. In addition, our results for $N_{\rm eff}$ for just the Standard Model cosmology, or equivalently, when the axion decays far before neutrino decoupling, is $3.040$. This value slightly differs from the true Standard Model value of $N_{\rm eff}^{\rm SM} = 3.044$ \cite{Bennett:2020zkv,Froustey:2020mcq,Zyla:2020zbs} by $\sim 0.2 \%$ because we do not include effects from QED corrections or neutrino oscillations.

The dynamical timescale ($\sim 1/H$, where $H$ is Hubble) involved in the cosmological evolution of the Boltzmann equation spans many orders of magnitude from the end of the QCD phase transition to past neutrino decoupling. Hence, we solve the system of Boltzmann equations in terms of the logarithmic timescale $y = \ln (t/t_{\chi \rm PT})$, where $t_{\chi \rm PT} = 1/2H(T_{\chi \rm PT})$ is the starting time of the Boltzmann code.

Last, we determine the extra QCD contributions to the energy density ($\rho_{\delta_{\rm QCD}}$) and pressure ($ P_{\delta_{\rm QCD}}$) arising from heavy mesons and from ideal gas law deviations using the calculations of \cite{Saikawa:2018rcs}, which tabulated $g_*(T)$ and $g_{*S}(T)$ for the Standard Model across the QCD phase transition.
\begin{figure}[tb]
    \centering
    \includegraphics[width=.45\textwidth]{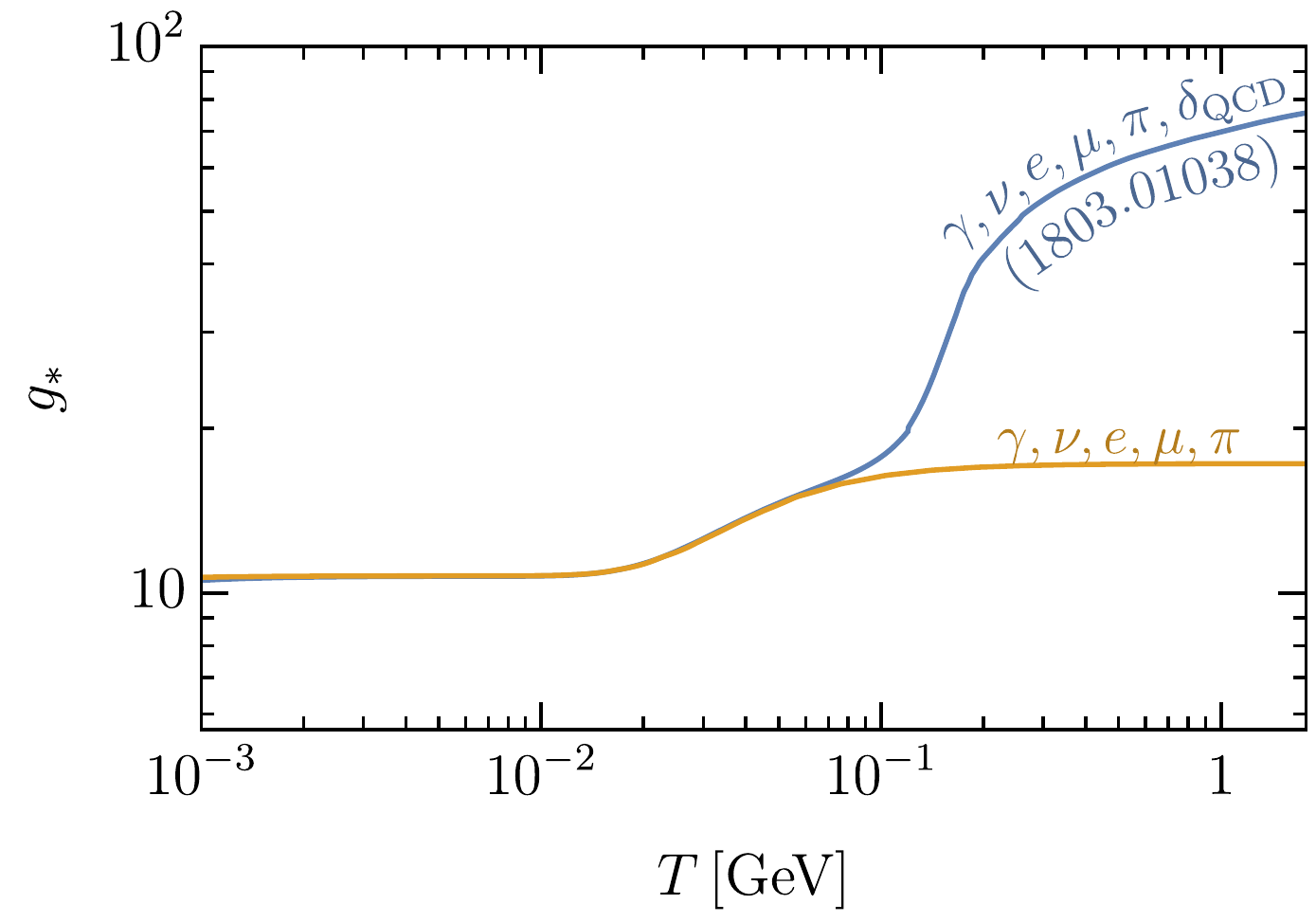}\hfill
    \includegraphics[width=.47\textwidth]{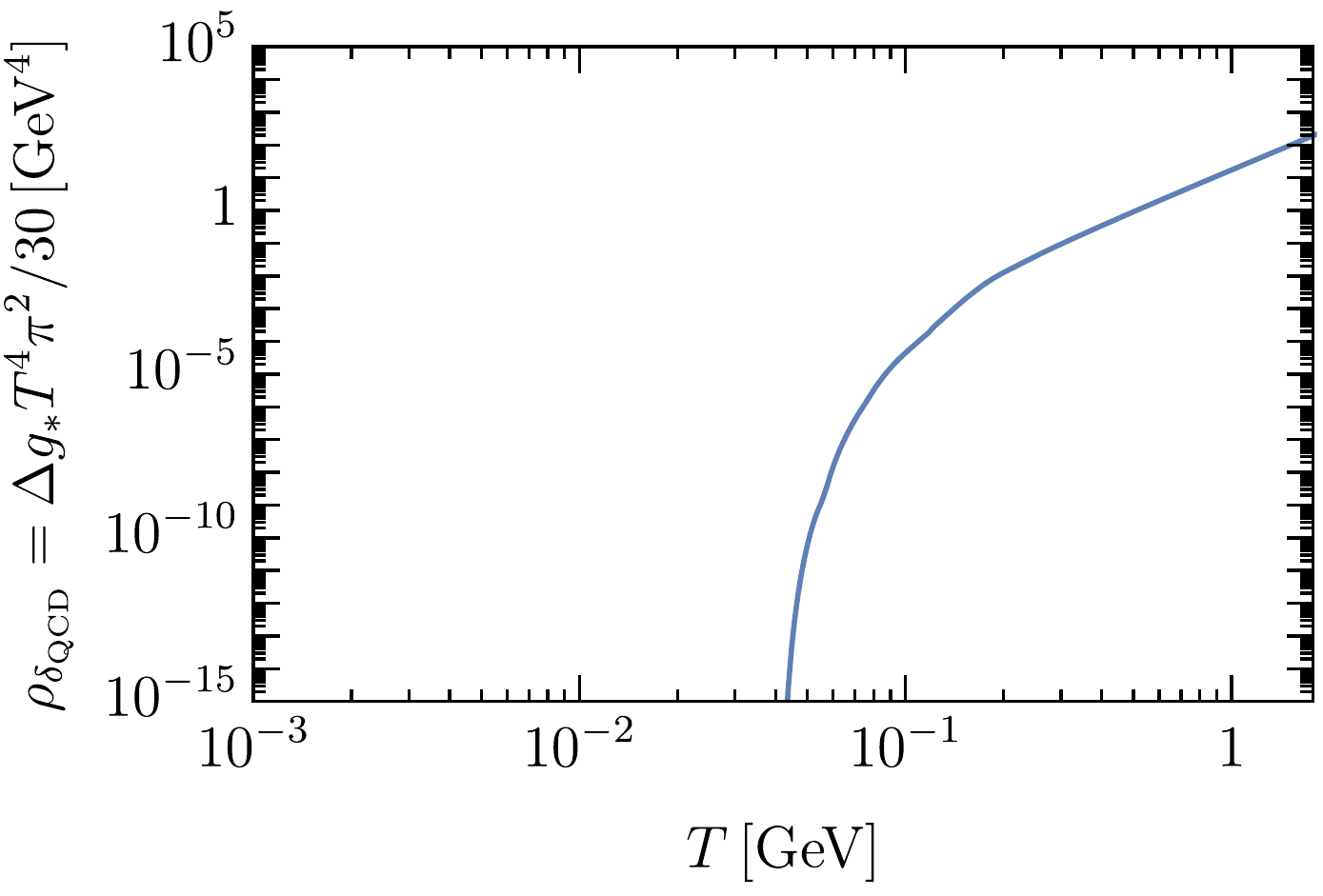}\hfill
    \includegraphics[width=.45\textwidth]{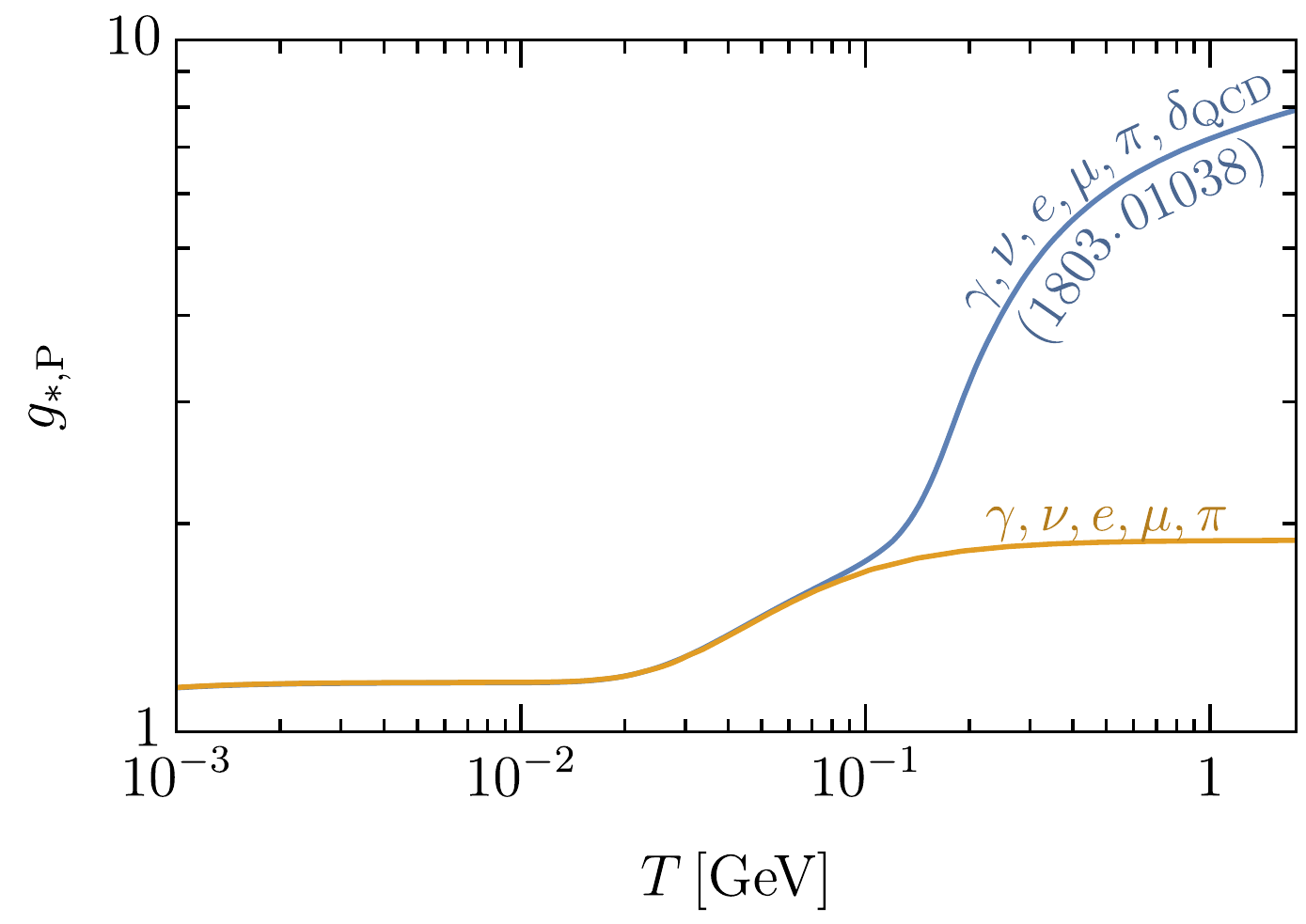}\hfill
    \includegraphics[width=.47\textwidth]{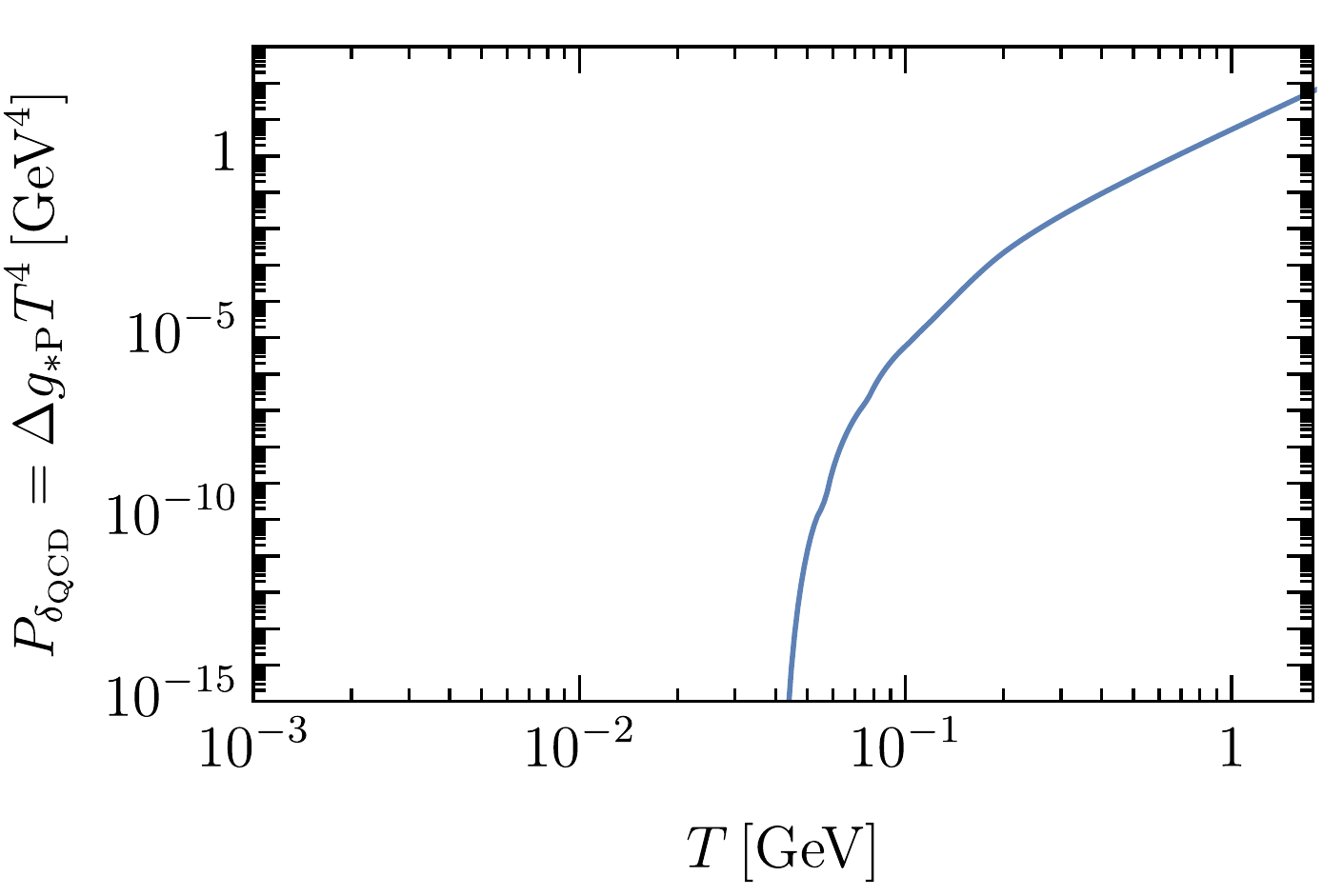}\hfill
    \caption{
    Top left: The orange contour shows $g_*$ as a function of temperature assuming an ideal gas of photons, neutrinos, electrons, muons, and pions, while the blue contour shows $g_*$ for the Standard Model including heavy mesons and deviations from the ideal gas law \cite{Saikawa:2018rcs}. Top right: $\rho_{\delta \rm QCD}$ as a function of temperature, extracted from the difference in the blue and orange $g_*$ contours on the top left panel. Bottom panels: Same as top panels but for $g_{*P} \equiv P/T^4$.
    }
    \label{fig:gstarComparison}
\end{figure}
The top left panel of Fig.~$\ref{fig:gstarComparison}$ shows $g_*$ as a function of temperature $T$. The orange contour shows the value of $g_*$ assuming an ideal gas comprised of photons, neutrinos, electrons, muons, and pions in thermal equilibrium. The blue contour, which diverges from the orange above $T \sim 100$ MeV, is the Standard Model value of $g_*$ from \cite{Saikawa:2018rcs}, which includes contributions, $\delta_{\rm QCD}$, from heavy mesons and from ideal gas law deviations. We extract $\rho_{\delta_{\rm QCD}}$ by taking the difference between the blue and orange contours, $\Delta g_*$, and multiplying this result by $\pi^2 T^4/30$, as shown by the top right panel.

Similarly, the bottom left panel of Fig.~$\ref{fig:gstarComparison}$ shows the ratio $P/T^4 = g_{*P}$ as a function of temperature $T$, where $P$ is the total pressure of the relevant species. As before, the orange contour shows the value of $g_{*P}$ assuming an ideal gas comprised of photons, neutrinos, electrons, muons, and pions in thermal equilibrium. The blue contour, which diverges from the orange above $T \sim 100$ MeV, is the Standard Model value of $g_{*P}$ from \cite{Saikawa:2018rcs}, which we infer through the relationship $g_{*P} = \frac{\pi^2}{30}(\frac{4}{3}g_{*S} - g_*)$. We extract $P_{\delta_{\rm QCD}}$ by taking the difference between the blue and orange contours, $\Delta g_{*P}$, and multiplying this result by $T^4$, as shown by the bottom right panel.

\bibliographystyle{JHEP}
\bibliography{HeavyAxion}

\providecommand{\href}[2]{#2}\begingroup\raggedright\begin{thebibliography}{10}

\bibitem{Nelson:1983zb}
A.~E. Nelson, \emph{{Naturally Weak CP Violation}},
  \href{https://doi.org/10.1016/0370-2693(84)92025-2}{\emph{Phys. Lett. B}
  {\bfseries 136} (1984) 387}.

\bibitem{Barr:1984qx}
S.~M. Barr, \emph{{Solving the Strong CP Problem Without the Peccei-Quinn
  Symmetry}}, \href{https://doi.org/10.1103/PhysRevLett.53.329}{\emph{Phys.
  Rev. Lett.} {\bfseries 53} (1984) 329}.

\bibitem{Beg:1978mt}
M.~A.~B. Beg and H.~S. Tsao, \emph{{Strong P, T Noninvariances in a Superweak
  Theory}}, \href{https://doi.org/10.1103/PhysRevLett.41.278}{\emph{Phys. Rev.
  Lett.} {\bfseries 41} (1978) 278}.

\bibitem{Mohapatra:1978fy}
R.~N. Mohapatra and G.~Senjanovic, \emph{{Natural Suppression of Strong p and t
  Noninvariance}},
  \href{https://doi.org/10.1016/0370-2693(78)90243-5}{\emph{Phys. Lett. B}
  {\bfseries 79} (1978) 283}.

\bibitem{Babu:1989rb}
K.~S. Babu and R.~N. Mohapatra, \emph{{A Solution to the Strong {CP} Problem
  Without an Axion}},
  \href{https://doi.org/10.1103/PhysRevD.41.1286}{\emph{Phys. Rev. D}
  {\bfseries 41} (1990) 1286}.

\bibitem{Peccei:1977hh}
R.~D. Peccei and H.~R. Quinn, \emph{{CP Conservation in the Presence of
  Instantons}}, \href{https://doi.org/10.1103/PhysRevLett.38.1440}{\emph{Phys.
  Rev. Lett.} {\bfseries 38} (1977) 1440}.

\bibitem{Peccei:1977ur}
R.~D. Peccei and H.~R. Quinn, \emph{{Constraints Imposed by CP Conservation in
  the Presence of Instantons}},
  \href{https://doi.org/10.1103/PhysRevD.16.1791}{\emph{Phys. Rev.} {\bfseries
  D16} (1977) 1791}.

\bibitem{Georgi:1981pu}
H.~M. Georgi, L.~J. Hall and M.~B. Wise, \emph{{Grand Unified Models With an
  Automatic {Peccei-Quinn} Symmetry}},
  \href{https://doi.org/10.1016/0550-3213(81)90433-8}{\emph{Nucl. Phys. B}
  {\bfseries 192} (1981) 409}.

\bibitem{Harlow:2018tng}
D.~Harlow and H.~Ooguri, \emph{{Symmetries in quantum field theory and quantum
  gravity}}, \href{https://doi.org/10.1007/s00220-021-04040-y}{\emph{Commun.
  Math. Phys.} {\bfseries 383} (2021) 1669}
  [\href{https://arxiv.org/abs/1810.05338}{{\ttfamily 1810.05338}}].

\bibitem{Banks:2010zn}
T.~Banks and N.~Seiberg, \emph{{Symmetries and Strings in Field Theory and
  Gravity}}, \href{https://doi.org/10.1103/PhysRevD.83.084019}{\emph{Phys. Rev.
  D} {\bfseries 83} (2011) 084019}
  [\href{https://arxiv.org/abs/1011.5120}{{\ttfamily 1011.5120}}].

\bibitem{Holman:1992us}
R.~Holman, S.~D.~H. Hsu, T.~W. Kephart, E.~W. Kolb, R.~Watkins and L.~M.
  Widrow, \emph{{Solutions to the strong CP problem in a world with gravity}},
  \href{https://doi.org/10.1016/0370-2693(92)90491-L}{\emph{Phys. Lett. B}
  {\bfseries 282} (1992) 132}
  [\href{https://arxiv.org/abs/hep-ph/9203206}{{\ttfamily hep-ph/9203206}}].

\bibitem{Barr:1992qq}
S.~M. Barr and D.~Seckel, \emph{{Planck scale corrections to axion models}},
  \href{https://doi.org/10.1103/PhysRevD.46.539}{\emph{Phys. Rev. D} {\bfseries
  46} (1992) 539}.

\bibitem{Kamionkowski:1992mf}
M.~Kamionkowski and J.~March-Russell, \emph{{Planck scale physics and the
  Peccei-Quinn mechanism}},
  \href{https://doi.org/10.1016/0370-2693(92)90492-M}{\emph{Phys. Lett. B}
  {\bfseries 282} (1992) 137}
  [\href{https://arxiv.org/abs/hep-th/9202003}{{\ttfamily hep-th/9202003}}].

\bibitem{Dine:1992vx}
M.~Dine, \emph{{Problems of naturalness: Some lessons from string theory}},  in
  \emph{{Conference on Topics in Quantum Gravity}}, 7, 1992,
  \href{https://arxiv.org/abs/hep-th/9207045}{{\ttfamily hep-th/9207045}}.

\bibitem{Choi:1992xp}
K.-w. Choi, D.~B. Kaplan and A.~E. Nelson, \emph{{Is CP a gauge symmetry?}},
  \href{https://doi.org/10.1016/0550-3213(93)90082-Z}{\emph{Nucl. Phys. B}
  {\bfseries 391} (1993) 515}
  [\href{https://arxiv.org/abs/hep-ph/9205202}{{\ttfamily hep-ph/9205202}}].

\bibitem{Dine:1992ya}
M.~Dine, R.~G. Leigh and D.~A. MacIntire, \emph{{Of CP and other gauge
  symmetries in string theory}},
  \href{https://doi.org/10.1103/PhysRevLett.69.2030}{\emph{Phys. Rev. Lett.}
  {\bfseries 69} (1992) 2030}
  [\href{https://arxiv.org/abs/hep-th/9205011}{{\ttfamily hep-th/9205011}}].

\bibitem{Choi:1985cb}
K.~Choi and J.~E. Kim, \emph{{DYNAMICAL AXION}},
  \href{https://doi.org/10.1103/PhysRevD.32.1828}{\emph{Phys. Rev. D}
  {\bfseries 32} (1985) 1828}.

\bibitem{Cadamuro:2011fd}
D.~Cadamuro and J.~Redondo, \emph{{Cosmological bounds on pseudo
  Nambu-Goldstone bosons}},
  \href{https://doi.org/10.1088/1475-7516/2012/02/032}{\emph{JCAP} {\bfseries
  02} (2012) 032} [\href{https://arxiv.org/abs/1110.2895}{{\ttfamily
  1110.2895}}].

\bibitem{Planck:2018vyg}
{\scshape Planck} collaboration, \emph{{Planck 2018 results. VI. Cosmological
  parameters}},
  \href{https://doi.org/10.1051/0004-6361/201833910}{\emph{Astron. Astrophys.}
  {\bfseries 641} (2020) A6}
  [\href{https://arxiv.org/abs/1807.06209}{{\ttfamily 1807.06209}}].

\bibitem{CMB-S4:2016ple}
{\scshape CMB-S4} collaboration, \emph{{CMB-S4 Science Book, First Edition}},
  \href{https://arxiv.org/abs/1610.02743}{{\ttfamily 1610.02743}}.

\bibitem{Chang:1993gm}
S.~Chang and K.~Choi, \emph{{Hadronic axion window and the big bang
  nucleosynthesis}},
  \href{https://doi.org/10.1016/0370-2693(93)90656-3}{\emph{Phys. Lett. B}
  {\bfseries 316} (1993) 51}
  [\href{https://arxiv.org/abs/hep-ph/9306216}{{\ttfamily hep-ph/9306216}}].

\bibitem{Ferreira:2018vjj}
R.~Z. Ferreira and A.~Notari, \emph{{Observable Windows for the QCD Axion
  Through the Number of Relativistic Species}},
  \href{https://doi.org/10.1103/PhysRevLett.120.191301}{\emph{Phys. Rev. Lett.}
  {\bfseries 120} (2018) 191301}
  [\href{https://arxiv.org/abs/1801.06090}{{\ttfamily 1801.06090}}].

\bibitem{Arias-Aragon:2020qtn}
F.~Arias-Arag\'on, F.~D'eramo, R.~Z. Ferreira, L.~Merlo and A.~Notari,
  \emph{{Cosmic Imprints of XENON1T Axions}},
  \href{https://doi.org/10.1088/1475-7516/2020/11/025}{\emph{JCAP} {\bfseries
  11} (2020) 025} [\href{https://arxiv.org/abs/2007.06579}{{\ttfamily
  2007.06579}}].

\bibitem{Arias-Aragon:2020shv}
F.~Arias-Arag\'on, F.~D'Eramo, R.~Z. Ferreira, L.~Merlo and A.~Notari,
  \emph{{Production of Thermal Axions across the ElectroWeak Phase
  Transition}},
  \href{https://doi.org/10.1088/1475-7516/2021/03/090}{\emph{JCAP} {\bfseries
  03} (2021) 090} [\href{https://arxiv.org/abs/2012.04736}{{\ttfamily
  2012.04736}}].

\bibitem{Ferreira:2020bpb}
R.~Z. Ferreira, A.~Notari and F.~Rompineve,
  \emph{{Dine-Fischler-Srednicki-Zhitnitsky axion in the CMB}},
  \href{https://doi.org/10.1103/PhysRevD.103.063524}{\emph{Phys. Rev. D}
  {\bfseries 103} (2021) 063524}
  [\href{https://arxiv.org/abs/2012.06566}{{\ttfamily 2012.06566}}].

\bibitem{DEramo:2021psx}
F.~D'Eramo, F.~Hajkarim and S.~Yun, \emph{{Thermal axion production at low
  temperatures: a smooth treatment of the QCD phase transition}},
  \href{https://arxiv.org/abs/2108.04259}{{\ttfamily 2108.04259}}.

\bibitem{DEramo:2021lgb}
F.~D'Eramo, F.~Hajkarim and S.~Yun, \emph{{Thermal QCD Axions across
  Thresholds}},  \href{https://arxiv.org/abs/2108.05371}{{\ttfamily
  2108.05371}}.

\bibitem{DEramo:2021usm}
F.~D'Eramo and S.~Yun, \emph{{Flavor violating axions in the early Universe}},
  \href{https://doi.org/10.1103/PhysRevD.105.075002}{\emph{Phys. Rev. D}
  {\bfseries 105} (2022) 075002}
  [\href{https://arxiv.org/abs/2111.12108}{{\ttfamily 2111.12108}}].

\bibitem{Hannestad:2003ye}
S.~Hannestad and G.~Raffelt, \emph{{Cosmological mass limits on neutrinos,
  axions, and other light particles}},
  \href{https://doi.org/10.1088/1475-7516/2004/04/008}{\emph{JCAP} {\bfseries
  04} (2004) 008} [\href{https://arxiv.org/abs/hep-ph/0312154}{{\ttfamily
  hep-ph/0312154}}].

\bibitem{Giare:2020vzo}
W.~Giar\`e, E.~Di~Valentino, A.~Melchiorri and O.~Mena, \emph{{New cosmological
  bounds on hot relics: axions and neutrinos}},
  \href{https://doi.org/10.1093/mnras/stab1442}{\emph{Mon. Not. Roy. Astron.
  Soc.} {\bfseries 505} (2021) 2703}
  [\href{https://arxiv.org/abs/2011.14704}{{\ttfamily 2011.14704}}].

\bibitem{Caloni:2022uya}
L.~Caloni, M.~Gerbino, M.~Lattanzi and L.~Visinelli, \emph{{Novel cosmological
  bounds on thermally-produced axion-like particles}},
  \href{https://arxiv.org/abs/2205.01637}{{\ttfamily 2205.01637}}.

\bibitem{Dimopoulos:1979pp}
S.~Dimopoulos, \emph{{A Solution of the Strong {CP} Problem in Models With
  Scalars}}, \href{https://doi.org/10.1016/0370-2693(79)91233-4}{\emph{Phys.
  Lett. B} {\bfseries 84} (1979) 435}.

\bibitem{Tye:1981zy}
S.~H.~H. Tye, \emph{{A Superstrong Force With a Heavy Axion}},
  \href{https://doi.org/10.1103/PhysRevLett.47.1035}{\emph{Phys. Rev. Lett.}
  {\bfseries 47} (1981) 1035}.

\bibitem{Holdom:1982ex}
B.~Holdom and M.~E. Peskin, \emph{{Raising the Axion Mass}},
  \href{https://doi.org/10.1016/0550-3213(82)90228-0}{\emph{Nucl. Phys. B}
  {\bfseries 208} (1982) 397}.

\bibitem{Holdom:1985vx}
B.~Holdom, \emph{{Strong QCD at High-energies and a Heavy Axion}},
  \href{https://doi.org/10.1016/0370-2693(85)90371-5}{\emph{Phys. Lett. B}
  {\bfseries 154} (1985) 316}.

\bibitem{Flynn:1987rs}
J.~M. Flynn and L.~Randall, \emph{{A Computation of the Small Instanton
  Contribution to the Axion Potential}},
  \href{https://doi.org/10.1016/0550-3213(87)90089-7}{\emph{Nucl. Phys. B}
  {\bfseries 293} (1987) 731}.

\bibitem{Agrawal:2017ksf}
P.~Agrawal and K.~Howe, \emph{{Factoring the Strong CP Problem}},
  \href{https://doi.org/10.1007/JHEP12(2018)029}{\emph{JHEP} {\bfseries 12}
  (2018) 029} [\href{https://arxiv.org/abs/1710.04213}{{\ttfamily
  1710.04213}}].

\bibitem{Gherghetta:2020keg}
T.~Gherghetta, V.~V. Khoze, A.~Pomarol and Y.~Shirman, \emph{{The Axion Mass
  from 5D Small Instantons}},
  \href{https://doi.org/10.1007/JHEP03(2020)063}{\emph{JHEP} {\bfseries 03}
  (2020) 063} [\href{https://arxiv.org/abs/2001.05610}{{\ttfamily
  2001.05610}}].

\bibitem{Rubakov:1997vp}
V.~A. Rubakov, \emph{{Grand unification and heavy axion}},
  \href{https://doi.org/10.1134/1.567390}{\emph{JETP Lett.} {\bfseries 65}
  (1997) 621} [\href{https://arxiv.org/abs/hep-ph/9703409}{{\ttfamily
  hep-ph/9703409}}].

\bibitem{Berezhiani:2000gh}
Z.~Berezhiani, L.~Gianfagna and M.~Giannotti, \emph{{Strong CP problem and
  mirror world: The Weinberg-Wilczek axion revisited}},
  \href{https://doi.org/10.1016/S0370-2693(00)01392-7}{\emph{Phys. Lett. B}
  {\bfseries 500} (2001) 286}
  [\href{https://arxiv.org/abs/hep-ph/0009290}{{\ttfamily hep-ph/0009290}}].

\bibitem{Weinberg:1977ma}
S.~Weinberg, \emph{{A New Light Boson?}},
  \href{https://doi.org/10.1103/PhysRevLett.40.223}{\emph{Phys. Rev. Lett.}
  {\bfseries 40} (1978) 223}.

\bibitem{Wilczek:1977pj}
F.~Wilczek, \emph{{Problem of Strong $P$ and $T$ Invariance in the Presence of
  Instantons}}, \href{https://doi.org/10.1103/PhysRevLett.40.279}{\emph{Phys.
  Rev. Lett.} {\bfseries 40} (1978) 279}.

\bibitem{Fukuda:2015ana}
H.~Fukuda, K.~Harigaya, M.~Ibe and T.~T. Yanagida, \emph{{Model of visible QCD
  axion}}, \href{https://doi.org/10.1103/PhysRevD.92.015021}{\emph{Phys. Rev.
  D} {\bfseries 92} (2015) 015021}
  [\href{https://arxiv.org/abs/1504.06084}{{\ttfamily 1504.06084}}].

\bibitem{Hook:2019qoh}
A.~Hook, S.~Kumar, Z.~Liu and R.~Sundrum, \emph{{High Quality QCD Axion and the
  LHC}}, \href{https://doi.org/10.1103/PhysRevLett.124.221801}{\emph{Phys. Rev.
  Lett.} {\bfseries 124} (2020) 221801}
  [\href{https://arxiv.org/abs/1911.12364}{{\ttfamily 1911.12364}}].

\bibitem{Kelly:2020dda}
K.~J. Kelly, S.~Kumar and Z.~Liu, \emph{{Heavy axion opportunities at the DUNE
  near detector}},
  \href{https://doi.org/10.1103/PhysRevD.103.095002}{\emph{Phys. Rev. D}
  {\bfseries 103} (2021) 095002}
  [\href{https://arxiv.org/abs/2011.05995}{{\ttfamily 2011.05995}}].

\bibitem{Kim:1979if}
J.~E. Kim, \emph{{Weak Interaction Singlet and Strong CP Invariance}},
  \href{https://doi.org/10.1103/PhysRevLett.43.103}{\emph{Phys. Rev. Lett.}
  {\bfseries 43} (1979) 103}.

\bibitem{Shifman:1979if}
M.~A. Shifman, A.~I. Vainshtein and V.~I. Zakharov, \emph{{Can Confinement
  Ensure Natural CP Invariance of Strong Interactions?}},
  \href{https://doi.org/10.1016/0550-3213(80)90209-6}{\emph{Nucl. Phys. B}
  {\bfseries 166} (1980) 493}.

\bibitem{Millea:2015qra}
M.~Millea, L.~Knox and B.~Fields, \emph{{New Bounds for Axions and Axion-Like
  Particles with keV-GeV Masses}},
  \href{https://doi.org/10.1103/PhysRevD.92.023010}{\emph{Phys. Rev. D}
  {\bfseries 92} (2015) 023010}
  [\href{https://arxiv.org/abs/1501.04097}{{\ttfamily 1501.04097}}].

\bibitem{Depta:2020wmr}
P.~F. Depta, M.~Hufnagel and K.~Schmidt-Hoberg, \emph{{Robust cosmological
  constraints on axion-like particles}},
  \href{https://doi.org/10.1088/1475-7516/2020/05/009}{\emph{JCAP} {\bfseries
  05} (2020) 009} [\href{https://arxiv.org/abs/2002.08370}{{\ttfamily
  2002.08370}}].

\bibitem{Aloni:2018vki}
D.~Aloni, Y.~Soreq and M.~Williams, \emph{{Coupling QCD-Scale Axionlike
  Particles to Gluons}},
  \href{https://doi.org/10.1103/PhysRevLett.123.031803}{\emph{Phys. Rev. Lett.}
  {\bfseries 123} (2019) 031803}
  [\href{https://arxiv.org/abs/1811.03474}{{\ttfamily 1811.03474}}].

\bibitem{ParticleDataGroup:2020ssz}
{\scshape Particle Data Group} collaboration, \emph{{Review of Particle
  Physics}}, \href{https://doi.org/10.1093/ptep/ptaa104}{\emph{PTEP} {\bfseries
  2020} (2020) 083C01}.

\bibitem{Salvio:2013iaa}
A.~Salvio, A.~Strumia and W.~Xue, \emph{{Thermal axion production}},
  \href{https://doi.org/10.1088/1475-7516/2014/01/011}{\emph{JCAP} {\bfseries
  01} (2014) 011} [\href{https://arxiv.org/abs/1310.6982}{{\ttfamily
  1310.6982}}].

\bibitem{DiLuzio:2021vjd}
L.~Di~Luzio, G.~Martinelli and G.~Piazza, \emph{{Breakdown of chiral
  perturbation theory for the axion hot dark matter bound}},
  \href{https://doi.org/10.1103/PhysRevLett.126.241801}{\emph{Phys. Rev. Lett.}
  {\bfseries 126} (2021) 241801}
  [\href{https://arxiv.org/abs/2101.10330}{{\ttfamily 2101.10330}}].

\bibitem{Graf:2010tv}
P.~Graf and F.~D. Steffen, \emph{{Thermal axion production in the primordial
  quark-gluon plasma}},
  \href{https://doi.org/10.1103/PhysRevD.83.075011}{\emph{Phys. Rev. D}
  {\bfseries 83} (2011) 075011}
  [\href{https://arxiv.org/abs/1008.4528}{{\ttfamily 1008.4528}}].

\bibitem{Cadamuro:2010cz}
D.~Cadamuro, S.~Hannestad, G.~Raffelt and J.~Redondo, \emph{{Cosmological
  bounds on sub-MeV mass axions}},
  \href{https://doi.org/10.1088/1475-7516/2011/02/003}{\emph{JCAP} {\bfseries
  02} (2011) 003} [\href{https://arxiv.org/abs/1011.3694}{{\ttfamily
  1011.3694}}].

\bibitem{weldon1982covariant}
H.~A. Weldon, \emph{Covariant calculations at finite temperature: the
  relativistic plasma}, {\emph{Physical Review D} {\bfseries 26} (1982) 1394}.

\bibitem{Hannestad_2005}
S.~Hannestad, A.~Mirizzi and G.~Raffelt, \emph{A new cosmological mass limit on
  thermal relic axions},
  \href{https://doi.org/10.1088/1475-7516/2005/07/002}{\emph{Journal of
  Cosmology and Astroparticle Physics} {\bfseries 2005} (2005) 002}.

\bibitem{Saikawa:2018rcs}
K.~Saikawa and S.~Shirai, \emph{{Primordial gravitational waves, precisely: The
  role of thermodynamics in the Standard Model}},
  \href{https://doi.org/10.1088/1475-7516/2018/05/035}{\emph{JCAP} {\bfseries
  05} (2018) 035} [\href{https://arxiv.org/abs/1803.01038}{{\ttfamily
  1803.01038}}].

\bibitem{methodOfLines}
``{The Numerical Method of Lines}.''
  {\url{https://reference.wolfram.com/language/tutorial/NDSolveMethodOfLines.html}}.

\bibitem{Ertas:2020xcc}
F.~Ertas and F.~Kahlhoefer, \emph{{On the interplay between astrophysical and
  laboratory probes of MeV-scale axion-like particles}},
  \href{https://doi.org/10.1007/JHEP07(2020)050}{\emph{JHEP} {\bfseries 07}
  (2020) 050} [\href{https://arxiv.org/abs/2004.01193}{{\ttfamily
  2004.01193}}].

\bibitem{Dobrich:2015jyk}
B.~D\"obrich, J.~Jaeckel, F.~Kahlhoefer, A.~Ringwald and K.~Schmidt-Hoberg,
  \emph{{ALPtraum: ALP production in proton beam dump experiments}},
  \href{https://doi.org/10.1007/JHEP02(2016)018}{\emph{JHEP} {\bfseries 02}
  (2016) 018} [\href{https://arxiv.org/abs/1512.03069}{{\ttfamily
  1512.03069}}].

\bibitem{Dolan:2017osp}
M.~J. Dolan, T.~Ferber, C.~Hearty, F.~Kahlhoefer and K.~Schmidt-Hoberg,
  \emph{{Revised constraints and Belle II sensitivity for visible and invisible
  axion-like particles}},
  \href{https://doi.org/10.1007/JHEP12(2017)094}{\emph{JHEP} {\bfseries 12}
  (2017) 094} [\href{https://arxiv.org/abs/1709.00009}{{\ttfamily
  1709.00009}}].

\bibitem{NA64:2020qwq}
{\scshape NA64} collaboration, \emph{{Search for Axionlike and Scalar Particles
  with the NA64 Experiment}},
  \href{https://doi.org/10.1103/PhysRevLett.125.081801}{\emph{Phys. Rev. Lett.}
  {\bfseries 125} (2020) 081801}
  [\href{https://arxiv.org/abs/2005.02710}{{\ttfamily 2005.02710}}].

\bibitem{FASER:2018eoc}
{\scshape FASER} collaboration, \emph{{FASER\textquoteright{}s physics reach
  for long-lived particles}},
  \href{https://doi.org/10.1103/PhysRevD.99.095011}{\emph{Phys. Rev. D}
  {\bfseries 99} (2019) 095011}
  [\href{https://arxiv.org/abs/1811.12522}{{\ttfamily 1811.12522}}].

\bibitem{Gori:2020xvq}
S.~Gori, G.~Perez and K.~Tobioka, \emph{{KOTO vs. NA62 Dark Scalar Searches}},
  \href{https://doi.org/10.1007/JHEP08(2020)110}{\emph{JHEP} {\bfseries 08}
  (2020) 110} [\href{https://arxiv.org/abs/2005.05170}{{\ttfamily
  2005.05170}}].

\bibitem{Mariotti:2017vtv}
A.~Mariotti, D.~Redigolo, F.~Sala and K.~Tobioka, \emph{{New LHC bound on
  low-mass diphoton resonances}},
  \href{https://doi.org/10.1016/j.physletb.2018.06.039}{\emph{Phys. Lett. B}
  {\bfseries 783} (2018) 13}
  [\href{https://arxiv.org/abs/1710.01743}{{\ttfamily 1710.01743}}].

\bibitem{Chakraborty:2021wda}
S.~Chakraborty, M.~Kraus, V.~Loladze, T.~Okui and K.~Tobioka, \emph{{Heavy QCD
  axion in b\textrightarrow{}s transition: Enhanced limits and projections}},
  \href{https://doi.org/10.1103/PhysRevD.104.055036}{\emph{Phys. Rev. D}
  {\bfseries 104} (2021) 055036}
  [\href{https://arxiv.org/abs/2102.04474}{{\ttfamily 2102.04474}}].

\bibitem{Bertholet:2021hjl}
E.~Bertholet, S.~Chakraborty, V.~Loladze, T.~Okui, A.~Soffer and K.~Tobioka,
  \emph{{Heavy QCD axion at Belle II: Displaced and prompt signals}},
  \href{https://doi.org/10.1103/PhysRevD.105.L071701}{\emph{Phys. Rev. D}
  {\bfseries 105} (2022) L071701}
  [\href{https://arxiv.org/abs/2108.10331}{{\ttfamily 2108.10331}}].

\bibitem{Dunsky:2019upk}
D.~Dunsky, L.~J. Hall and K.~Harigaya, \emph{{Dark Matter, Dark Radiation and
  Gravitational Waves from Mirror Higgs Parity}},
  \href{https://doi.org/10.1007/JHEP02(2020)078}{\emph{JHEP} {\bfseries 02}
  (2020) 078} [\href{https://arxiv.org/abs/1908.02756}{{\ttfamily
  1908.02756}}].

\bibitem{Dunsky:2018mqs}
D.~Dunsky, L.~J. Hall and K.~Harigaya, \emph{{CHAMP Cosmic Rays}},
  \href{https://doi.org/10.1088/1475-7516/2019/07/015}{\emph{JCAP} {\bfseries
  07} (2019) 015} [\href{https://arxiv.org/abs/1812.11116}{{\ttfamily
  1812.11116}}].

\bibitem{Xu:2021rwg}
W.~L. Xu, J.~B. Mu\~noz and C.~Dvorkin, \emph{{Cosmological Constraints on
  Light (but Massive) Relics}},
  \href{https://arxiv.org/abs/2107.09664}{{\ttfamily 2107.09664}}.

\bibitem{ciaran_o_hare_2020_3932430}
C.~O'HARE, \emph{cajohare/axionlimits: Axionlimits},  July, 2020.
\newblock 10.5281/zenodo.3932430.

\bibitem{Preskill:1982cy}
J.~Preskill, M.~B. Wise and F.~Wilczek, \emph{{Cosmology of the Invisible
  Axion}}, \href{https://doi.org/10.1016/0370-2693(83)90637-8}{\emph{Phys.
  Lett. B} {\bfseries 120} (1983) 127}.

\bibitem{Abbott:1982af}
L.~F. Abbott and P.~Sikivie, \emph{{A Cosmological Bound on the Invisible
  Axion}}, \href{https://doi.org/10.1016/0370-2693(83)90638-X}{\emph{Phys.
  Lett. B} {\bfseries 120} (1983) 133}.

\bibitem{Dine:1982ah}
M.~Dine and W.~Fischler, \emph{{The Not So Harmless Axion}},
  \href{https://doi.org/10.1016/0370-2693(83)90639-1}{\emph{Phys. Lett. B}
  {\bfseries 120} (1983) 137}.

\bibitem{Chang:2018rso}
J.~H. Chang, R.~Essig and S.~D. McDermott, \emph{{Supernova 1987A Constraints
  on Sub-GeV Dark Sectors, Millicharged Particles, the QCD Axion, and an
  Axion-like Particle}},
  \href{https://doi.org/10.1007/JHEP09(2018)051}{\emph{JHEP} {\bfseries 09}
  (2018) 051} [\href{https://arxiv.org/abs/1803.00993}{{\ttfamily
  1803.00993}}].

\bibitem{Ayala:2014pea}
A.~Ayala, I.~Dom\'\i{}nguez, M.~Giannotti, A.~Mirizzi and O.~Straniero,
  \emph{{Revisiting the bound on axion-photon coupling from Globular
  Clusters}}, \href{https://doi.org/10.1103/PhysRevLett.113.191302}{\emph{Phys.
  Rev. Lett.} {\bfseries 113} (2014) 191302}
  [\href{https://arxiv.org/abs/1406.6053}{{\ttfamily 1406.6053}}].

\bibitem{CAST:2007jps}
{\scshape CAST} collaboration, \emph{{An Improved limit on the axion-photon
  coupling from the CAST experiment}},
  \href{https://doi.org/10.1088/1475-7516/2007/04/010}{\emph{JCAP} {\bfseries
  04} (2007) 010} [\href{https://arxiv.org/abs/hep-ex/0702006}{{\ttfamily
  hep-ex/0702006}}].

\bibitem{CAST:2017uph}
{\scshape CAST} collaboration, \emph{{New CAST Limit on the Axion-Photon
  Interaction}}, \href{https://doi.org/10.1038/nphys4109}{\emph{Nature Phys.}
  {\bfseries 13} (2017) 584}
  [\href{https://arxiv.org/abs/1705.02290}{{\ttfamily 1705.02290}}].

\bibitem{Vinyoles:2015aba}
N.~Vinyoles, A.~Serenelli, F.~L. Villante, S.~Basu, J.~Redondo and J.~Isern,
  \emph{{New axion and hidden photon constraints from a solar data global
  fit}}, \href{https://doi.org/10.1088/1475-7516/2015/10/015}{\emph{JCAP}
  {\bfseries 10} (2015) 015}
  [\href{https://arxiv.org/abs/1501.01639}{{\ttfamily 1501.01639}}].

\bibitem{Kawasaki:1999na}
M.~Kawasaki, K.~Kohri and N.~Sugiyama, \emph{{Cosmological constraints on late
  time entropy production}},
  \href{https://doi.org/10.1103/PhysRevLett.82.4168}{\emph{Phys. Rev. Lett.}
  {\bfseries 82} (1999) 4168}
  [\href{https://arxiv.org/abs/astro-ph/9811437}{{\ttfamily
  astro-ph/9811437}}].

\bibitem{Kawasaki:2000en}
M.~Kawasaki, K.~Kohri and N.~Sugiyama, \emph{{MeV scale reheating temperature
  and thermalization of neutrino background}},
  \href{https://doi.org/10.1103/PhysRevD.62.023506}{\emph{Phys. Rev. D}
  {\bfseries 62} (2000) 023506}
  [\href{https://arxiv.org/abs/astro-ph/0002127}{{\ttfamily
  astro-ph/0002127}}].

\bibitem{Hasegawa:2019jsa}
T.~Hasegawa, N.~Hiroshima, K.~Kohri, R.~S.~L. Hansen, T.~Tram and S.~Hannestad,
  \emph{{MeV-scale reheating temperature and thermalization of oscillating
  neutrinos by radiative and hadronic decays of massive particles}},
  \href{https://doi.org/10.1088/1475-7516/2019/12/012}{\emph{JCAP} {\bfseries
  12} (2019) 012} [\href{https://arxiv.org/abs/1908.10189}{{\ttfamily
  1908.10189}}].

\bibitem{Hannestad:1995rs}
S.~Hannestad and J.~Madsen, \emph{{Neutrino decoupling in the early universe}},
  \href{https://doi.org/10.1103/PhysRevD.52.1764}{\emph{Phys. Rev. D}
  {\bfseries 52} (1995) 1764}
  [\href{https://arxiv.org/abs/astro-ph/9506015}{{\ttfamily
  astro-ph/9506015}}].

\bibitem{Hannestad:2005df}
S.~Hannestad, A.~Mirizzi and G.~Raffelt, \emph{{New cosmological mass limit on
  thermal relic axions}},
  \href{https://doi.org/10.1088/1475-7516/2005/07/002}{\emph{JCAP} {\bfseries
  07} (2005) 002} [\href{https://arxiv.org/abs/hep-ph/0504059}{{\ttfamily
  hep-ph/0504059}}].

\bibitem{Bennett:2020zkv}
J.~J. Bennett, G.~Buldgen, P.~F. De~Salas, M.~Drewes, S.~Gariazzo, S.~Pastor
  et~al., \emph{{Towards a precision calculation of $N_{\rm eff}$ in the
  Standard Model II: Neutrino decoupling in the presence of flavour
  oscillations and finite-temperature QED}},
  \href{https://doi.org/10.1088/1475-7516/2021/04/073}{\emph{JCAP} {\bfseries
  04} (2021) 073} [\href{https://arxiv.org/abs/2012.02726}{{\ttfamily
  2012.02726}}].

\bibitem{Froustey:2020mcq}
J.~Froustey, C.~Pitrou and M.~C. Volpe, \emph{{Neutrino decoupling including
  flavour oscillations and primordial nucleosynthesis}},
  \href{https://doi.org/10.1088/1475-7516/2020/12/015}{\emph{JCAP} {\bfseries
  12} (2020) 015} [\href{https://arxiv.org/abs/2008.01074}{{\ttfamily
  2008.01074}}].

\bibitem{Zyla:2020zbs}
{\scshape Particle Data Group} collaboration, \emph{{Review of Particle
  Physics}}, \href{https://doi.org/10.1093/ptep/ptaa104}{\emph{PTEP} {\bfseries
  2020} (2020) 083C01}.

\end{thebibliography}\endgroup

\end{document}